\documentclass[lettersize,journal]{IEEEtran}
\normalsize
\usepackage{comment}
\usepackage{xcolor}
\usepackage{graphicx}
\usepackage{lipsum}
\usepackage{hyperref}
\usepackage[caption=false]{subfig}
\usepackage{cite}
\usepackage{pifont}
\usepackage{enumitem}

\usepackage{amsmath}
\usepackage{atbegshi}% http://ctan.org/pkg/atbegshi
\usepackage{algorithmic}
\usepackage[ruled, vlined,linesnumbered]{algorithm2e}
\usepackage{bbding}
\usepackage{pifont}
\usepackage{wasysym}
\usepackage{amssymb}
\def\BibTeX{{\rm B\kern-.05em{\sc i\kern-.025em b}\kern-.08em
    T\kern-.1667em\lower.7ex\hbox{E}\kern-.125emX}}
\usepackage{balance}
\begin{document}

%
%----------------TITLE--------------------------------------
\title{\huge{Data-Aided CSI Estimation Using Affine-Precoded Superimposed Pilots in Orthogonal Time Frequency Space Modulated MIMO Systems}}
%-----------------------------------------------------------
%----------------AUTHORS NAME-------------------------------
\author{{Anand~Mehrotra{~\IEEEmembership{Student Member,~IEEE}}, Suraj~Srivastava{~\IEEEmembership{Member,~IEEE}}, 
Aditya~K.~Jagannatham{~\IEEEmembership{Senior~Member,~IEEE}},
Lajos~Hanzo{~\IEEEmembership{Life~Fellow,~IEEE}}}
%----------------------------------------------------------- %-------------------THANKS----------------------------------
\thanks{L. Hanzo would like to acknowledge the financial support of the 
Engineering and Physical Sciences Research Council projects EP/W016605/1 
and EP/X01228X/1 as well as of the European Research Council's Advanced 
Fellow Grant QuantCom (Grant No. 789028). The work of Aditya K. Jagannatham was supported in part by the Qualcomm Innovation Fellowship, and in part by the Arun Kumar Chair.\\
A.Mehrotra, S. Srivastava and A. K. Jagannatham are with the Department of Electrical Engineering, Indian Institute of Technology Kanpur, UP 208016, India (e-mail: anandme@iitk.ac.in, ssrivast@iitk.ac.in, adityaj@iitk.ac.in).\\ L. Hanzo is with the School of Electronics and Computer Science, University of Southampton, Southampton SO17 1BJ, U.K. (e-mail: lh@ecs.soton.ac.uk).
}}
%------------------------------------------------------------
\maketitle
\begin{abstract}
An orthogonal affine-precoded superimposed pilot (AP-SIP)-based architecture is developed for the cyclic prefix (CP)-aided single input single output (SISO) and multiple input multiple output (MIMO) orthogonal time frequency space (OTFS) systems relying on  arbitrary transmitter-receiver (Tx-Rx) pulse shaping. The data and pilot symbol matrices are affine-precoded and superimposed in the delay Doppler (DD)-domain followed by the development of an end-to-end DD-domain relationship for the input-output symbols. At the receiver, the decoupled pilot and data symbol are extracted by employing orthogonal precoder matrices, which eliminates the mutual interference. Furthermore, a novel pilot-aided Bayesian learning (PA-BL) technique is conceived for the channel state information (CSI) estimation of SISO OTFS systems based on the expectation-maximization (EM) technique. Subsequently, a data-aided Bayesian learning (DA-BL)-based joint CSI estimation and data detection technique is proposed, which beneficially harnesses the estimated data symbols for improved CSI estimation. In this scenario our sophisticated data detection rule also integrates the CSI uncertainty of channel estimation into our the linear  minimum mean square error (LMMSE) detectors. The AP-SIP framework is also extended to MIMO OTFS systems, wherein the DD-domain input matrix is affine-precoded for each transmit antenna (TA). Then an EM algorithm-based PA-BL scheme is derived for simultaneous row-group sparse CSI estimation for this system, followed also by our data-aided DA-BL scheme that performs joint CSI estimation and data detection. Moreover, the Bayesian Cramer-Rao bounds (BCRBs) are also derived for both SISO as well as MIMO OTFS systems. Finally, simulation results are presented for characterizing the performance of the proposed CSI estimation techniques in a range of typical settings along with their bit error rate (BER) performance in comparison to an ideal system having perfect CSI.
\end{abstract}
%------------------------------------------------------------
%-------------------KEYWORDS---------------------------------
%------------------------------------------------------------
%\IEEEpeerreviewmaketitle
%\IEEEPARstart 
%-------------------INTRODUCTION-----------------------------
\section{Introduction} \label{sec:intro_section}Next-generation wireless communication systems are expected to support reliable communication in high (350-400 Km/Hr) \cite{high_speed1}, \cite{high_speed2} to extremely high mobility (800-900 Km/Hr) \cite{high_speed3} scenarios, which poses significant challenges. In such high mobility use cases, the existing multicarrier modulation (MCM) schemes such as orthogonal frequency division multiplexing (OFDM) experience significant performance degradation owing to the inter-carrier interference (ICI) that arises due to the high Doppler shift.

To overcome this challenge, Hadani \textit{et al}. proposed the novel orthogonal time frequency space (OTFS) modulation technique \cite{hadani2017orthogonal}, which relies on the delay and Doppler (DD)-domain, rather than on the conventional time-frequency (TF)-domain for modulation of the information symbols. This leads to significantly improved performance in high Doppler systems, since the DD-domain channel tends to remain approximately constant for a much larger duration than its conventional TF-domain counterpart, potentially facilitating improved channel state information (CSI) estimation and prediction. As a result, OTFS has gained immense popularity toward application in diverse domains such as radar \cite{application2019otfs_radar}, the massive Internet of Things (mIoT) \cite{application2021otfs_iot}, and in the vehicle-to-everything (V2X)\cite{applicationotfs_v2x} scenarios. Needless to say, the availability of accurate CSI is key toward realizing the performance benefits promised by such systems. Therefore, several researchers have developed CSI estimation techniques for unleashing the full potential of OTFS. These contributions are reviewed next.
%-------------------LITERATURE REVIEW STARTS----------------
\subsection{Literature Review}
The seminal work in \cite{monk2016otfs} developed an explicit end-to-end input-output DD-domain model for a single input and single output (SISO) OTFS system. Bespoke pilot transmission scheme were conceived for DD-domain CSI estimation in \cite{hadani2017orthogonal}, \cite{mehrotra_anand}, \cite{Ramachandran2020OTFSAN}. However, a major drawback of these techniques is that they use an entire OTFS frame to transmit the pilot symbols, which leads to a high pilot overhead. Moreover, the quality of the CSI estimate acquired by this procedure is very sensitive to the value of the associated empirically chosen threshold. Nonetheless, this approach has been extended to multiple input and multiple output (MIMO) OTFS systems by the authors of \cite{MIMO_OTFS_CHOCKALINGAM}, wherein the pilot impulses used for the different transmit antennas (TAs) are sufficiently well separated to result in interference-free outputs at the receive antennas (RAs) in the DD-domain.
However, the pilot overhead required by this scheme is once again excessive due to the substantial guard interval required between the impulses to avoid their interference.
Raviteja \textit{et al.}, in \cite{raviteja2018embedded}, designed a further advanced scheme where the data symbols, pilots, and guard bands are embedded in the same OTFS frame. The interference between the pilot and data symbols is avoided by using suitable guard bands. An important benefit of the embedded design proposed in \cite{raviteja2018embedded} is that both CSI estimation and data detection are performed using the same OTFS frame. Although the scheme of \cite{raviteja2018embedded} is capable of successfully reducing the overhead in comparison to the impulse pilot-based scheme, the pilots and guard intervals together still lead to significant throughput loss. 

Then Shan \textit{et al.} \cite{shan_otfs_antenna_array}, conceived a scheme that utilized a large-scale antenna array at the receiver to decrease the guard bands, resulting in reduced overhead. Their approach involved the reception of signals at various angles by RAs, followed by a decoupling stage to extract multiple parallel streams. However, compensating for phase became complicated due to the signal rotation in delay and Doppler coordinates caused by different angles of arrival, especially when data and pilot signals were superimposed. A key shortcoming of both the impulse based and embedded pilot schemes discussed is that the required overhead is high. Therefore, the efficiency is reduced. Furthermore, the performance of the aforementioned schemes heavily depends on a threshold that has to be tuned empirically.
\begin{table*}[t]
\begin{center}
\caption{Contrasting our novelty to the existing techniques}
\label{table:1}
\begin{tabular}{|l|c|c|c|c|c|c|c|c|c|c|}\hline
Features 
&\cite{Ramachandran2020OTFSAN} 
&\cite{MIMO_OTFS_CHOCKALINGAM}
&\cite{raviteja2018embedded} 
&\cite{prem2021superimposed} 
&\cite{yuan2021superimposed} 
&\cite{data_aided_OFDM}*
&\cite{suraj2021bayesian}
&\cite{suraj2021row_group}
&\cite{Data_aided_MIMO_OFDM}*
& Proposed\\ 
\hline
OTFS & \checkmark & \checkmark & \checkmark & \checkmark & \checkmark &  & \checkmark & \checkmark &  & \checkmark \\
\hline 
MIMO system & \checkmark & \checkmark &  &  &   & & & & \checkmark & \checkmark\\
\hline
Affine precoding &  &  &  &  &   & & & & \checkmark & \checkmark \\
\hline
Superimposed signalling &  &  &  &  \checkmark & \checkmark &   & & & \checkmark & \checkmark \\ 
\hline
Flexible pilot slot &  &  &  &  & \checkmark & &\checkmark &\checkmark&\checkmark&\checkmark\\ 
\hline
Limited pilot subcarriers & & \checkmark
&  & \checkmark &\checkmark & \checkmark &  &\checkmark &  &\checkmark\\
\hline
CSI estimation in domain & DD & DD & DD & DD  & DD & F & TF & TF & F & DD \\
\hline
Prior information (Dominant paths) & A & A & NA  & A & A & A & NA & NA & A & NA \\
\hline
Simultaneous row-group sparse & &  &   &  &  &  &  & \checkmark &  & \checkmark \\ 
\hline
Data aided CSI estimation &  &  & &  \checkmark & \checkmark & \checkmark & &  & \checkmark & \checkmark\\
\hline
CSI uncertainty for data detection
&  &  &  & &  &  & &  & \checkmark & \checkmark\\
\hline 
ZF based modified detection rule &  &  &  & & &  & &  & \checkmark & \checkmark\\
\hline
MMSE based modified detection rule &  &  & & &  &  & &  & & \checkmark\\
\hline
Tx-Rx Pulse Shaping & I & I & P & P & I & I & P & P & I & P\\
\hline
\multicolumn{11}{l}{Note: * Considers OFDM scenario, A: Available, NA: Not available, P: Practical, I: Ideal}
\end{tabular}
\end{center}
\end{table*}
The accuracy of pilot-based CSI estimation schemes can be significantly improved by superposing the data and pilots, as shown by the authors of \cite{prem2021superimposed} and \cite{yuan2021superimposed}. {A pivotal drawback of the techniques described in \cite{prem2021superimposed}, \cite{yuan2021superimposed}, which are based on direct superposition of the data and pilot symbols, is the resultant mutual interference that degrades the performance. Moreover, the above studies employ the MAP and message passing-based detection rules, respectively, and require the number of active dominant reflectors to be known at the receiver, which renders them impractical. More importantly, these existing works have not considered MIMO OTFS systems, which are addressed in this treatise}. An alternative technique of superimposing data over pilots is to employ the affine precoding principle, which was proposed by Tran \textit{et al.} \cite{affine_precoding_MIMO}, \cite{superimposed_affine_precoding}. A major advantage of using this approach is that the interference arising due to the superposition of the data can be removed entirely at any SNR, thanks to the orthogonality of precoder matrices. However, the schemes proposed in \cite{affine_precoding_MIMO}, \cite{superimposed_affine_precoding} consider a time-invariant channel and are hence not suitable for high Doppler systems, in which the channel varies rapidly in time. 

One of the major constraints in adopting such techniques in OTFS-based systems is the need to design hardware that is also compatible with the existing systems. The authors of \cite{ofdmbasedotfs2018}, \cite{Ahmad2018MIMO_OFDMbasedOTFS} have successfully addressed this issue via development of OFDM-based OTFS systems. Their solution is capable of realizing OTFS based on current OFDM hardware by applying bespoke pre- and post-processing blocks, which renders it remarkably easy to implement in practical systems. 
Consequently, we propose an affine-precoded-superimposed pilots (AP-SIP)-based DD-domain technique for CSI estimation in both SISO and MIMO OTFS systems. Another novelty of our work is that we exploit the DD-domain sparsity arising due to having a low number of active reflectors, which can result in significant performance improvement, without requiring prior statistical knowledge of the channel. {Recently, the authors of \cite{ai_chokalingam,ai2,ai3} demonstrated the excellent gains achieved by AI-based approaches in terms of improving the accuracy of OTFS channel estimation. These promising schemes can be further explored in our future works}. Table I provides a tabular representation of the contributions of the various works discussed above in relation to the current one. The main contributions of this paper are defined next in a clear and structured format.
%------------------------------------------------------------
%-------------CONTRIBUTION STARTS----------------------------
\subsection{Contributions}
\begin{itemize}
\item This study develops an AP-SIP model for CP-aided SISO OTFS systems along with arbitrary Tx-Rx pulse shaping. As a consequence of orthogonal affine precoding, the pilots and data outputs can be readily separated at the receiver by post-multiplying the received signal with the respective precoder matrices, which suppresses the mutual interference.
\item A novel pilot-aided Bayesian learning (PA-BL) technique is conceived for SISO OTFS systems that exploits the sparsity of the DD-domain channel for improved CSI estimation. This framework employs the classic expectation-maximization (EM) framework that updates the DD-domain CSI iteratively and has a much lower complexity than the conventional maximum likelihood (ML) technique.
\item A data-aided joint CSI estimation and data detection technique, termed as DA-BL, is also proposed for this system, which employs a modified data detection rule together with the BL principle for joint channel estimation and data detection in order to further improve the accuracy of the channel estimate obtained. The improved LMMSE receiver proposed for this system also integrates the CSI uncertainty arising due to channel estimation into the data detection process.
\item {A data-aided joint CSI estimation and data detection technique, termed as DA-BL, is also proposed for this system, which employs a modified decoding rule together with the BL principle, and further improves the accuracy of the channel estimate. The LMMSE receiver developed for this system also integrates the CSI uncertainty arising due to estimation imperfections into the data detection process.}
\item {Additionally, a simultaneous row-group sparse DA-BL scheme is also developed for joint CSI estimation and data detection in MIMO OTFS systems. This algorithm maximizes the accuracy of CSI estimation by leveraging the data symbols decoded via an improved MMSE rule that accounts for the CSI imperfections}.
\item In addition, Bayesian Cramer-Rao bounds (BCRBs) have been derived for both SISO and MIMO OTFS systems in order to benchmark the CSI estimation performance.
%\item To demonstrate the efficacy of the suggested methods, extensive simulation results are shown for CSI estimation and data detection in AP-SIP SISO and MIMO OTFS systems under different settings.
\end{itemize}
\subsection{Organization of the Paper}
The remaining sections of the paper are structured as follows.
Section \ref{sec:siso_sys_mdl} proposes the AP-SIP SISO OTFS model. Section-\ref{sec:sparse_estimation_model_siso} formulates the CSI estimation problem of SISO OTFS systems and develops the PA-BL scheme that exclusively uses the pilot symbols and subsequently the DA-BL scheme that also exploits the data symbols in addition to pilots. Section-\ref{sec:mimo_sys_mdl} presents the AP-SIP MIMO OTFS CSI estimation model, followed by the PA-BL and DA-BL schemes that exploit the inherent simultaneous row-group sparsity for CSI estimation in Section-\ref{sec:sparse_estimation_model_mimo}. Furthermore, the Bayesian Cramer-Rao bounds of both SISO and MIMO AP-SIP-based OTFS channel estimation are derived in Section-\ref{sec:BCRB} to benchmark the CSI estimation performance.  Section-\ref{sec:sim_results} presents our simulation results, and Section-\ref{sec:concl_paper} concludes the paper. For ease of exposition, the proofs of some of the results are given in the appendices toward the end of the paper.
\subsection{Notation} \label{subsec:notation}
Upper case $(\mathbf{A})$ and lower case $(\mathbf{a})$ boldface letters are used to denote matrices and vectors, respectively. The quantity $\mathrm{vec}(\mathbf A)$ denotes the vector obtained by stacking the columns of matrix $\mathbf{A}$ while $\mathrm{vec}^{-1}(\mathbf a)$ denotes the inverse operation of constructing matrix $\mathbf{A}$ from $\mathbf{a}$. The notation $[\cdot]_{M} $ represents the modulo-$M$ operation, while $\text{diag}\left(\mathbf{a}\right)$ denotes a diagonal matrix with elements of vector $\mathbf{a}$ on the principal diagonal. Finally, the paper frequently uses the $\mathrm{vec}$ operator property $\mathrm{vec}\left( \mathbf{ABC} \right) = \left( \mathbf{C}^T \otimes \mathbf{A} \right) \mathrm{vec}\left( \mathbf{B} \right)$, where $\otimes$ stands for the matrix Kronecker product.
%-----------------------Block diagram-------------------------
\begin{figure*}[t]
\centering
{\includegraphics[scale=0.50]{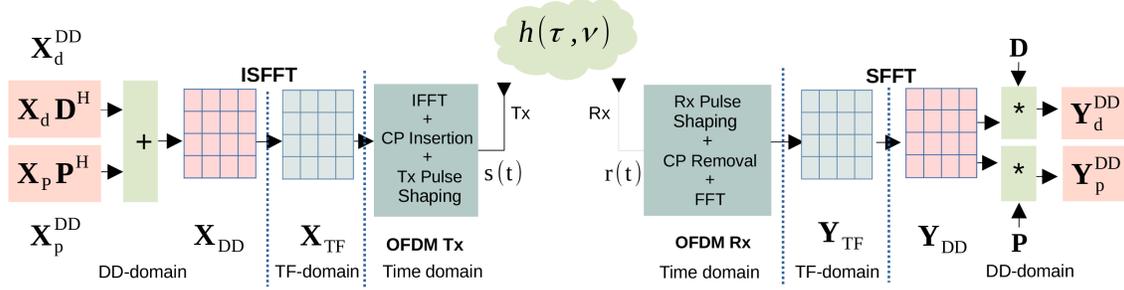}}
\caption{Block diagram of AP-SIP-based OTFS system implemented using OFDM. }
\label{fig:otfs_super}
\vspace{-10pt}
\end{figure*}
\section{AP-SIP based SISO OTFS system model} \label{sec:siso_sys_mdl}
Consider an AP-SIP-based SISO OTFS system constructed upon an OFDM-based multicarrier system, as shown in Fig. \ref{fig:otfs_super}.  Let the subcarrier spacing and the OFDM symbol duration be denoted by $\Delta f$, and $T$, respectively, so that $T \Delta f = 1$. In addition, let $M$ and $N$ be the number of symbols arranged along the frequency and time axes, respectively, in the TF-domain grid of the OFDM system. These TF-domain symbols are generated through an appropriate transformation from an equivalent DD-domain grid, wherein $M, N$ symbols are placed along the delay and Doppler axes, respectively. Consequently, the resultant OTFS system has a bandwidth of $M\Delta f$ and a frame duration of $NT$. Accordingly, the delay and Doppler axes of the DD-domain grid are sampled at integer multiples of $\Delta \tau= \frac{1}{M\Delta f}$ and $\Delta \nu = \frac{1}{NT}$, respectively. Let $ \mathbf{X}_{d} \in \mathbb{C}^{M \times K_1} $ and $ \mathbf{X}_{p} \in \mathbb{C}^{M \times K_2} $ denote the information and pilot symbol matrices, respectively, where $ K_1+K_2=N $. The elements of both $\mathbf{X}_{d}$ and $\mathbf{X}_{p}$ are drawn from a suitable constellation with average powers of $\sigma _d^2$ and $\sigma_p^2$, respectively, i.e., $ \mathbb{E}\lbrace \mathbf{X}_{d}\mathbf{X}^{H}_{d}\rbrace =\sigma _d^2 \mathbf{I}_{M} \text{ and } \sigma_p^2 = \frac{\mathrm{Tr}\left(\mathbf {X}_{p}\mathbf{X}_{p}^H \right)}{MK_2},$
so that $\sigma _d^2+\sigma_p^2=1$. In the proposed AP-SIP based system, the data input $\mathbf{X}_{d}$ and pilot input $\mathbf{X}_{p}$ are affine-precoded using the transmit precoder (TPC) matrices $ \mathbf{D}\in \mathbb{C}^{N \times K_1} $ and $ \mathbf{P}\in \mathbb{C}^{N \times K_2} $, respectively, so that they are semi-orthogonal, i.e.,
\begin{align}
\mathbf{P}^{H}\mathbf{P} &= \mathbf{I}_{K_2\times K_2},  \mathbf{D}^{H} \mathbf{D}=\mathbf{I}_{K_1\times K_1},\nonumber\\\mathbf{P}^{H}\mathbf{D} &=\mathbf{0}_{K_2\times K_1},\mathbf{D}^{H}\mathbf{P}=\mathbf{0}_{K_1\times K_2}.
\label{eq:semiorthogonal_property}
\end{align}
Interestingly, such TPC matrices as $\mathbf{P}$ and $\mathbf{D}$ can be readily obtained from an arbitrary unitary matrix $ \mathbf{U}\in \mathbb{C}^{N \times N}$, so that $\mathbf{P}=\mathbf{U}(:,1:K_1)$ and $\mathbf{D}=\mathbf{U}(:,K_1+1:N)$. Hence, as shown in the Fig. \ref{fig:otfs_super}, the AP-SIP based DD-domain data matrix $  \mathbf{X}_{\text{DD}} \in \mathbb{C}^{M \times N}$ is obtained as \begin{equation}
\mathbf{X}_{\text{DD}}=\mathbf{X}_d.\mathbf{D}^H+\mathbf{X}_p.\mathbf{P}^H
\label{eq:superimposed_data_SISO}
\end{equation}
The procedure of OTFS modulation is described next.

The OTFS system places the elements of the $M \times N$ AP-SIP data symbol matrix $\mathbf{X}_{\text{DD}}$ over the DD-domain grid. Subsequently, the inverse symplectic finite Fourier transform (ISFFT) is employed to map these DD-domain symbols to the TF grid at the transmitter. This operation is formulated as 
%\begin{equation} 
$\mathbf{X}_{\text{TF}} = {\mathbf{F}}_M \mathbf{X}_{\text{DD}}  {\mathbf{F}}_N^H \in \mathbb{C}^{M \times N}$, %\label{eq:mat_mod} 
%\end{equation}
where $\mathbf{F}_M$ and $\mathbf{F}_N$ represent the $M$th and $N$th order discrete Fourier transform (DFT) matrices, respectively. The OFDM modulator evaluates the $M$-point inverse fast Fourier transform (IFFT) of each column of the TF-domain symbol matrix $\mathbf{X}_{\text{TF}}$ followed by transmit pulse-shaping in order to obtain the time-domain (TD) symbol matrix $\mathbf{S} \in \mathbb{C}^{M \times N}$, which can be formulated as  
%\begin{align}
    $\mathbf{S} = \mathbf{P}_{\text{tx}} \mathbf{F}_M^H \mathbf{X}_{\text{TF}} =  \mathbf{P}_{\text{tx}} \mathbf{X}_{\text{DD}} \mathbf{F}^H_N$, %\label{eq:S_mat_dd}
%\end{align}
where diagonal matrix $\mathbf{P}_{\text{tx}}$ contain the $M$-samples of the transmit pulse $p_{\text{tx}}(t)$, i.e., $\mathbf{P}_{\text{tx}} = \mathrm{diag} \left\{p_{\text{tx}}\left(\frac{pT}{M}\right)\right\}_{p=0}^{M-1}$.
Furthermore, the conventional OFDM system transmits each column\footnote{By contrast, many OTFS schemes described in \cite{raviteja2018practical},\cite{suraj2021bayesian}, initially vectorize the TD symbol matrix $\mathbf{S}$ as 
$
\mathbf{s} =\mathrm{vec}\left( \mathbf{P}_{\text{tx}} \mathbf{X}_{\text{DD}} \mathbf{F}^H_N \right) = \left( \mathbf{F}_N^{H} \otimes \mathbf{P}_{\text{tx}} \right) \mathbf{x}_{\text{DD}}, 
$
where $\mathbf{x}_{\text{DD}} = \mathrm{vec} \left( \mathbf{X}_{\text {DD}} \right) \in \mathbb{C}^{MN \times 1}$, and subsequently append a CP of length $L$ for eliminating the inter-frame interference.} of the matrix $\mathbf{S}$ separately by performing parallel-to-serial (P/S) conversion, followed by appending a cyclic prefix (CP) of length $L$ to each column. The resultant TD signal is transmitted over a doubly-dispersive wireless channel, as described next. 

The DD-domain wireless channel $h(\tau,\nu)$ is formulated as \cite{raviteja2018interference, ramachandran2020otfs, hadani2017orthogonal,Hanzo_book}
\begin{equation} 
h(\tau,\nu) = \sum_{i=1}^{L_p} h_{i} \delta(\tau-\tau_{i}) \delta(\nu-\nu_{i}),
\end{equation}
which is a 2-dimensional function of the delay variable $\tau$ and Doppler variable $\nu$. 
{In the channel model above, the quantities $\tau_i$ and $\nu_i$ represent the delay and Doppler shifts, respectively, introduced by the $i$th reflector, $h_i$ represents the associated complex path gain, whereas $L_p$ is the number of multipath components. The quantity $\nu _i$ is expressed as $\nu _i = \frac{k_{i}}{NT}$, where $k_i$ is the fractional index corresponding to the Doppler shift $\nu_i$ and is given as $k_{i}=\text{round}(k_{i})+\mathcal{K}_i$, where $\vert{\mathcal{K}_i}\vert<0.5$. 
On the other hand for a typical wideband system associated with $M=32$ and $\Delta f=15 KHz$; the delay resolution obtained is $\Delta \tau=\frac{1}{M\Delta f}=2.08\mu sec$, which is low. Hence, one can safely assume that the delays of the multipath components are integer multiples of the delay resolution \cite{raviteja2018interference, tse_viswanath, raviteja_embedded}}. Note that for a typical \textit{underspread} channel \cite{raviteja2018interference,raviteja_embedded}, we have $l_{i} << M$ and $k_i << N$. The signal received at the output of this DD-domain channel is formulated next. Let $\mathbf{s}_{n} \in \mathbb{C}^{M \times 1}, 0 \leq n \leq N-1,$ denote the $n$th column of the TD symbol matrix $\mathbf{S}$, and $\mathbf{r}_{n} \in \mathbb{C}^{M \times 1}$ represent the corresponding output after CP removal. The $p$th sample of $\mathbf{r}_{n}$, denoted by $r_n(p), 0 \leq p \leq M-1,$ is expressed as
\begin{align} 
r_n(p) = \sum _{i=1}^{L_p} h_i e^{j2\pi \frac{{k_{i}}(p-l_{i})}{MN}} s_n\big([p-l_{i}]_{M}\big) + w_n(p), \label{eq:rn} 
\end{align}

where $\mathbf{s}_n(p)$ denotes the $p$th element of the vector $\mathbf{s}_{n}$ and $w_n(p)$ represents the noise samples. Let $\mathbf{r}_n \in \mathbb{C}^{M \times 1}$ be arranged as $\mathbf{r}_n = \left[ r_n(0), r_n(1), \cdots, r_n(M-1) \right]^T$  and $\mathbf{w}_n \in \mathbb{C}^{M \times 1}$ be arranged as $    \mathbf{w}_n = \left[ w_n(0), w_n(1), \cdots, w_n(M-1) \right]^T$. Using \eqref{eq:rn} and $\mathbf{r}_n$, the received signal vector $\mathbf{r}_n$ can be formulated as 
\begin{align} 
\mathbf{r}_n = \sum_{i=1}^{L_p} h_i \left(\mathbf{\bar\Pi}\right)^{l_{i}} \left(\mathbf{\bar\Delta}_{l_i}\right)^{k_{i}} \mathbf{s}_n + \mathbf{w}_n = \mathbf{r}_n = \mathbf{\bar{H}}\mathbf{s}_n + \mathbf{w}_n.  \label{eq:r_mat} 
\end{align}
where the matrix $\mathbf{\bar H} \in \mathbb{C}^{M \times M}$ be defined as $\mathbf{\bar H} = \sum_{i=1}^{L_p} h_i \left(\mathbf{\bar\Pi}\right)^{l_{i}} \left(\mathbf{\bar\Delta}_{l_i}\right)^{k_{i}}$, $\mathbf{\bar\Pi}$ denotes a permutation matrix of order $M$
and $\mathbf{\bar\Delta}_{l_i} \in \mathbb{C}^{M \times M}$ represents the diagonal matrix as shown in \cite{suraj2021row_group},\cite{suraj2021bayesian} 
\begin{align}
\bar{\mathbf{\Delta}}_{l_i} = \begin{cases}
\mathrm{diag}\left\{1, \omega, \cdots, \omega^{M-l_i-1}, \omega^{-l_i}, \cdots, \omega^{-1} \right\}, \text{if } l_i \neq 0,\\
\mathrm{diag}\left\{1, \omega, \cdots, \omega^{M-1} \right\}, \text{for } l_i=0,
\end{cases}
\end{align}
with $\omega = e^{j2\pi\frac{1}{MN}}$. 
Furthermore, upon concatenating the outputs $\mathbf{r}_n, 0 \leq n \leq N-1,$ as $\mathbf{R} = \left[ \mathbf{r}_0, \mathbf{r}_1, \cdots, \mathbf{r}_{N-1}\right] \in \mathbb{C}^{M \times N}$, we have 
\begin{equation} 
\mathbf{R} = \mathbf{\bar{H}}\mathbf{S} + \mathbf{W}, \label{eq:r_mat_2}
\end{equation}
where $\mathbf{W} = \left[ \mathbf{w}_0, \mathbf{w}_1, \cdots, \mathbf{w}_{N-1}\right] \in \mathbb{C}^{M \times N}$ represents the concatenated noise matrix. Subsequently, OTFS demodulation is applied to TD sample matrix $\mathbf{R}$ as follows.

The OFDM demodulator first applies a receive pulse shaping filter $p_{\text{rx}}(t)$ of duration $T$, followed by performing the $M$-point FFT over each column of the received signal matrix $\mathbf{R}$ for obtaining the TF-domain demodulated symbol matrix $\mathbf{Y}_{\text{TF}} \in \mathbb{C}^{M \times N}$. These operations are given by
%\begin{align}
    $\mathbf{Y}_{\text{TF}} = \mathbf{F}_M \mathbf{P}_{\text{rx}} \mathbf{R}$, %\label{eq:Y_TF_mat}
%\end{align}
where $\mathbf{P}_{\text{rx}} = \mathrm{diag} \left\{p_{\text{rx}}^{*}\left(\frac{pT}{M}\right)\right\}_{p=0}^{M-1}$. Next, the DD-domain demodulated OTFS signal $\mathbf{Y}_{\text{DD}} \in \mathbb{C}^{M \times N}$ is obtained by performing the SFFT of the OFDM-demodulated signal $\mathbf{Y}_{\text{TF}}$, which is expressed as 
%\begin{align}
$ \mathbf{Y}_\text{DD}= \mathbf{F}_M^{H} \mathbf {Y}_\text{TF} \mathbf{F}_N = \mathbf{P}_\text{rx}\mathbf{R}\mathbf {F}_N$.
%\label{eq:TF2DD_mat} 
%\end{align}
Finally, upon substituting $\mathbf{R}$, %from \eqref{eq:r_mat_2} 
and in turn substituting $\mathbf{S}$ % from \eqref{eq:S_mat_dd} 
into $\mathbf{Y}_\text{DD}$ %\eqref{eq:TF2DD_mat}
, the simplified input-output DD-domain relationship of the AP-SIP SISO OTFS system is derived as
\begin{align}
    \mathbf{Y}_{\text{DD}} = \mathbf{\bar{H}}_{\text{DD}}  \mathbf{X}_{\text{DD}} + \mathbf{W}_{\text{DD}},
\label{eq:TF2DD_mat1_siso} 
\end{align}
where $\mathbf{\bar{H}}_{\text{DD}} = {\mathbf P}_{\text {rx}} \mathbf{\bar{H}} {\mathbf P}_{\text {tx}} \in \mathbb{C}^{M \times M}$, which can be further expressed as
\begin{align}
    \mathbf{\bar{H}}_{\text{DD}} = \sum_{i=1}^{L_p} h_i {\mathbf P}_{\text {rx}} \left(\mathbf{\bar\Pi}\right)^{l_{i}} \left(\mathbf{\bar\Delta}_{l_i}\right)^{k_{i}} {\mathbf P}_{\text {tx}}, \label{eq:h_dd}
\end{align}
and $\mathbf{W}_{\text{DD}} = {\mathbf P}_{\text {rx}} \mathbf{W}{\mathbf F}_N$. 
By contrast, the input-output relationship derived in other OTFS studies, such as \cite{raviteja2018practical,raviteja2018embedded}, is of the form $     \mathbf{y}_{\text{DD}}= \mathbf{H}_{\text{DD}} \mathbf{x}_{\text{DD}} + \mathbf{w}_{\text{DD}}, $ where we have $\mathbf{y}_{\text{DD}} = \mathrm{vec} \left({\mathbf Y_{\text{DD}}} \right) \in \mathbb{C}^{MN \times 1}$, and the end-to-end channel matrix $\mathbf{H}_{\text{DD}}$ is of size $MN \times MN$. Note that the simplified relationship of \eqref{eq:TF2DD_mat1_siso} can be attributed to the CP concatenation to each column of the TD symbol matrix $\mathbf{S}$. It follows from \eqref{eq:superimposed_data_SISO} that the AP-SIP matrix $\mathbf{X}_\text{DD}$ comprises the data matrix $\mathbf{X}_d$ and pilot matrix $\mathbf{X}_p$. Owing to the semi-orthogonality of the TPC matrices $\mathbf{P}$ and $\mathbf{D}$, one can decouple the resultant output corresponding to the data matrix  $\mathbf{X}_d$ during signal detection, as described subsequently. Upon substituting (\ref{eq:superimposed_data_SISO}) in (\ref{eq:TF2DD_mat1_siso}), we have
\begin{equation}
\mathbf{Y}_{\text{DD}}= \mathbf{\bar{H}}_{\text{DD}}(\mathbf{X}_d.\mathbf{D}^H+\mathbf{X}_p.\mathbf{P}^H)    +\mathbf{W}_{\text{DD}}.
\label{eq:superimposed_channel_model_SISO}
\end{equation}
One can now decouple the data output $\mathbf{Y}_{\text{DD},d}\in \mathbb{C}^{M \times K_1}$ from $\mathbf{Y}_{\text{DD}}$ of \eqref{eq:superimposed_channel_model_SISO} by post-multiplying it by the data TPC matrix $\mathbf{D}$, i.e ${\mathbf{Y}}_{\text{DD},d} = \mathbf{Y}_{\text{DD}}\mathbf{D}$, followed by further simplification exploiting the
properties stated in \eqref{eq:semiorthogonal_property}, as 
\begin{align}
        {\mathbf{Y}}_{\text{DD},d} &= 
        \mathbf{\bar{H}}_{\text{DD}} \left({{\mathbf{X}_d\mathbf{D}^H}\mathbf{D} +\mathbf{X}_p{\mathbf{P}^H}} \mathbf{D}\right)
        +\mathbf{W}_{\text{DD}}\mathbf{D}
\nonumber\\&=\mathbf{\bar{H}}_{\text{DD}} \mathbf{X}_d 
        +{\mathbf{W}}_{\text{DD},d}.
\label{eq:data_decoupling}
\end{align}
Furthermore, considering $\mathbf{w}_{\text{DD},d} = \mathrm{vec} \left({\mathbf W_{\text{DD},d}} \right) \in \mathbb{C}^{MK_1 \times 1}$, its  covariance matrix $\mathbf{R}_{\text{w},d}$ is given by
$   \mathbf{R}_{\text{w},d} = 
   \sigma^2 \big[ \mathbf{I}_{K_1} \otimes \left( \mathbf{P}_{\text{rx}} \mathbf{P}_{\text{rx}}^H  \right) \big].
$
This in turn implies that any column of the noise matrix $\mathbf{W}_{\text{DD},d}$ has the covariance of $ \mathbf{R}_{\text{W},d} = \sigma^2 \left( \mathbf{P}_{\text{rx}} \mathbf{P}_{\text{rx}}^H  \right)\in \mathbb{C}^{M \times M}$. 
Thus, considering the average symbol power of $\mathbf{X}_{d}$ as $\sigma_d^2$, the LMMSE detector is formulated as 
\begin{align}
   \mathbf{X}_{{d}}^{\text{MMSE}} = \left( \mathbf{\bar{H}}_{\text{DD}}^H \mathbf{R}_{\text{W},d}^{-1} \mathbf{\bar H}_{\text{DD}}  + \frac{1}{\sigma_d^2}\mathbf{I}_{M} \right)^{-1} \mathbf{\bar{H}}_{\text{DD}}^H \mathbf{R}_{\text{W},d}^{-1} \mathbf{Y}_{\text{DD},d},
   \label{eq:mmse_detector_SISO}
\end{align}
which is finally demodulated according to the specific transmit constellation symbols using the  nearest neighbor decoding rule \cite{Proakis2007}.
\section{Sparse DD-domain CSI estimation model for AP-SIP SISO OTFS systems}\label{sec:sparse_estimation_model_siso}
For CSI estimation, the pilot output ${\mathbf{Y}_{\text{DD},p}}\in \mathbb{C}^{M \times K_2}$ is decoupled from the data upon post-multiplying $\mathbf{Y}_{\text{DD}}$ of (\ref{eq:superimposed_channel_model_SISO}) by the TPC matrix $\mathbf{P}$, i.e ${\mathbf{Y}_{\text{DD},p}}=\mathbf{Y}_{\text{DD}}\mathbf{P} $, followed by further simplification exploiting the properties stated in \eqref{eq:semiorthogonal_property}, as
\begin{align}
       {\mathbf{Y}_{\text{DD},p}}&=\mathbf{\bar{H}}_{\text{DD}} \left({{\mathbf{X}_d\mathbf{D}^H}\mathbf{P} +\mathbf{X}_p{\mathbf{P}^H}} \mathbf{P}\right)
        +\mathbf{W}_{\text{DD}}\mathbf{P}
\nonumber\\&=\bar{\mathbf{H}}_{\text{DD}} \mathbf{X}_p + {\mathbf{W}_{\text{DD},p}}.
       \label{eq:pilot_decoupling}
\end{align}
Furthermore, considering $\mathbf{w}_{\text{DD},p}=\mathrm{vec}\left({\mathbf{W}_{\text{DD,p}}}\right)\in\mathbb{C}^{MK_2\times1}$, its  covariance matrix $\mathbf{R}_{\text{w},p}$ is given by $\mathbf{R}_{\text{w},p}=\sigma^2\big[\mathbf{I}_{K_2}\otimes\left(\mathbf{P}_{\text{rx}}\mathbf{P}_{\text{rx}}^H \right)\big]$. Upon employing the pilot output model above, the DD-domain channel can be estimated as discussed next when considering a typical underspread wireless channel, one can denote its maximum delay spread by $M_\tau$ and Doppler spread by $N_\nu$. This implies that the delay and Doppler shifts introduced by each multipath component of the wireless channel obey: $ l_{max}=\text{max}(l_i) < M_\tau << M$, and $k_{max}=\text{max}(k_i) < N_\nu << N, \forall i$.
\begin{figure}[t]
    \centering
\includegraphics[scale=0.35]{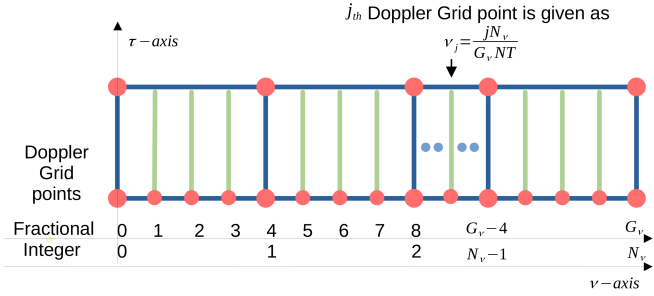}
\caption{Fractional and integer Doppler grids}
\vspace{-10pt}
\label{fig:factional_Doppler_grid}
\end{figure} 
{Toward introducing fractional Doppler, consider a grid of size $G_\tau \times G_\nu$, such that $G_\nu>>N_\nu$, i.e., each grid interval corresponding to integer Doppler shift is divided into multiple intervals as shown in Fig. 2. The $j$th Doppler-grid point, $0 \leq j \leq G_{\nu}$, corresponds to a Doppler-shift of $\nu_j = \frac{jN_{\nu}}{G_{\nu}NT}$ Hz. For $G_\nu = N_\nu$, it reduces to the previous model with integer Doppler shift. Employing this framework, the channel ${h}(\tau,\nu)$ is given by
\begin{equation} 
{h}(\tau,\nu) = \sum_{i=0}^{M_\tau-1}\sum_{j=0}^{{G_{\nu}-1}
} h_{i,j} \delta(\tau-\tau_{i}) \delta(\nu-\nu_{j}). 
\label{eq:h_dd_bar}
\end{equation}}
Owing to having only a few dominant multipath components, 
most of the coefficients $h_{i,j}$ are zero, and only a small number of coefficients $L_p$ are non zero, which correspond to the $L_p$ dominant reflectors, associated with $L_p<< M_\tau G_\nu$. {Equivalently, the sparse representation of the matrix $\bar{\mathbf{H}}_{\text{DD}}$ of \eqref{eq:h_dd_bar} can be obtained as
\begin{align}
    \mathbf{\bar{H}}_{\text{DD}} = \sum_{i=0}^{M_\tau-1}\sum_{j=0}^{G_\nu-1} {h_{i,j}} {\mathbf P}_{\text {rx}} \left(\mathbf{\bar\Pi}\right)^{{i}} \left(\mathbf{\bar\Delta}_{i}\right)^{{j}} {\mathbf P}_{\text {tx}}.
    \label{eq:SISO_channel_unknown_taps_final}
\end{align}}
{Now, upon substituting (\ref{eq:SISO_channel_unknown_taps_final}) into (\ref{eq:pilot_decoupling}) yields
\begin{equation}
{\mathbf{Y}_{\text{DD},p}}=\sum_{i=0}^{M_\tau-1}\sum_{j=0}^{G_\nu-1} {h_{i,j}} {\mathbf P}_{\text {rx}} \left(\mathbf{\bar\Pi}\right)^{{i}} \left(\mathbf{\bar\Delta}_{i}\right)^{{j}} {\mathbf P}_{\text {tx}} \mathbf{X}_p + {\mathbf{W}_{\text{DD},p}}.
\end{equation}
Furthermore, vectorizing ${\mathbf{Y}_{\text{DD},p}}$ as ${\mathbf{y}_{\text{DD},p}}=\mathrm{vec}({\mathbf{Y}_{\text{DD},p}})$, yields
\begin{align}
&\mathbf{y}_{\text{DD},p}=\nonumber\\&\sum_{i=0}^{M_\tau-1}\sum_{j=0}^{G_\nu-1}{\Bigg[\mathbf{I}_{K_2}\otimes{\bigg({\mathbf{P}}_\text{rx}(\bar{\mathbf{\Pi}})^i ({\bar{\mathbf{\Delta}}_i})^j \mathbf{P}_\text{tx}\bigg)}\Bigg] \mathbf{x}_{p}}  h_{i,j}+{\mathbf{w}}_{\text{DD},p},
\end{align}
where $\mathbf{x}_{p}=\mathrm{vec}(\mathbf{X}_{p}) \in \mathbb{C}^{M K_2 \times 1}$. This can be further simplified as 
\begin{equation}
\mathbf{y}_{\text{DD},p}=\sum_{i=0}^{M_\tau-1}\sum_{J=0}^{G_\nu-1}
    { \mathbf{{w}}_{p,i,j}}h_{i,j}+{\mathbf{w}}_{\text{DD},p},
\end{equation}
where $ \mathbf{w}_{p,i,j}
={\big[{\mathbf{I}_{K_2}\otimes{\left({\mathbf{P}}_\text{rx}(\bar{\mathbf{\Pi}})^i ({\bar{\mathbf{\Delta}}_i})^j \mathbf{P}_\text{tx}\right)}\big]}\mathbf{x}_{p}} \in \mathbb{C}^{M K_2 \times 1} $. The above relationship can be expressed as the canonical sparse signal recovery problem of
\begin{equation}
\mathbf{y}_{\text{DD},p}=\boldsymbol{\Omega}_p\mathbf{h}+{\mathbf{w}}_{\text{DD},p},
\label{eq:linear_sparse_SISO_model}
\end{equation}
where the dictionary matrix $\boldsymbol{\Omega}_p\in \mathbb{C}^{M K_2 \times M_\tau N_\nu}$ and sparse channel vector $\mathbf{h}\in {\mathbb{C}^{M_\tau N_\nu \times 1}}$ are given by
\begin{align}
     \boldsymbol{\Omega}_p &= \left[\mathbf{w}_{p,0,0}\hdots \mathbf{w}_{p,0,G_\nu-1}\hdots \mathbf{w}_{p,M_\tau-1,0}\hdots  \mathbf{w}_{p,M_\tau-1,{G_\nu-1}}\right].
\label{eq:dictionary_matrix_pilot}\\
  \mathbf{h} &=\left[h_{0,0}  \hdots, h_{0,{G_\nu-1}},\hdots,h_{{M_\tau-1},0} \hdots, h_{{M_\tau-1},{G_\nu-1}}\right]^T.
\label{eq:sparse_channel_coeeficient_vector}
\end{align}
For linear estimation model shown in (\ref{eq:linear_sparse_SISO_model}), the conventional MMSE estimate $\widehat{\mathbf{h}}_{\text{MMSE}}$ is given by
\begin{align} \widehat{\mathbf {h}}_{\text{MMSE}} = \left({\boldsymbol{\Omega}_p^H} \mathbf {R}_{\text{w},p}^{-1} {\boldsymbol{\Omega}_p } + \mathbf {R}_h^{-1}\right)^{-1} {\boldsymbol{\Omega}_p^H} \mathbf {R}_{\text{w},p}^{-1} \mathbf {y}_{\text{DD},p}
\label{eq:pilot_mmse}. \end{align}
Note that the conventional MMSE estimate shown above requires knowledge of the channel's covariance matrix $\mathbf{R}_h \in \mathbb {C}^{M_{\tau }G_{\nu } \times M_{\tau }G_{\nu }} $, which is unknown in practice. }Additionally, another significant drawback of using the conventional MMSE-based method for DD-domain CSI estimation is that it does not exploit the sparse characteristics of the DD-domain channel, which can lead to a significantly improved performance. To overcome these drawbacks, we conceive a Bayesian learning (BL) procedure for enhanced sparse DD-domain CSI estimation, as described in the next subsection.
\subsection{Pilot aided Bayesian learning (PA-BL) for AP-SIP SISO OTFS systems}
In the PA-BL framework, a parameterized Gaussian prior is assigned to the sparse channel vector $\mathbf{h}$ \cite{suraj2021bayesian}, as
%\begin{align} 
$f(\mathbf{h}; \boldsymbol{\Lambda}) = \prod _{i=0}^{M_{\tau }G_{\nu }-1}\frac{1}{(\pi \lambda _i)} \exp \left(-\displaystyle \frac{\vert \mathbf{h}(i)\vert ^2}{\lambda _i}\right)$.
%\label{eq:parameterized_gaussian}
%\end{align}
Here, $\lambda _i$ represents the unknown hyperparameter associated with the $i$th component of vector $\mathbf{h}$ and $\boldsymbol{\Lambda} = \mathrm{diag}\left(\lbrace \lambda _i \rbrace _{i=0}^{M_{\tau }G_{\nu }-1}\right) \in \mathbb {R}^{+ M_{\tau }G_{\nu }\times M_{\tau }G_{\nu }}$ denotes the hyperparameter matrix. %Employing the parameterized prior assignment above, the MMSE estimate of \eqref{eq:pilot_mmse} reduces to
%\begin{align} \widehat{\mathbf {h}}_{\text{MMSE}} = \left({\boldsymbol{\Omega}_p^H} \mathbf {R}_{\text{w},p}^{-1} {\boldsymbol{\Omega}_p } + \boldsymbol{\Lambda}^{-1}\right)^{-1} {\boldsymbol{\Omega}_p^H} \mathbf {R}_{\text{w},p}^{-1} \mathbf {y}_{\text{DD},p}.
%\label{eq:MMSE_estimate}
%\end{align}
%Hence, it can be observed that the MMSE estimate $\widehat{\mathbf {h}}_{\text{MMSE}}$ depends on estimation of the hyperparameters $\lambda_i, 0\leq i \leq M_\tau N_\nu-1$. 
In the proposed PA-BL method, the hyperparameter estimation is performed using the well-known expectation maximization (EM)-procedure which is detailed in Algorithm 1.

\begin{algorithm}
\DontPrintSemicolon
\caption{PA-BL-based sparse CSI estimation in SISO OTFS systems}\label{algo_PA_BL_siso}
\KwIn{Decoupled pilot output vector ${\mathbf{y}}_{\text{DD},p}$ form (19) and pilot dictionary matrix ${\boldsymbol{\Omega}_p}$ from (20), noise covariance matrix $\mathbf{R}_{\text{w},p}$, stopping parameters $\epsilon \text{\ and\ } N_{\text{max}}$}
\SetKwInput{kwInit}{Initialization}\vspace{10pt}
\kwInit{$\widehat{\lambda_i}^{(0)}=1, \forall\ 0 \leq i \leq {M_\tau G_\nu}-1, \widehat{\boldsymbol{\Lambda}}^{(0)}=\mathbf{I}_{M_\tau G_\nu} $, $\widehat{\boldsymbol{\Lambda}}^{(-1)} = \mathbf{0}$, $j = 0$}\vspace{10pt}
\KwOut{$\widehat{\mathbf{h}}^{\text{PA-BL}} = \boldsymbol{\mu}^{(j)}$}\vspace{10pt}
\SetKwBlock{Begin}{while}{end function}
\Begin(\(\)$\|\widehat{\boldsymbol{\Lambda}}^{(j)} - \widehat{\boldsymbol{\Lambda}}^{(j-1)} \|^{2}_{F}>\epsilon$ and $j<N_{\text{max}}$ \textbf{do}){
$j \gets j+1$\;
\textbf{E-step:}\; Compute the \textit{aposteriori} covariance and mean
 \begin{align*}
 \boldsymbol{\Sigma}^{(j)} &= [\boldsymbol{\Omega}_p^H \mathbf {R}_{\text{w},p}^{-1} \boldsymbol{\Omega}_p + (\widehat{\boldsymbol{\Lambda}}^{(j-1)})^{-1}]^{-1},\\
       \boldsymbol{\mu}^{(j)} &= \boldsymbol{\Sigma}^{(j)} \boldsymbol{\Omega}_p^H \mathbf {R}_{\text{w},p}^{-1} \mathbf {y}_{\text{DD},p}\end{align*}\;
\textbf{M-step:}\; Compute the hyperparameter estimates\;
\textbf{{for}} $i\gets 0 \text{ to } M_\tau G_\nu-1$ \textbf{ do}  {\begin{align*}\widehat{\lambda}_i^{(j)} = \boldsymbol{\Sigma}^{(j)}(i,i) + | \boldsymbol{\mu}^{(j)}(i) |^2\end{align*}.}
}
\end{algorithm}
Upon convergence of the EM procedure, the estimate of the sparse vector $\mathbf{h}$ is given by the \textit{aposteriori} mean $\mathbf{\mu}^{(j)}$ computed in step-2. Finally, the estimated CSI $\widehat{\mathbf{H}}_\text{DD}^{\text{PA-BL}}$ of our AP-SIP based SISO-OTFS system is given by
\begin{align}
\mathbf{\widehat{H}}_{\text{DD}}^{\text{PA-BL}} = \sum_{i,j} \widehat{h}_{i,j}^\text{PA-BL} {\mathbf P}_{\text {rx}} \left(\mathbf{\bar\Pi}\right)^{{i}} \left(\mathbf{\bar\Delta}_{i}\right)^{{j}} {\mathbf P}_{\text {tx}}.
    \label{eq:estimated_channel}
\end{align}
Subsequently, the estimated CSI $\mathbf{\widehat{H}}_{\text{DD}}^\text{PA-BL}$ (\ref{eq:estimated_channel}) is utilized in the MMSE detector defined by (\ref{eq:mmse_detector_SISO}) for data detection. Note that the CSI estimation framework developed above employs only the pilot output $\mathbf{y}_{\text{DD},p}$ for channel estimation. In order to further improve the CSI estimation, one can exploit the estimate of the data symbol matrix $\mathbf{X}_{\text{DD}}$ for iteratively refining the CSI estimate $\mathbf{\widehat{H}}_{\text{DD}}^\text{PA-BL}$. The resultant uncertainty in the CSI estimates, which is characterized by the \textit{aposteriori} covariance matrix $\boldsymbol{\Sigma}^{(j)}$, can be exploited for developing the optimal MMSE detector. This motivates us to develop a data-aided CSI estimation framework for AP-SIP SISO-OTFS systems, which is described next.
\subsection{Data aided Bayesian learning (DA-BL) for AP-SIP SISO OTFS systems} \label{sec:J-SIP}

Consider the decoupled data output $\mathbf{Y}_{\text{DD},d}$ expression obtained in (\ref{eq:data_decoupling}). Upon substituting $ \mathbf{\bar{H}}_{\text{DD}}$ from \eqref{eq:SISO_channel_unknown_taps_final} into the decoupled data output is given by 
\begin{equation}
    {\mathbf{Y}_{\text{DD},d}}=\sum_{i=0}^{M_\tau-1}\sum_{j=0}^{G_\nu-1} {h_{i,j}} {\mathbf P}_{\text {rx}} \left(\mathbf{\bar\Pi}\right)^{{i}} \left(\mathbf{\bar\Delta}_{i}\right)^{{j}} {\mathbf P}_{\text {tx}} \mathbf{X}_d + {\mathbf{W}_{\text{DD},d}}.
     \label{eq:sparse_siso_otfs_data}
\end{equation}
Upon, vectorizing $\mathbf{Y}_{\text{DD},d}$ as $\mathbf{y}_{\text{DD},d}=\mathrm{vec}(\mathbf{Y}_{\text{DD},d})$, the resulting expression is
\begin{align}
  &\mathbf{y}_{\text{DD},d}\\&=\sum_{i=0}^{M_\tau-1}\sum_{j=0}^{G_\nu-1}
  {\Bigg[\mathbf{I}_{K_1}\otimes{\bigg({\mathbf{P}}_\text{rx}(\bar{\mathbf{\Pi}})^i ({\bar{\mathbf{\Delta}}_i})^j \mathbf{P}_\text{tx}\bigg)}\Bigg] \mathbf{x}_{d} h_{i,j}}+{\mathbf{w}}_{\text{DD},d}\nonumber,
\end{align}
where $\mathbf{x}_{d}=\mathrm{vec}(\mathbf{X}_{d}) \in \mathbb{C}^{M K_1 \times 1}$. This can be further simplified as
\begin{equation}
   \mathbf{y}_{\text{DD},d}=\sum_{i=0}^{M_\tau-1}\sum_{J=0}^{G_\nu-1}
    { \mathbf{w}_{d,i,j}}h_{i,j}+{\mathbf{w}}_{\text{DD},d},
\end{equation}
where $ \mathbf{w}_{d,i,j}
={\big[{\mathbf{I}_{K_1}\otimes{\left({\mathbf{P}}_\text{rx}(\bar{\mathbf{\Pi}})^i ({\bar{\mathbf{\Delta}}_i})^j \mathbf{P}_\text{tx}\right)}\big]}\mathbf{x}_{d}} \in \mathbb{C}^{M K_1 \times 1} $. By casting the above relationship into the sparse signal recovery model, we get
\begin{equation}
  \mathbf{y}_{\text{DD},d}={\boldsymbol{\Omega}}_d \mathbf{h}+{\mathbf{w}}_{\text{DD},d},
  \label{eq:linear_sparse_DABL_SISO_model}
\end{equation}
where the dictionary matrix $\boldsymbol{\Omega}_d\in \mathbb{C}^{M K_1 \times M_\tau N_\nu}$ is given by 
\begin{align}
\boldsymbol{\Omega}_d&= \left[\mathbf{w}_{d,0,0} \hdots \mathbf{w}_{d,0,G_\nu-1}\hdots  \mathbf{w}_{d,M_\tau-1,0}\hdots  \mathbf{w}_{d,M_\tau-1,{G_\nu-1}}\right].
\label{eq:dictionary_matrix_data}
\end{align}
The sparse channel vector $\mathbf{h}\in {\mathbb{C}^{M_\tau G_\nu \times 1}}$ is given by \eqref{eq:sparse_channel_coeeficient_vector}. For DA-BL-based sparse CSI estimation and detection, the data output from \eqref{eq:linear_sparse_SISO_model} and pilot output from \eqref{eq:linear_sparse_DABL_SISO_model} are stacked as
\begin{align} {\underbrace{\begin{bmatrix}\mathbf{y}_{\text{DD},d}\\ \mathbf{y}_{\text{DD},p} \end{bmatrix}}_{{\bf y} \in \mathbb{C}^{MN\times 1} }}&={\underbrace{\begin{bmatrix}\boldsymbol{\Omega}_d\\ \boldsymbol{\Omega}_p\\ \end{bmatrix}}_{\boldsymbol{\Phi } \in \mathbb {C}^{MN \times M_\tau G_\nu} }}{\bf h} + {\underbrace{\begin{bmatrix}\mathbf {w}_{\text{DD},d}\\ \mathbf {w}_{\text{DD},p} \end{bmatrix}}_{\mathbf{v} \in \mathbb {C}^{MN\times 1} }}.
\label{eq:joint_equation_SISO_intermediate}
\end{align}
Thus, the resultant system model corresponding to the data-aided system is given by \begin{align}
\mathbf{y} ={\boldsymbol{\Phi }}\mathbf{h}+ \mathbf{v},
\label{eq:joint_equation_SISO}
\end{align}
where the covariance matrix of the noise vector $\mathbf{v}$ is $\mathbf{R}_v=\mathrm{blkdiag}(\mathbf{R}_{\text{w},d} , \mathbf{R}_{\text{w},p} ) \in \mathbb{C}^{MN \times MN}$. The DA-BL technique proposed for joint CSI estimation and data detection proceeds as follows.

For the DA-BL framework, the parameterized Gaussian prior is assigned to the sparse DD-domain CSI $\mathbf{h}$ given as
%\begin{align} 
$f(\mathbf{h}; \boldsymbol{\Lambda}) = \prod _{i=0}^{M_{\tau }G_{\nu }-1}\frac{1}{(\pi \lambda _i)} \exp \left(-\displaystyle \frac{\vert \mathbf{h}(i)\vert ^2}{\lambda _i}\right)$, 
%\label{eq:parameterized_gaussian}
%\end{align}
similar to the PA-BL framework. The proposed DA-BL framework jointly and iteratively estimates the hyperparameter matrix $\boldsymbol \Lambda$ and data symbol matrix $\mathbf{X}_{d}$  using the EM algorithm. Here, the complete information set is represented as $\lbrace \mathbf {y},\mathbf {h}\rbrace$, where $\mathbf{h}$ is the hidden variable, and $\mathbf{y}$ is the observation variable. Let $\{\mathbf X_d,\boldsymbol{\Lambda }\}$ represent the unknown parameter set and let $\widehat{\boldsymbol{\Lambda }}^{(j-1)} = \bigg\{ \widehat{\mathbf {X}}_{d}^{(j-1)}, \widehat{\boldsymbol{\Lambda }}^{(j-1)}\bigg\}$, where $\widehat{\mathbf {X}}_{d}^{(j-1)} 
$ and $\widehat{\boldsymbol{\Lambda }}^{(j-1)}$  denote the estimates of the data symbol matrix ${\mathbf {X}}_{d}$ and the hyperparameter matrix ${\boldsymbol{\Lambda }}$ gleaned from the $(j-1)$th EM iteration. For the current $j$th iteration, the E-step computes the log-likelihood function $\mathcal {L}(\boldsymbol{\Lambda }\mid \widehat{\boldsymbol{\Lambda }}^{(j-1)})$ of the complete data set, which is given by
\begin{align}
&\mathcal {L}(\boldsymbol{\Lambda }\mid \widehat{\boldsymbol{\Lambda }}^{(j-1)})
=\mathbb {E}_{{\bf h}\mid {\mathbf{y}};\widehat{\boldsymbol{\Lambda }}^{(j-1)}} \left\lbrace \log \left[ p\left({\bf y},{\bf h}; \boldsymbol{\Lambda }\right)\right]\right\rbrace\\
&=\mathbb {E}\left\lbrace \log \left[ p\left({\bf y}\mid {\mathbf{h}};\mathbf{X}_{d}\right)\right]\right\rbrace
+\mathbb {E}\left\lbrace \log \left[p \left({\bf h};\boldsymbol{\Lambda }\right)\right] \right\rbrace. 
\label{eq:log_likelihood_EM}
\end{align}
In the M-step, the log-likelihood $\mathcal {L}(\boldsymbol{\Lambda }\mid \widehat{\boldsymbol{\Lambda }}^{(j-1)})$ computed is jointly maximized with respect to the unknown parameter set $\boldsymbol{\Lambda}$. Note that the first term  $\log \left[ p\left({\bf y}\mid {\mathbf{h}}; \mathbf{X}_{d}\right)\right]$ in \eqref{eq:log_likelihood_EM} depends only on the data matrix $\mathbf{X}_{d}$, and the second term $ \log \left[p \left({\mathbf {h}};\boldsymbol{\Lambda }\right)\right]$ depends exclusively on the hyperparameter matrix $\boldsymbol{\Lambda}$. Therefore, the joint maximization of $\mathcal {L}(\boldsymbol{\Lambda }\mid \widehat{\boldsymbol{\Lambda }}^{(j-1)})$ with respect to $ \mathbf {X}_{d}$ and $\boldsymbol{\Lambda}$ reduces to two independent maximizations of the first and second terms with respect to $\mathbf{X}_{d}$ and $\boldsymbol{\Lambda}$, respectively. The M-step1 for the hyperparameter update $\widehat{{\lambda }}^{(j)}_{i}$  can be expressed as
\begin{align}
\widehat{{\lambda }}^{(j)}_{i} =\arg \max _{\substack{{{\lambda }_i}\geq 0}}\mathbb {E}_{{\bf h}\mid {\bf y};\widehat{\boldsymbol{\Lambda }}^{\left(j-1\right)}}\left\lbrace \log \left[ {p}\left(\mathbf {h};{\boldsymbol{\Lambda }}\right)\right]\right\rbrace,
\end{align}
for $0 \leq i \leq M_\tau G_\nu-1$. This can be simplified similar to Step3 of Algorithm 1 as
\begin{align} \widehat{{\lambda }}_i^{(j)}=\boldsymbol{\Sigma }^{(j)}(i,i) + \vert \widehat{\mathbf {h}}^{(j)}(i)\vert ^2,
\label{eq:joint_hyperparameter_update}
\end{align}
where the \textit{a posteriori} mean vector $\widehat{\mathbf{h}}^{(j)}\in \mathbb {C}^{M_\tau G_\nu \times 1}$ and the associated covariance matrix $\boldsymbol{\Sigma }^{(j)}\in \mathbb {C}^{M_\tau G_\nu \times M_\tau G_\nu}
$
are determined as
\begin{align}
\widehat{\mathbf{h}}^{(j)}&=\boldsymbol{\Sigma }^{(j)}({\boldsymbol{\Phi }}^{\left(j-1\right)})^H \mathbf{R}_{v}^{-1}\mathbf{y},\nonumber\\ 
\boldsymbol{\Sigma }^{(j)}&=[({\boldsymbol{\Phi }}^{\left(j-1\right)})^{H} \mathbf {R}_{v}^{-1}{\boldsymbol{\Phi }}^{(j-1)} + (\widehat{\boldsymbol{\Lambda }}^{(j-1)})^{-1}]^{-1}.
\end{align}
Here, the quantity ${\boldsymbol{\Phi}}^{(j-1)}$ is constructed from the estimate of the data input $\widehat{\mathbf{X}}_{d}^{(j-1)}$ at the $(j-1)$th iteration. Subsequently, in M-step2, the update $\widehat{\mathbf {X}}_{d}^{(j)}$ of the data matrix is determined as
\begin{align} \widehat{\mathbf{X}}_{d}^{(j)} &= \arg \max _{\mathbf{X}_{d}} \mathbb {E}  \{  \log [ p(\mathbf {y}_{\text{DD},d}\mid \mathbf {h}; \mathbf {X}_{d}) ] \}, 
\end{align}
which can be further formulated as
\begin{align}
\widehat{\mathbf{X}}_{d}^{(j)}&= \arg \min _{\substack{\mathbf {X}_{d}}} \mathbb{E}\{ || \mathbf{y}_{\text{DD},d}-{\boldsymbol{\Omega}}_d \mathbf{h}||_2^2 \}\nonumber\\
&\equiv \arg \min _{\substack{\mathbf {X}_{d}}} \mathbb{E}\{ || {\mathbf{Y}}_{\text{DD},d} -\mathbf{\bar{H}}_{\text{DD}} \mathbf{X}_d || _F^2 \}.
\label{eq:cost_function_LS}
\end{align}
\begin{algorithm}
\DontPrintSemicolon
\caption{DA-BL based sparse CSI estimation in SISO OTFS systems}\label{algo_DA_BL_siso}
\KwIn{Decoupled joint output vector ${\mathbf{y}}$ from (30), joint dictionary matrix ${\boldsymbol{\Phi}}$ from (29), joint noise covariance matrix $\mathbf{R}_v=\mathrm{blkdiag}(\mathbf{R}_{\text{w},d} , \mathbf{R}_{\text{w},p} )$, threshold $\epsilon$ and $N_{\text{max}}$}\vspace{10pt}
\SetKwInput{kwInit}{Initialization}
\kwInit{$ \widehat{\boldsymbol{\Lambda}}^{(0)}=\widehat{\boldsymbol{\Lambda}}_{\text{PA-BL}}^{(j)}$, $\widehat{\boldsymbol{\Lambda}}^{(-1)} = \mathbf{0}$, $j = 0$, $\boldsymbol{\Omega}_d^{(-1)}=\widehat{\boldsymbol{\Omega}}_d^\text{PA-BL}$, $\mathbf{\Phi}^{(-1)}=
{\begin{bmatrix}{\boldsymbol{\Omega}}_{d}^{(-1)}\\ \boldsymbol{\Omega}_p\\ \end{bmatrix}}$}\vspace{10pt}
\KwOut{$\widehat{\mathbf{h}}^{\text{DA-BL}} = \boldsymbol{\mu}^{(j)}$ and $\widehat{\mathbf{X}}^{\text{DA-BL}}=\widehat{\mathbf{X}}_d^{(j)}$}\vspace{10pt}
\SetKwBlock{Begin}{while}{end function}
\Begin(\(\)$\|\widehat{\boldsymbol{\Lambda}}^{(j)} - \widehat{\boldsymbol{\Lambda}}^{(j-1)} \|^{2}_{F}>\epsilon$ and $j<N_{\text{max}}$ \textbf{do}){
$j \gets j+1$\;
\textbf{E-step:}\; Compute the \textit{aposteriori} covariance and mean
\begin{align*}\boldsymbol{\Sigma}^{(j)} &= [\boldsymbol{\Phi}^{(j-1) H} \mathbf {R}_{v}^{-1} \boldsymbol{\Phi}^{(j-1)H} + (\boldsymbol{\Lambda}^{(j-1)})^{-1}]^{-1},\\ \boldsymbol{\mu}^{(j)} &= \boldsymbol{\Sigma}^{(j)} \boldsymbol{\Phi}^{(j-1)H} \mathbf {R}_{v}^{-1} \mathbf{y}
\end{align*}\;
\textbf{M-step1:} \;
Compute the hyperparameter estimates\;
\textbf{for} $i\gets 0 \text{ to } M_\tau N_\nu-1$ \textbf{ do} {\begin{align*}\widehat{\lambda}_i^{(j)} = \boldsymbol{\Sigma }^{(j)}(i,i) + | \boldsymbol{\mu}^{(j)}(i) |^2\end{align*}.}\;
\textbf{M-step2:}\;
Update estimate of the data matrix $\widehat{\mathbf {X}}_{d}^{(j)}$ by using LMMSE estimate of \eqref{eq:MMSE_joint_siso} or ZF estimate of \eqref{eq:zero_forcing_siso}.
}
\end{algorithm}
Upon simplifying the cost function of \eqref{eq:cost_function_LS}, the expression obtained is \begin{align}
 &= \mathbb{E}[\mathrm{Tr} \{ (\mathbf{Y}_{\text{DD},d} -\mathbf{\bar{H}}_{\text{DD}} \mathbf{X}_d)^H(\mathbf{Y}_{\text{DD},d} -\mathbf{\bar{H}}_{\text{DD}} \mathbf{X}_d)  \} ]\nonumber \\
&=\mathrm{Tr} \{ \mathbf{Y}_{\text{DD},d}^H \mathbf{Y}_{\text{DD},d}- \mathbf{Y}_{\text{DD},d}^H \mathbf{\widehat{H}}_{\text{DD}}^{(j)}\mathbf{X}_{d}\nonumber\\ &- \mathbf {X}_{d}^H ({\mathbf {\widehat{H}}}_{\text{DD}}^{(j)})^H \mathbf {Y}_{\text{DD},d} - \mathbf {X}_{d}^H \mathbb{E}[\mathbf{\bar{H}}_{\text{DD}}^H \mathbf {\bar{H}}_{\text{DD}}]\mathbf {X}_{d}\},
\label{eq:simplified_cost_function_LS}
\end{align}
where ${\mathbf {\widehat{H}}}_{\text{DD}}^{(j)}$ is the CSI estimate for the $j$th iteration, given as
\begin{equation}
    {\mathbf {\widehat{H}}}_{\text{DD}}^{(j)}=\mathbb{E} \big[ {\mathbf{\bar{H}}_{\text{DD}}} \big]
 = \sum_{u,v} \widehat{h}_{u,v}^{(j)}{\mathbf P}_{\text{rx}} \left(\mathbf{\bar\Pi}\right)^{u} \left(\mathbf{\bar\Delta}_{u}\right)^{v}{\mathbf {P}}_{\text{tx}}.
 \end{equation}
The quantity $\widehat{h}_{u,v}^{(j)}$ is the estimate of ${h}_{u,v}$ for the $j$th iteration. As shown in Appendix-A, the quantity $\mathbb{E}[\mathbf{\bar{H}}_{\text{DD}}^H \mathbf {\bar{H}}_{\text{DD}}]$ can be simplified to
\begin{equation}
\mathbb{E}[\mathbf{\bar{H}}_{\text{DD}}^H \mathbf {\bar{H}}_{\text{DD}}]= (\widehat{\mathbf{H}}_{\text{DD}}^{(j)})^H \widehat{\mathbf{H}}_{\text{DD}}^{(j)} + \boldsymbol{\Xi}^{(J)},
\end{equation}
where $\boldsymbol{\Xi}$ represents the channel estimation uncertainty, which is defined in terms of the error covariance matrix $\boldsymbol{\Sigma }_{h}$ of $\mathbf{\widehat{h}}_{\text{DD}}=\mathrm{vec} (\widehat{\mathbf {H}}_\text{DD})$ as
\begin{align}
    \boldsymbol{\Xi}(p,q)= \mathrm{Tr}\left[ \boldsymbol{\Sigma }_{h} (\tilde{p}-M+1: \tilde{p},\tilde{q}-M+1: \tilde{q}) \right],
\end{align}
where $\tilde{p}=pM, \tilde{q}=qM$. 
The detailed derivation of $\boldsymbol{\Xi}$ is given in Appendix B. 

Hence, M-step2 can be expressed as
\begin{align} 
\widehat{\mathbf{X}}_{d}^{(j)}=&\arg \min_{\mathbf {X}_{d}} \{ ||\mathbf {Y}_{\text{DD},d}-\widehat{\mathbf{H}}_{\text{DD}}^{(j)}\mathbf{X}_{d}||_F^2 +|| \left(\boldsymbol{\Xi }^{(j)}\right)^{\frac{1}{2}} \mathbf{X}_{d} ||_F^2 \}\\
=&\arg \min_{\mathbf {X}_{d}}\Bigg\{ \Bigg\Vert \begin{bmatrix}\mathbf {Y}_{\text{DD},d} \\ \mathbf {0} \end{bmatrix} -\begin{bmatrix}\widehat{\mathbf {H}}_{\text{DD}}^{(j)}\\ \left(\boldsymbol{\Xi }^{(j)}\right)^{\frac{1}{2}}\\ \end{bmatrix}\mathbf{X}_{d}\Bigg\Vert _F^2 \Bigg\}. 
\label{eq:zero_forcing_siso}
\end{align}
Note that the above detection rule is equivalent to zero forcing (ZF) signal detection for the data matrix $\mathbf{X}_d$. In order to further improve the detection performance, we also present an LMMSE-based detection procedure, which is derived in Appendix A. 

The simplified LMMSE-based data detection rule is formulated as
\begin{align}
 \widehat{\mathbf{X}}_d^{(j)}= {\widehat{\mathbf {H}}_{\text{DD}}^{(j) H}}\Big[\widehat{\mathbf {H}}_{\text{DD}}^{(j)}{\widehat{\mathbf {H}}_{\text{DD}}^{(j) H}}+ \boldsymbol{\Xi}^{(j)} +\frac{\sigma^2}{\sigma _d^2} \left( \mathbf{P}_{\text{rx}} \mathbf{P}_{\text{rx}}^H  \right) \Big]^{-1}\mathbf{Y}_{\text{DD},d}.    \label{eq:MMSE_joint_siso}
\end{align}
\section{AP-SIP based MIMO OTFS system model}
\label{sec:mimo_sys_mdl}
Consider an AP-SIP MIMO OTFS system constructed upon an OFDM-based multicarrier system. Let the number of transmitter antennas (TAs) in the MIMO system be $N_t$, and the number of receiver antennas (RAs) be $N_r$. The data and pilot inputs corresponding to the $t$th TA are $\mathbf{X}_t^d \in \mathbb{C}^{M \times K_1} $ and $ \mathbf{X}_t^p \in \mathbb{C}^{M \times K_2} $, respectively, for $1\leq t \leq N_t$. Similar to AP-SIP-based SISO OTFS systems, for each TA, the data and pilot inputs are affine-precoded using the semi-orthogonal matrices $\mathbf{D}$ and $\mathbf{P}$, respectively. Thus, the input $\mathbf{X}_t^\text{DD} \in \mathbb{C}^{M \times N}$  corresponding to each TA in the DD-domain is given by
\begin{equation}
    \mathbf{X}_t^\text{DD}=\mathbf{X}_t^d\mathbf{D}^{H}+\mathbf{X}_t^p\mathbf{P}^{H}.
    \label{eq:superimposed_data_MIMO}
\end{equation}
The DD-domain representation of the MIMO wireless channel corresponding to the $r$th RA and $t$th TA, where $1 \leq r \leq N_r, 1 \leq t \leq N_t$, can be modeled  as
\begin{equation} 
h_{r,t}(\tau,\nu) = \sum_{i=1}^{L_p} h_{i,r,t} \delta(\tau-\tau_{i}) \delta(\nu-\nu_{i}), \label{eq:ch_mimo}
\end{equation} 
where $\tau_i$ and $\nu_i$ represent the delay and Doppler shifts  introduced by the $i$th reflector. %Note that many of the coefficients $h_{i,r,t}$ will be zero owing to the significantly lower number of dominant reflectors in the channel.

Let the $t$th TA transmit the  symbol matrix $\mathbf{X}_t^\text{DD}$ over the DD-domain wireless channel  $\mathbf{H}_{r,t}^\text{DD}
\in \mathbb{C}^{M \times M}$ corresponding to $r$th RA and $t$th TA, where $ \mathbf{H}_{r,t}^\text{DD} =\mathbf{P}_\text{rx} \mathbf{H}_{r,t} \mathbf{P}_\text{tx}$,
At the $r$th RA the received signal $\mathbf{Y}_r^\text{DD} \in \mathbb{C}^ { M \times N}$ has a contribution from all the transmitted symbols $\mathbf{X}_t^\text{DD}$, and can be expressed as
\begin{equation}
    \mathbf{Y}_r^\text{DD}=\sum_{t=1}^{N_t}{\mathbf{H}_{r,t}^\text{DD}\mathbf{X}_t^\text{DD}+\mathbf{W}_r^\text{DD}}.
\label{eq:input_output_eq_MIMO}    
\end{equation}
By stacking the outputs corresponding to all the RAs, the output $\mathbf{Y}_{\text{DD}} \in \mathbb{C}^{MN_r \times N}$ is given as ${\mathbf{Y}}_{\text{DD}}=[ (\mathbf{Y}_1^{\text{DD}})^T   (\mathbf{Y}_2^{\text{DD}})^T  \hdots   (\mathbf{Y}_{N_r}^{\text{DD}})^T]^T$. Thus, the input-output relationship is given by
\begin{equation}
  {\mathbf{Y}}_{\text{DD}}={\widetilde{\mathbf{H}}}_{\text{DD}}{\mathbf{X}}_{\text{DD}}+{\mathbf{W}}
    _{\text{DD}},
    \label{eq:TF2DD_mat1_mimo} 
\end{equation}
where ${\mathbf{X}}_{\text{DD}}[(\mathbf{X}_1^{\text{DD}})^T  (\mathbf{X}_1^{\text{DD}})^T \hdots (\mathbf{X}_{N_t}^{\text{DD}})^T]^T\in \mathbb{C}^{M N_t \times N}$, ${\mathbf{W}}_{\text{DD}}=[ (\mathbf{W}_1^{\text{DD}})^T  (\mathbf{W}_2^{\text{DD}})^T \hdots  (\mathbf{W}_{N_r}^{\text{DD}})^T]^T\in \mathbb{C}^{MN_r \times N}$ and $\widetilde{{\mathbf{H}}}_{\text{DD}} \in \mathbb{C}^{MN_r \times MN_t}$ represents the DD-domain MIMO-OTFS channel given as
\begin{align}
\widetilde{{\mathbf{H}}}_{\text{DD}} 
&= \mathrm{blkmtx} \big\lbrace{\mathbf{H}_{r,t}^\text{DD}\big\rbrace}_{r=1, t=1}^{N_r, N_t}\nonumber \\
&= (\mathbf{I}_{N_r} \otimes \mathbf{P}_{\text{rx}})\Big[\mathrm{blkmtx}\big\lbrace{\mathbf{H}_{r,t}\big\rbrace}_{r=1,t=1}^{N_r, N_t}\Big] ({\mathbf{I}_{N_t} \otimes \mathbf{P}_{\text{tx}}}),
\label{eq:H_DD_mat_mimo_bar}
\end{align}
where channel matrix between the $r$th RA and $t$th TA is given by $\mathbf{H}_{r,t}=\sum_{i=1}^{L_p} h_{i,r,t} \left(\mathbf{\bar\Pi}\right)^{l_{i}} \left(\mathbf{\bar\Delta}_{l_i}\right)^{k_{i}}$ from \eqref{eq:ch_mimo}. %It follows from \eqref{eq:superimposed_data_MIMO} that the AP-SIP matrix $\mathbf{X}_{t}^\text{DD}$ comprises the data and pilot matrices and can be decoupled owing to the semi-orthogonality property of the precoder matrices $\mathbf{P}$ and $\mathbf{D}$, as described next. 
Substituting $\mathbf{X}_{t}^\text{DD}$ from \eqref{eq:superimposed_data_MIMO} in \eqref{eq:input_output_eq_MIMO} yields
\begin{equation}
    \mathbf{Y}_r^\text{DD}=\sum_{t=1}^{N_t}{\mathbf{H}_{r,t}^\text{DD}\left(\mathbf{X}_t^d\mathbf{D}^{H}+\mathbf{X}_t^p\mathbf{P}^{H}\right)+\mathbf{W}_r^\text{DD}}.
     \label{eq:receiver_MIMO}
\end{equation}
The data and pilot matrices can be decoupled by exploiting the semi-orthogonality of the TPC matrices $\mathbf{P}$ and $\mathbf{D}$, as described next.

One can now decouple the data output $\mathbf{Y}_r^{\text{DD},d}$ from $\mathbf{Y}_r^\text{DD}$ of \eqref{eq:receiver_MIMO} upon post-multiplying it by the data TPC matrix $\mathbf{D}$ as 
\begin{align}
    \mathbf{Y}_r^{\text{DD},d}=\mathbf{Y}_r^\text{DD}\mathbf{D}
    &= \sum_{t=1}^{N_t}{\mathbf{H}_{r,t}^\text{DD}\left(\mathbf{X}_t^d \mathbf{D}^{H}\mathbf{D}+\mathbf{X}_t^p\mathbf{P}^{H}\mathbf{D}\right)+\mathbf{W}_r^\text{DD}}\mathbf{D}
 \nonumber\\  &= \sum_{t=1}^{N_t}{\mathbf{H}_{r,t}^\text{DD}\mathbf{X}_t^d +\mathbf{W}_r^{\text{DD},d}}.
    \label{eq:data_decoupling_MIMO}
\end{align}
The simplification of  \eqref{eq:data_decoupling_MIMO} follows upon
exploiting the properties in \eqref{eq:semiorthogonal_property}. Furthermore, stacking the outputs corresponding to all RAs, the output of the receiver $\widetilde{\mathbf{Y}}_{\text{DD},d}\in \mathbb{C}^{MN_r \times K_1}$ is expressed as
\begin{equation}
    \widetilde{\mathbf{Y}}_{\text{DD},d}=\left[ (\mathbf{Y}_1^{\text{DD},d})^T   (\mathbf{Y}_2^{\text{DD},d})^T  \hdots   (\mathbf{Y}_{N_r}^{\text{DD},d})^T\right]^T.
\end{equation}
The system model of decoupled data detection is given by
\begin{equation}
    \widetilde{\mathbf{Y}}_{\text{DD},d}=\widetilde{\mathbf{H}}_\text{DD}\widetilde{\mathbf{X}}_d+\widetilde{\mathbf{W}}
    _{\text{DD},d},
\end{equation}
where \begin{align*}\widetilde{{\mathbf{X}}}_d&=\left[(\mathbf{X}_1^d)^T  (\mathbf{X}_2^d)^T \hdots (\mathbf{X}_{N_t}^d)^T\right]^T\in \mathbb{C}^{M N_t \times N},\\ \widetilde{\mathbf{W}}_{\text{DD},d}&=\left[ (\mathbf{W}_1^{\text{DD},d})^T  (\mathbf{W}_2^{\text{DD},d})^T \hdots  (\mathbf{W}_{N_r}^{\text{DD},d})^T\right]^T\in \mathbb{C}^{MN_r \times N}\end{align*}. 
Once again, the MMSE-based detector can be formulated as
\begin{align}
\widetilde{\mathbf{X}}_d^{\text{MMSE}} &= \left( \widetilde{\mathbf{H}}_{\text{DD}}^H \widetilde{\mathbf{R}}_{\text{W},d}^{-1} \widetilde{\mathbf{H}}_{\text{DD}}  + \frac{1}{\sigma_d^2}\mathbf{I}_{MN_t} \right)^{-1} \widetilde{\mathbf{H}}_{\text{DD}}^H \widetilde{\mathbf{R}}_{\text{W},d}^{-1} \widetilde{\mathbf{Y}}_{\text{DD},d}, \label{eq:mmse_detector_MIMO}
\end{align}
where $\widetilde{\mathbf{R}}_{\text{W},d} = \left(\mathbf{I}_{N_r} \otimes \mathbf{R}_{\text{W},d}\right) \in \mathbb{C}^{MN_r \times MN_r}$ denotes the noise covariance matrix. 
\section{Sparse DD-domain CSI estimation model for AP-SIP MIMO OTFS systems }
\label{sec:sparse_estimation_model_mimo}
When relying in the received signal in \eqref{eq:receiver_MIMO}, one can now decouple the pilot output $\mathbf{Y}_r^{\text{DD},p}$ from $\mathbf{Y}_r^\text{DD}$ of \eqref{eq:receiver_MIMO} by post multiplying it by the TPC matrix $\mathbf{P}$ for CSI estimation as 
\begin{align}
    \mathbf{Y}_r^{\text{DD},p}=\mathbf{Y}_r^\text{DD}\mathbf{P}
    &= \sum_{t=1}^{N_t}{\mathbf{H}_{r,t}^\text{DD}\left(\mathbf{X}_t^d \mathbf{D}^{H}\mathbf{P}+\mathbf{X}_t^p\mathbf{P}^{H}\mathbf{P}\right)+\mathbf{W}_r^\text{DD}}\mathbf{P}
  \nonumber\\&= \sum_{t=1}^{N_t}{\mathbf{H}_{r,t}^\text{DD}\mathbf{X}_t^p +\mathbf{W}_r^{\text{DD},p}}.
    \label{eq:pilot_decoupling_MIMO}
\end{align}
The simplification of \eqref{eq:pilot_decoupling_MIMO} follows upon exploiting the results of \eqref{eq:semiorthogonal_property}.
Let the maximum delay spread of the DD-domain MIMO OTFS channel be $M_\tau$ and the maximum Doppler spread be denoted by $N_\nu$. For an under-spread channel, the parameters obey $l_{max}=\text{max}(l_i) < M_\tau << M$ and $k_{max}=\text{max}(k_i) < N_\nu << N$.
{
Toward introducing fractional Doppler, consider a grid of size $G_\tau \times G_\nu$, such that $G_\nu>>N_\nu$, similar to SISO OTFS system defined in Section-III}. Let furthermore $h_{i,j,r,t}$ represent the path gain of the ${i}$th delay tap and $j$th Doppler tap for the $r$th RA and $t$th TA. The channel ${h}_{r,t}(\tau,\nu)$ between the $r$th RA and $t$th TA is given by
\begin{equation} 
{h}_{r,t}(\tau,\nu) = \sum_{i=0}^{M_\tau-1}\sum_{j=0}^{{G_{\nu}-1}
} h_{i,j,r,t} \delta(\tau-\tau_{i}) \delta(\nu-\nu_{j}). 
\end{equation}  
%or it can be expressed as
%,\begin{equation} \mathbf{h}_{r,t} =\sum_{i=1}^{M_\tau}\sum_{j=1}^{N_{\nu}} h_{i,j,r,t} {{\bar{\mathbf{\Pi}}}^i{{\bar{\mathbf{\Delta}}}}^j}\label{eq:ch_mimo_estimation}\end{equation} 
Upon substituting the expression for $\mathbf{H}_{r,t}^\text{DD}$ into (\ref{eq:pilot_decoupling_MIMO}) followed by vectorization, the resulting expression is
\begin{align}
    &\mathbf{y}_r^{\text{DD},p}\nonumber\\&\!\!=\!\!\sum_{t=1}^{N_t}
    \mathrm{vec}\left(\mathbf{P}_\text{rx} {\sum_{i=0}^{M_\tau-1}\sum_{j=0}^{G_{\nu}-1} {h_{i,j,r,t}(\bar{\mathbf{\Pi}})^i ({\bar{\mathbf{\Delta}}_i})^j}}\mathbf{P}_\text{tx} \mathbf{X}_t^p\right)\!\!+\!\!\mathbf{w}_r^{\text{DD},p},
    \label{eq:pilot_decoupling_MIMO_sparse}
\end{align}
where $\mathbf{y}_r^{\text{DD},p}=\mathrm{vec}(\mathbf{Y}_r^\text{DD}\mathbf{P}) \in \mathbb{C}^{MK_2 \times 1}$ ,  $\mathbf{w}_r^{\text{DD},p}=\mathrm{vec}(\mathbf{W}_r^\text{DD}\mathbf{P})\in \mathbb{C}^{MK_2 \times 1}$. Simplifying the previous expression yields
\begin{equation}
\mathbf{y}_r^{\text{DD},p}=\sum_{t=1}^{N_t}\sum_{i=0}^{M_\tau-1}\sum_{J=0}^{G_{\nu}-1}{\mathbf{w}}_{i,j,t}^{p}h_{i,j,r,t}+\mathbf{w}_{r}^{\text{DD},p},
\end{equation}
where $ \mathbf{w}_{i,j,t}^{p}
=\left(\mathbf{I}_{K_2}\otimes{\mathbf{P}}_\text{rx}(\bar{\mathbf{\Pi}})^i ({\bar{\mathbf{\Delta}}_i})^j\mathbf{P}_\text{tx}\right)\mathbf{x}_t^p 
\in \mathbb{C}^{MK_2 \times 1} $ and $\mathbf{x}_t^{p}=\mathrm{vec}\left(\mathbf{X}_t^p\right) \in \mathbb{C}^{MK_2 \times 1}$. The above relationship may also be  expressed in vectorized form as
\begin{equation}
\mathbf{y}_r^{\text{DD},p}=\sum_{t=1}^{N_t}\boldsymbol{\Omega}_t^{p}\mathbf{h}_{r,t}+\mathbf{w}_r^{{\text{DD},p}},
\label{eq:single_antenna_MIO}
\end{equation}
where $\mathbf{y}_r^{\text{DD},p}$ represents the observation vector for the decoupled pilots at the $r$th RA corresponding to all the TAs, $\boldsymbol{\Omega}_t^p \in \mathbb{C}^{MK_2 \times M_\tau G_\nu}$ represents the dictionary matrix for the $t$th TA, which is given by 
$
    \boldsymbol{\Omega}_t^p = [\mathbf{w}_{0,0,t}^{p}\hdots \mathbf{w}_{0,G_\nu-1,t}^{p}\hdots \mathbf{w}_{M_\tau-1,0,t}^{p}\hdots \mathbf{w}_{M_\tau-1,G_\nu-1,t}^{p}].
$
The quantity $\mathbf{h}_{r,t} \in \mathbb{C}^{M_\tau G_\nu \times 1}$ is the channel coefficient vector for the RA-TA pair $(r,t)$, which is given by 
\begin{align}
\mathbf{h}_{r,t}=&[h_{0,0,r,t}\hdots h_{0,G_\nu-1,r,t}\hdots\nonumber\\ &\hdots h_{M_\tau-1,0,r,t}\hdots h_{M_\tau-1,G_\nu-1,r,t}]^T.
\label{eq:sparse_channel_coefficient_vector_MIMO_r_t}
\end{align}
The above result can be expressed as 
\begin{equation}
\mathbf{y}_r^{\text{DD},p}=\widetilde{\boldsymbol{\Omega}}_p\mathbf{h}_r+\mathbf{w}_{r}^{\text{DD},p},
  \label{eq:output_RA}
  \end{equation}
where the dictionary matrix $\widetilde{\boldsymbol{\Omega}}_p \in \mathbb{C}^{MK_2 \times M_\tau G_\nu Nt}$
is given as $    \widetilde{\boldsymbol{\Omega}}_p =\left[\boldsymbol{\Omega}_1^p, \boldsymbol{\Omega}_2^p, \hdots, \boldsymbol{\Omega}_{N_t}^p \right],
$ and the channel coefficient vector $\mathbf{h}_r\in \mathbb{C}^{M_\tau G_\nu N_t \times 1 } $ for a particular $r$ is given as
$ \mathbf{h}_r=[ \mathbf{h}_{r,1}^T, \mathbf{h}_{r,2}^T, \hdots, \mathbf{h}_{r,Nt}^T ]^T
\label{eq:sparse_channel_coefficient_vector_MIMO_r}$. Upon concatenating the outputs corresponding to all the RAs using \eqref{eq:output_RA}, the resultant observation matrix $\widetilde{\mathbf{Y}}_{\text{DD},p}=\left[\mathbf{y}_1^{\text{DD},p}, \mathbf{y}_2^{\text{DD},p}, \hdots, \mathbf{y}_{Nr}^{\text{DD},p}\right]$ is given by
\begin{equation}
\widetilde{\mathbf{Y}}_{{\text{DD},p}}=  \widetilde{\boldsymbol{\Omega}}_p \widetilde{\mathbf{H}} + \widetilde{\mathbf{W}}_p,
\label{eq:MIMO_sparse_problem_pilot}
\end{equation}
where the channel coefficient matrix $\widetilde{\mathbf{H}}\in \mathbb{C}^{M_\tau G_\nu N_t \times N_r}$ across all the RAs is given by 
$\widetilde{\mathbf{H}}=\left[\mathbf{h}_1, \mathbf{h}_2, \hdots, \mathbf{h}_{Nr}\right]$, and the corresponding noise matrix $\widetilde{\mathbf{W}}_p$ is $\widetilde{\mathbf{W}}_p=\left[ \mathbf{w}_1^{\text{DD},p}, \mathbf{w}_{2}^{\text{DD},p}, \hdots, \mathbf{w}_{N_r}^{\text{DD},p}\right]\in \mathbb{C}^{M k_2\times N_r}$.
For the model shown in \eqref{eq:MIMO_sparse_problem_pilot}, the conventional MMSE estimate of the CSI, denoted by $\widehat{\mathbf{H}}_{\text{MMSE}}$, is given as 
\begin{align} \widehat{\mathbf {H}}_{\text{MMSE}} = \left(\widetilde{\boldsymbol{\Omega}}_p^H \mathbf {R}_{w_p}^{-1} \widetilde{\boldsymbol{\Omega}}_p + \mathbf {R}_h^{-1}\right)^{-1} \widetilde{\boldsymbol{\Omega}}_p^H \mathbf {R}_{w_p}^{-1} \widetilde{\mathbf{Y}}_{\text{DD},p}, \end{align}
where the channel's covariance matrix $\mathbf {R}_h \in \mathbb {C}^{M_{\tau }G_{\nu }N_t \times M_{\tau }G_{\nu }N_t} $ is unknown and hence it is set as $\mathbf{I}_{M_\tau G_\nu N_t}$, while the noise covariance matrix is  ${R}_{\text{w},p}=\sigma^2 ({\mathbf{I}_{K_2} \otimes \mathbf{P}_\text{rx} \mathbf{P}_\text{rx}^H}) \in \mathbb {C}^{MK_2 \times MK_2}$. The next section describes the BL-based estimation technique conceived for MIMO OTFS systems, where we exploit the sparsity in addition to determining the channel's covariance, in order to improve the performance of the LMMSE estimate.
\subsection{Pilot aided Bayesian learning (PA-BL) for AP-SIP MIMO OTFS
systems} 
As for the MIMO OTFS channel, the delay and Doppler shifts corresponding to the multipath components are identical for all TA/RA pairs. Thus, the sparsity profile of the vectors $h_{r,t}\in \mathbb{C}^{M_\tau G_\nu \times 1}$ is identical for all pairs $(r,t)$. The vector $\mathbf{h}_r \in \mathbb{C}^{M_\tau G_\nu N_t \times 1}$ obtained by stacking $h_{r,t}$ for all values of $t$, exhibits a group sparse structure, since the ${i}$th group of coefficients, $0\leq i \leq M_{\tau}G_{\nu}-1$, given by \vspace{-5pt} \begin{align*}\bigg \{ \mathbf {h}_{r}\big (\left [{(t-1) M_\tau G_\nu }\right] + i \big) \bigg \}_{t=1}^{N_{t}},\end{align*} are either all zero or non-zero. Moreover, the matrix $\widetilde{\mathbf{H}}$, which is obtained by concatenating vector $\mathbf{h}_r$ for all values of $r$, exhibits a row sparse structure, since the $i$th group of rows denoted by the row indices $\big \{ \left [{(t-1) M_\tau G_\nu }\right]+i \big \}_{t=1}^{N_{t}}$ are either all zero or non-zero. Thus, $\widetilde{\mathbf{H}}$ has a simultaneously row-group sparse structure. The PA-BL framework designed for the estimation of this row-group sparse matrix begins by assigning a parameterized Gaussian prior to the sparse channel vector $\mathbf{h}_{r,t}$ as
\begin{equation*}
f(\mathbf {h}_{r,t}; \boldsymbol{\Lambda}) = \prod _{i=0}^{M_\tau G_\nu-1}\frac {1}{(\pi \lambda _{i})} \exp \left({-\displaystyle \frac {\vert \mathbf {h}_{r,t}(i)\vert ^{2}}{\lambda _{i}}}\right).
%\label{eq:parameterized_gaussian_mimo}
\end{equation*}
For the PA-BL framework, the sparsity profile of $\mathbf{h}_{r,t}$ is identical for each RA/TA pair. The update equations for the $j$th iteration are summarized in Algorithm \ref{algo_PA_BL_mimo}. 
The estimate $\widehat{\mathbf{h}}_{i,j,r,t}$ is obtained as
$
\widehat {h}_{i,j,r,t}=\widehat {\mathbf {H}}^{{\text{SBS PA-BL}}}\left [{i(G_{\nu }+1)+j+(t-1)M_\tau G_\nu,r-1}\right].
$
The estimate $\widehat {\mathbf {H}}_{r,t}^\text{DD}$ of the channel matrix corresponding to the $r$th RA and $t$th TA is obtained as
\begin{align}
\widehat {\mathbf {H}}_{r,t}^\text{DD}= \mathbf{P}_{\text{rx}}\Big[{ \sum _{i=0}^{M_{\tau}-1} \sum _{j=0}^{G_{\nu}-1} \widehat {h}_{i,j,r,t}\left ({\bar {\boldsymbol{\Pi }}}\right)^{i} (\bar {\boldsymbol{\Delta}}_{i})^j}\Big] \mathbf{P}_{\text{tx}}.
\end{align}
\begin{algorithm}
\DontPrintSemicolon
\caption{PA-BL-based row-group sparse CSI estimation in MIMO OTFS systems}\label{algo_PA_BL_mimo}
\KwIn{Decoupled pilot output matrix $\widetilde{\mathbf{Y}}_{\text{DD},p}$, pilot dictionary matrix $\widetilde{\boldsymbol{\Omega}}_p$ from (62), noise covariance matrix $\mathbf{R}_{\widetilde{\mathbf{W}}_p}$, threshold $\epsilon \text{\ and maximum iteration\ } N_{\text{max}}$.}
\SetKwInput{kwInit}{Initialization}
\kwInit{$\widehat{\lambda_i}^{(0)}=1, \forall\ 0 \leq i \leq {M_\tau G_\nu}-1, \widehat{\boldsymbol{\Lambda}}^{(0)}=\mathbf{I}_{M_\tau G_\nu} $, $\widehat{\boldsymbol{\Lambda}}^{(-1)} = \mathbf{0}$, set $j = 0$}
\vspace{10pt}
\KwOut{$\widehat{\mathbf{H}}^{\text{SBS PA-BL}} = \boldsymbol{\mathcal{M}}^{(j)}$.}
\vspace{2pt}
\SetKwBlock{Begin}{while}{end function}
\Begin($\|\widehat{\boldsymbol{\Lambda}}^{(j)} - \widehat{\boldsymbol{\Lambda}}^{(j-1)} \|^{2}_{F}>\epsilon$ and $j<N_{\text{max}}$){
$j \gets j+1$\;
\textbf{E-step:}\; Compute the \textit{aposteriori} covariance and mean
\begin{align*} \boldsymbol{\Sigma}^{(j)} &= [\widetilde{\boldsymbol{\Omega}}_p^H \mathbf{R}_{\widetilde{\mathbf{W}}_p}^{-1} \widetilde{\boldsymbol{\Omega}}_p + (\mathbf{I}_{N_t} \otimes (\boldsymbol{\Lambda }^{(j-1)})^{-1})]^{-1},\\ \boldsymbol{\mathcal{M}}^{(j)} &= \boldsymbol{\Sigma}^{(j)} \widetilde{\boldsymbol{\Omega}}_p^H \mathbf{R}_{\widetilde{\mathbf{W}}_p}^{-1} \widetilde{\mathbf{Y}}_{\text{DD},p}
\end{align*}
\textbf{M-step:}\; Compute the hyperparameter estimates\;
\SetKwBlock{Begin}{for}{end function} 
\Begin($i\gets 0 \text{ to } M_\tau N_\nu-1$ \textbf{ do}) {
\begin{align*}
\widehat {\lambda }_{i}^{(j)} &=\frac {1}{N_{r}N_{t}} \sum _{r=1}^{N_{r}} \sum _{t=1}^{N_{t}} |{ \boldsymbol {\mathcal {M}}^{(j)}[(t-1)N_{t}+i,r-1]}|^{2}\\ &+\frac{1}{N_{t}} \sum _{t=1}^{N_{t}} {\boldsymbol{\Sigma }}^{(j)} [(t-1)N_{t}+i,(t-1)N_{t}+i]\end{align*}.}
}
\end{algorithm}
\subsection{Data aided joint CSI estimation and data detection for AP-SIP MIMO OTFS systems}
\begin{algorithm}[t]
\DontPrintSemicolon
\caption{DA-BL-based row-group sparse CSI estimation in MIMO OTFS systems}\label{algo_DA_BL_mimo}
\KwIn{Decoupled joint output matrix $\widetilde{\mathbf{Y}}$ from (71), joint dictionary matrix $\widetilde{\boldsymbol{\Phi}}$ from (70), joint noise covariance matrix $\mathbf{R}_v=\mathrm{blkdiag}(\mathbf{R}_{\text{w},d} , \mathbf{R}_{\text{w},p} )$, threshold $\epsilon$, $N_{\text{max}}$.}
\SetKwInput{kwInit}{Initialization}
\kwInit{$\widehat{\mathbf{\Lambda}}^{(0)}=\widehat{\boldsymbol{\Lambda}}_{\text{PA-BL}}^{(j)}$, $\widehat{\boldsymbol{\Lambda}}^{(-1)} = \mathbf{0}$, $j = 0$, $\widetilde{\boldsymbol{\Omega}}_d^{(-1)}=\widehat{\boldsymbol{\Omega}}_d^\text{PA-BL}$, $\widetilde{\mathbf{\Phi}}^{(-1)}=
{\begin{bmatrix}\widetilde{{\boldsymbol{\Omega}}}_{d}^{(-1)}\\ \widetilde{\boldsymbol{\Omega}}_p\\ \end{bmatrix}}$}
\vspace{10pt}
\KwOut{$\widehat{\mathbf{H}}^{\text{DA-BL}} = \boldsymbol{\mathcal{M}}^{(j)}$ and $\widehat{\mathbf{X}}^{\text{DA-BL}}=\widehat{\mathbf{X}}_d^{(j)}$.}
\vspace{2pt}
\SetKwBlock{Begin}{while}{end function}
\Begin(\(\)$\|\widehat{\boldsymbol{\Lambda}}^{(j)} - \widehat{\boldsymbol{\Lambda}}^{(j-1)} \|^{2}_{F}>\epsilon$ and $j<N_{\text{max}}$ \textbf{do}){
$j \gets j+1$\;
\textbf{E-step:}\; Compute the \textit{aposteriori} covariance and mean\;
\begin{align*} \boldsymbol{\Sigma}^{(j)} &= [\widetilde{\boldsymbol{\Phi}}^H \mathbf{R}_{\widetilde{\mathbf{v}}}^{-1} \widetilde{\boldsymbol{\Phi}} + (\mathbf{I}_{N_t} \otimes (\boldsymbol{\Lambda }^{(j-1)})^{-1})]^{-1},\\ \boldsymbol{\mathcal{M}}^{(j)} &= \boldsymbol{\Sigma}^{(j)} \widetilde{\boldsymbol{\Phi}}^H \mathbf{R}_{\widetilde{\mathbf{v}}}^{-1} \widetilde{\mathbf{Y}}.
\end{align*}
\textbf{M-step1:}\; Compute the hyperparameter estimates\;
\SetKwBlock{Begin}{for}{end function} 
\Begin($i\gets 0 \text{ to } M_\tau N_\nu-1$ \textbf{ do}) {
\begin{align*}\widehat {\lambda }_{i}^{(j)}&=\frac {1}{N_{r}N_{t}} \sum _{r=1}^{N_{r}} \sum _{t=1}^{N_{t}} | \boldsymbol {\mathcal {M}}^{(j)}[(t-1)N_{t}+i,r-1] |^{2} \\&+\frac {1}{N_{t}} \sum _{t=1}^{N_{t}} {\boldsymbol{\Sigma }}^{(j)}[(t-1)N_{t}+i,(t-1)N_{t}+i ]\end{align*}.}
\textbf{M-step2:}
Update estimate of data matrix $\widehat{\mathbf {X}}_{d}^{(j)}$ by using 
%ZF estimate of \eqref{eq:zero_forcing_mimo} or 
the LMMSE estimate of \eqref{eq:MMSE_joint_mimo}
}
\end{algorithm}
Commencing with the output $\mathbf{Y}_r^{\text{DD},d}$ in (\ref{eq:data_decoupling_MIMO}) and decoupling the data from the superimposed pilots, followed by substituting for $\mathbf{H}_{r,t}^\text{DD}$ and vectorizing, the resulting expression is
\begin{align}
    &\mathbf{y}_r^{\text{DD},d}
    \nonumber\\&\!\!=\!\!
    \sum_{t=1}^{N_t}
    \mathrm{vec}
    \left (\sum_{i=0}^{M_\tau-1}\sum_{J=0}^{G_{\nu}-1} 
    \mathbf{P}_\text{rx} h_{i,j,r,t}
    ({\bar {\boldsymbol{\Pi }}})^{i} (\bar {\boldsymbol{\Delta}}_{i})^j \mathbf{P}_\text{tx} \mathbf{X}_t^d
    \right) \!\!+\!\!\mathbf{w}_r^{\text{DD},d},
    \label{eq:data_decoupling_MIMO_sparse}
\end{align}
where $\mathbf{y}_r^{\text{DD},d}=\mathrm{vec}(\mathbf{Y}_r^\text{DD}.\mathbf{D}) \in \mathbb{C}^{MK_1 \times 1}$ ,  $\mathbf{w}_r^{\text{DD},d}=\mathrm{vec}(\mathbf{W}_r^\text{DD}.\mathbf{D})\in \mathbb{C}^{MK_1 \times 1}$. Further simplification of \eqref{eq:data_decoupling_MIMO_sparse} yields
\begin{equation}
   \mathbf{y}_r^{\text{DD},d}=\sum_{t=0}^{N_t-1}\sum_{i=0}^{M_\tau-1}\sum_{J=0}^{G_{\nu}-1}
    { \mathbf{w}}_{i,j,t}^{d}h_{i,j,r,t}+\mathbf{w}_{r}^{\text{DD},d},
\end{equation}
where $ \mathbf{w}_{i,j,t}^{d}
=({\mathbf{I}_{K_1}\otimes{\mathbf{P}}_\text{rx}\bar{\mathbf{\Pi}}}^i {\bar{\mathbf{\Delta}}}^j\mathbf{P}_\text{tx})\mathbf{x}_t^d \in \mathbb{C}^{MK_1 \times 1} $. The above relationship can be  expressed in the compact form of
\begin{equation}
  \mathbf{y}_r^{\text{DD},d}=\sum_{t=0}^{N_t-1}\mathbf{\Omega}_t^{d}\mathbf{h}_r+\mathbf{w}_r^{\text{DD},d},
\end{equation}
where the dictionary matrix $\mathbf{\Omega}_t^d \in \mathbb{C}^{MK_1 \times M_\tau N_\nu}$ corresponding to the $t$th TA is given by
$\mathbf{\Omega}_t^d =[\mathbf{w}_{0,0,t}^{d} \hdots \mathbf{w}_{0,G_\nu-1,t}^{d} \hdots \mathbf{w}_{M_\tau-1,0,t}^{d} \hdots \mathbf{w}_{M_\tau-1,G_\nu-1,t}^{d}]$.
%\label{eq:dictionary_matrix_r_d}
%\end{align}
The channel coefficient vector $\mathbf{h}_{r,t} \in \mathbb{C}^{M_\tau G_\nu \times 1}$ for a particular RA and TA pair $(r,t)$ is given by \eqref{eq:sparse_channel_coefficient_vector_MIMO_r_t}.
The above expression can be further simplified as 
\begin{equation}
  \mathbf{y}_r^{\text{DD},d}=\widetilde{\mathbf{\Omega}}_d.\mathbf{h}_r+\mathbf{w}_{r}^{\text{DD},d},  \end{equation}
where $\mathbf{y}_r^{\text{DD},d}$ represents the observation vector for the decoupled data at the $r$th RA corresponding to all the TAs. Furthermore, the dictionary matrix $\widetilde{\mathbf{\Omega}}_d \in \mathbb{C}^{MK_1 \times M_\tau G_\nu Nt}$ is given as $\widetilde{\mathbf{\Omega}}_d =[\mathbf{\Omega}_1^d, \mathbf{\Omega}_2^d, \hdots, \mathbf{\Omega}_{N_t}^d ]$, and $\mathbf{h}_r$ for a particular value of $r$ is given by $ \mathbf{h}_r=[ \mathbf{h}_{r,1}^T, \mathbf{h}_{r,2}^T, \hdots, \mathbf{h}_{r,Nt}^T ]^T$. Upon Concatenating the output vectors $\mathbf{y}_r^{\text{DD},d}$ corresponding to all the RAs, the resulting observation matrix  $\widetilde{\mathbf{Y}}_{\text{DD},d} =[\mathbf{y}_1^{\text{DD},d}, \mathbf{y}_2^{\text{DD},d}, \hdots, \mathbf{y}_{N_r}^{\text{DD},d}] \in \mathbb{C}^{MK_1 \times N_r}$ can be formulated out as
\begin{equation}
\widetilde{\mathbf{Y}}_{{\text{DD},d}}=  \widetilde{\mathbf{\Omega}}_d \widetilde{\mathbf{H}} + \widetilde{\mathbf{W}}_d,
\label{eq:MIMO_sparse_problem_data}
\end{equation}
where $\mathbf{W}_{\text{DD},d}=[ \mathbf{w}_1^{\text{DD},d}, \mathbf{w}_{2}^{\text{DD},d}, \hdots, \mathbf{w}_{Nr}^{\text{DD},d}]$.
When aiming for data-aided AP-SIP-based MIMO OTFS CSI estimation, one can stack the outputs from (\ref{eq:MIMO_sparse_problem_data}) and (\ref{eq:MIMO_sparse_problem_pilot}) to obtain the joint CSI estimation and detection model of
\begin{align} 
{\underbrace{\begin{bmatrix}\widetilde{\mathbf{Y}}_{\text{DD},d}\\ \widetilde{\mathbf{Y}}_{\text{DD},p} \end{bmatrix}}_{{\widetilde{\mathbf{Y}}} \in \mathbb {C}^{MN\times N_r} }}&={\underbrace{\begin{bmatrix}\bf \widetilde{\mathbf{\Omega}}_{d}\\ {\bf \widetilde{\mathbf{\Omega}}_{p}}\\ \end{bmatrix}}_{\widetilde{\boldsymbol{\Phi}} \in \mathbb {C}^{MN \times M_\tau G_\nu N_t} }}{\bf \widetilde{\mathbf{H}}} + {\underbrace{\begin{bmatrix}\mathbf {\widetilde{W}}_d\\ \mathbf {\widetilde{W}}_p \end{bmatrix}}_{\widetilde{\mathbf {V}} \in \mathbb {C}^{MN\times N_r} }}.\end{align}
Thus, our compact model of data-aided (DA) CSI estimation is given by
\begin{align}
\widetilde{\mathbf{Y}} =\widetilde{\mathbf{\Phi}}\widetilde{\mathbf{H}}+ \widetilde{\mathbf{V}}.
\label{eq:output_joint_mimo}
\end{align}
As for the DA-BL framework, the parameterized Gaussian prior given by $f(\mathbf {h}_{r,t}; \boldsymbol{\Lambda})$ is assigned to the sparse channel vector $\mathbf{h}_{r,t}$, similar to the PA-BL framework. The update $\widehat{\mathbf {X}}_{d}^{(j)}$ for the data matrix is determined as
$\widehat{\mathbf{X}}_{d}^{(j)} = \arg \max _{\mathbf{X}_{d}} \mathbb {E} \Big\{  \log [ p(\widetilde{\mathbf {Y}}_{\text{DD},d}\mid \widetilde{\mathbf {H}}; \mathbf {X}_{d})] \Big\},$
which can be further formulated as 
\begin{align}
\widehat{\mathbf{X}}_{d}^{(j)}=& \arg \min _{\substack{\mathbf {X}_{d}}} \mathbb{E}\left\{ || \widetilde{\mathbf{Y}}_{\text{DD},d}-\widetilde{{\boldsymbol{\Omega}}}_d \widetilde{\mathbf{H}}||_F^2 \right\}
\nonumber\\&\equiv \arg \min _{\substack{\mathbf {X}_{d}}} \mathbb{E}\left\{ || {\mathbf{Y}}_{\text{DD},d} -\widetilde{\mathbf{{H}}} \mathbf{X}_d ||_F^2\right \}.
\label{eq:cost_function_LS_mimo}
\end{align}
Let ${\mathbf{\widehat{H}}}^{(j)}$ denote the CSI estimate for the $j$th iteration given as ${\mathbf{\widehat{H}}}^{(j)}=\boldsymbol{\mathcal{M}}^{(j)}$. 
%upon simplifying the cost function similar to SISO case the ZF detector is given as 
%\begin{align} 
%\widehat{\mathbf{X}}_{d}^{(j)}=
%\arg \min_{\mathbf {X}_{d}}\Bigg\{ \Bigg\Vert \begin{bmatrix}\mathbf {Y}_{\text{DD},d} \\ \mathbf {0} \end{bmatrix} -\begin{bmatrix}\widehat{\mathbf {H}}^{(j)}\\ \left(\boldsymbol{\Xi }^{(j)}\right)^{\frac{1}{2}}\\ \end{bmatrix}\mathbf{X}_{d}\Bigg\Vert _F^2 \Bigg\}. 
%\label{eq:zero_forcing_mimo}
%\end{align}
%Whereas, 
Then the simplified LMMSE-based data detection rule is formulated as
\begin{align}
 \widehat{\mathbf{X}}_d^{(j)}= {(\widehat{\mathbf {H}}^{(j)})^ H}\left[\widehat{\mathbf {H}}^{(j)}{(\widehat{\mathbf {H}}^{(j) })^H}+ \boldsymbol{\Xi}^{(j)} +\frac{\sigma^2}{\sigma _d^2} \left( \mathbf{P}_{\text{rx}} \mathbf{P}_{\text{rx}}^H  \right) \right]^{-1}\!\!\!\!\!\!\!\!\!\widetilde{\mathbf{Y}}_{\text{DD},d},
\label{eq:MMSE_joint_mimo}
\end{align}
where the $(p,q)$th element of the matrix $\boldsymbol{\Xi}$  is given as
\begin{equation}
\boldsymbol{\Xi }^{(j)}(p,q)= \mathrm{Tr}\left[ \boldsymbol{\Sigma}_{h}^{(j)} (\tilde{p}-MN_r+1: \tilde{p},\tilde{q}-MN_t+1: \tilde{q}) \right].
\end{equation}
The quantity $\boldsymbol{\Sigma }_{h}^{(j)}=(\mathbf{I}_{N_r N_t}\otimes \boldsymbol{\zeta})(\mathbf{I}_{N_r} \otimes{\boldsymbol{\Sigma}}^{(j)})(\mathbf{I}_{N_r N_t}\otimes \boldsymbol{\zeta}^H)$ is the covariance matrix of  $\mathrm{vec}(\widetilde{\mathbf{H}}_\text{DD})$, $\boldsymbol{\Sigma}^{(j)}$ is the covariance matrix obtained from the E-step of Algorithm 4 and ${\boldsymbol{\zeta}} \in \mathbb {C}^{M^2\times M_\tau G\nu}$ is defined by \eqref{eq:zeta},
where $\boldsymbol{\varphi}_i^j=\mathrm{vec}\left[\mathbf{P}_\text{rx} 
({\bar {\boldsymbol{\Pi }}})^{i} (\bar {\boldsymbol{\Delta}}_{i})^j \mathbf{P}_\text{tx}\right] \in \mathbb{C}^{M^2 \times 1}$,  $\tilde{p}=pMN_r$ and $\tilde{q}=qMN_t$, as derived in Appendix B.
{
\subsection{Computational complexity and comparative efficiency analysis}
The computational complexities of the PA-BL and DA-BL techniques for the SISO OTFS system are of the order $O(M_\tau^3 G_\nu^3)$, which arise due to the matrix inversion of size-$M_\tau G_\nu \times M_\tau G_\nu$. Similarly, for the MIMO OTFS system, the computational complexities of both the proposed schemes can be shown to be of the order $O(M_\tau^3 G_\nu^3 N_t^3)$ due to the matrix inversion of size-$M_\tau N_\nu N_t \times M_\tau N_\nu N_t$ necessitated by each of them. On the other hand, the worst-case complexities of the OMP \cite{suraj2021bayesian} scheme for a SISO OTFS system, and its counterpart,  the RG-OMP \cite{suraj2021row_group} algorithm for a MIMO OTFS system, are of the order $O(M^3K_2^3)$, which is owing to the intermediate LS estimate computation required in each iteration.  
The efficiency of the proposed framework can be expressed as $S_{e}=1-\rho$, where $\rho$ is the overhead in each frame. It is worth noting that the proposed CSI estimation framework transmits $MK_2$ pilot symbols in a SISO OTFS frame comprising $MN$ symbols, and $MK_2N_t$ pilot symbols in a MIMO OTFS frame comprising $MNN_t$ symbols. Thus, the overhead is $\rho_{AP-SIP}=\frac{K_2}{N}$ and the respective efficiency is $S^{AP-SIP}_e=1-\frac{K_2}{N}=\frac{K_1}{N}$ for both SISO as well as MIMO OTFS systems.\\
\begin{table*}[]
    \caption{Comparative Efficiency Analysis}
       \centering
    \begin{tabular}{|c|c|c|c|}
         \hline
                & System 1 & System 2 & System 3\\   
         \hline
        $S^{AP-SIP}_e =\frac{K_1}{N}$ &   $0.9688$ & $0.9375$ & $0.9688$\\
        \hline
        $S^{EP-SISO}_e=1-\frac{(2M_\tau+1)(2N_\nu+1)}{MN}$ &  $0.7178$ & $0.4355$ & $0.7178$\\
        \hline
      $S^{EP-MIMO}_e=1-\frac{(N_t M_\tau+M_\tau+N_t)(2N_\nu+1)}{MNN_t}$ & $0.7842$ & $0.5684$ & $0.8174$\\
        \hline
    \end{tabular}
\end{table*}
On the other hand, the efficiencies for the conventional embedded pilot (EP)-based technique for SISO and MIMO scenarios are given by $S_e^{EP-SISO}$ and $S_e^{EP-MIMO}$, respectively, which can be determined as shown in Table-II for integer-Doppler based systems. The efficiencies are even lower for a fractional-Doppler system. The efficiency values obtained upon substituting the parameters from System 1, 2, and 3 are given in Table II, from which it can be readily observed that the proposed sparse CSI estimation framework has a significantly higher efficiency.
}
\section{BCRB for the DD-domain CSI estimator of SISO, MIMO OTFS systems}\label{sec:BCRB}
{For the SISO OTFS system in \eqref{eq:joint_equation_SISO}, the Fisher information matrices corresponding to the observation model and the channel prior are given by $\mathbf{J}_y$ and $\mathbf{J}_h$, respectively. Using the theory in \cite{cramer_bound_book}, these can be formulated as $\mathbf{J}_{y} = -\mathbb {E}_{\mathbf{y},\mathbf{h}} \lbrace \frac{\partial ^2 \log [ f(\mathbf{y} \mid \mathbf{h})]}{\partial \mathbf {h} \partial \mathbf {h}^H} \rbrace$, $\mathbf {J}_{h} = -\mathbb {E}_{\mathbf {h}} \lbrace \frac{\partial ^2 \log [ f(\mathbf {h}; \boldsymbol{\Lambda }) ] }{\partial \mathbf {h} \partial \mathbf {h}^H} \rbrace$. Thus, following some simplification, the overall Fisher information matrix $\mathbf{J}_{SISO}$ is given as
\begin{align}
\mathbf{J}_{SISO} = \mathbf{J}_y+\mathbf{J}_h= \bar{\mathbf{\Phi}}^H\mathbf{R}_v^{-1} \bar{\mathbf{\Phi}} + \boldsymbol{\Lambda}^{-1},
\end{align} where $\boldsymbol{\Lambda}$ is the true hyperparameter matrix. The BCRB for the error covariance matrix of $\mathbf{h}$ now follows as $\mathbf{J}^{-1}$. Furthermore, the BCRB for the MSE of the estimate of the DD-domain channel vector $\mathbf{h}_\text{DD}=\boldsymbol{\zeta}\mathbf{h}$ obeys \cite{cramer_bound_book}
\begin{equation}
    \mathrm{MSE}\geq \text{Tr}\{\boldsymbol{\zeta}\mathbf{J}_{SISO}^{-1}\boldsymbol{\zeta}^H\},
\end{equation}
where $\boldsymbol{\zeta}$ in the above expression is given by \eqref{eq:zeta} in Appendix B.\\
For the MIMO OTFS system model in \eqref{eq:output_joint_mimo}, the net Fisher information matrix $\mathbf{J}_{MIMO}$ corresponding to the observation $\widetilde{\mathbf{y}}$, where $\widetilde{\mathbf{y}}=\mathrm{vec}(\widetilde{\mathbf{Y}})$, and the prior distribution of the channel coefficient vector $\widetilde{\mathbf{h}}$, where $\widetilde{\mathbf{h}}=\mathrm{vec}(\widetilde{\mathbf{H}})$, is given as
$
    \mathbf{J}_{MIMO} = \widetilde{\mathbf{\Phi}}^H\mathbf{R}_v^{-1} \widetilde{\mathbf{\Phi}} + \boldsymbol{\Lambda}^{-1},
$
where $\boldsymbol{\Lambda}$ is the underlying true hyperparameter matrix .
Using \eqref{eq:MIMO_hdd_h} from Appendix \ref{appendeix_B}, the DD-domain channel satisfies the relationship $\mathbf{H}_\text{DD}=({\mathbf{I}_{N_r N_t}}
     \otimes {\boldsymbol{\zeta}})\mathbf{H}$. Therefore, the BCRB for the MSE of its estimate is given as
\begin{equation}
    \mathrm{MSE}\geq \text{Tr}\{({\mathbf{I}_{N_r N_t}}
     \otimes {\boldsymbol{\zeta}}) ({\mathbf{I}_{N_r}}
     \otimes\mathbf{J}_{MIMO}^{-1}) ({\mathbf{I}_{N_r N_t}}
     \otimes {\boldsymbol{\zeta}})^H\},
\end{equation}
where $\boldsymbol{\zeta}$ in the above expression is given by \eqref{eq:zeta} in Appendix \ref{appendeix_B}.}
\section{Simulation Results}
\label{sec:sim_results}
{ This section presents our results to characterize the performance of the proposed AP-SIP CSI estimation schemes. We consider three different systems, including a high-frequency millimeter wave system (System 1), and two sub-6 GHz band systems (System 2 and 3) with their detailed parameter values presented in Table III and IV, including the delay and Doppler values. The channel in System 3 is generated using the EVA model \cite{EVA} considering a maximum speed of 120 Km/Hr.}
\begin{table*}{h}
\caption{Design parameters for System 1,2 and 3}
\begin{center}
\begin{tabular}{ |l|c|c|c| } 
\hline
Parameters & System1 & System 2 & System 3\\ 
\hline
Carrier frequency (GHz) & 24 & 4 & 4\\
\hline
Subcarrier frequency (KHz) & 15 & 7.5 & 15 \\
\hline
$\#$ of Doppler-axis symbols $M \times$ Delay-axis symbols $N$ & 32 $\times$ 32 & 16 $\times$ 32 & 32 $\times$ 32\\ \hline
Max. doppler spread  $M_\tau \times$ Max. delay spread $N_\nu$ & 8 $\times$ 8 & 8 $\times$ 8 & 8 $\times$ 8 \\ \hline
Precoder matrix for Data $M \times K_1$ & $32\times31$ & $16\times15$ & $32\times31$\\  \hline
Precoder matrix for Pilot $M \times K_2$ & $32\times1$ & $16\times1$ & $32\times1$ \\ \hline
$\#$ of Dominant reflectors $L_p$ & 5 & 5 & 9\\  \hline
Modulation scheme & 4-PSK & 4-PSK & 16,64-QAM \\ \hline
Pulse-shape & Rectangular & Rectangular & Rectangular\\
\hline
\end{tabular}
\label{table:2}
\end{center}
\end{table*}
\begin{table*}
\caption{DD-profile for System 1 \; \; \; \;\; \; \; \;\; \; \; \; \;\;\; \; \; \;\; \; \; \;\; \; \; \; \;\;\; DD-profile for System 2}
    \centering
    \begin{tabular}{|l|c|c|c|c|c|}
    \hline
      Path-Index$(i)$   & 1 & 2 & 3 & 4 & 5 \\
      \hline
       Delay $\tau_i (\mu sec)$ & 2.08 & 4.16 & 6.24 & 8.32 & 10.41 \\
       \hline
       Doppler $\nu_i (Hz)$ & 0 & 470 & 940 & 1880 & 2820 \\
       \hline
       Speed (Km/Hr) & 0 & 21.1 & 42.2 & 84.4 & 126.6\\
       \hline
    \end{tabular}
    \centering
    \begin{tabular}{|l|c|c|c|c|c|}
    \hline
      Path-Index$(i)$   & 1 & 2 & 3 & 4 & 5 \\
      \hline
      Delay $\tau_i (\mu sec)$ & 4.16 & 8.33 & 12.49 & 16.66 & 20.83 \\
       \hline
       Doppler $\nu_i (Hz)$ & 0 & 470 & 940 & 1410 & 1880 \\
       \hline
       Speed (Km/Hr) & 0 & 126.9 & 253.6 & 380.4 & 507.2\\
       \hline
    \end{tabular}
    \label{tab:3}
\end{table*}
\subsection{NMSE performance analysis}
\begin{figure*}[t]
\centering
\subfloat[]
{\includegraphics[scale=0.2]{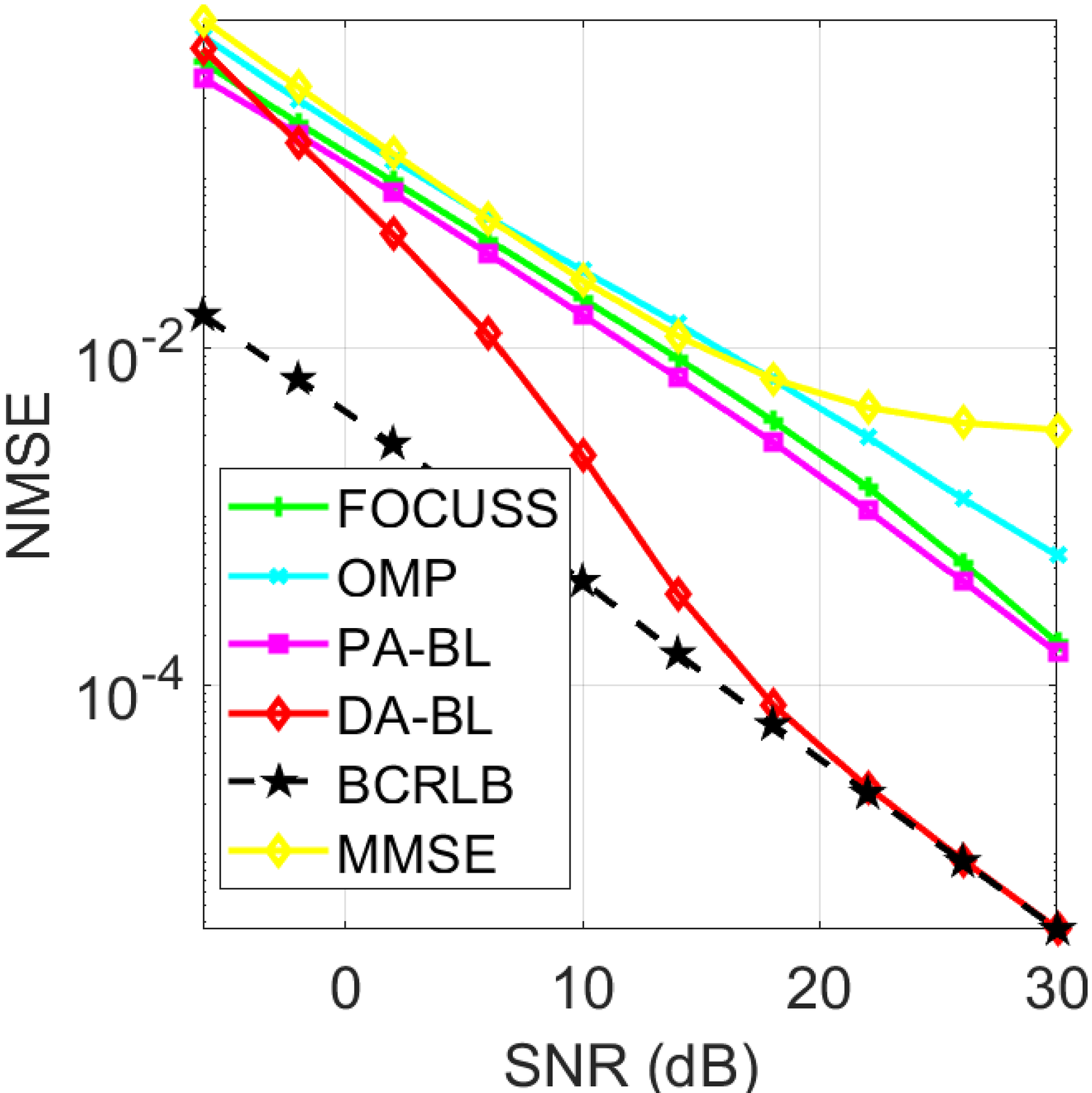}} 
\hfil
\subfloat[]
{\includegraphics[scale=0.2]{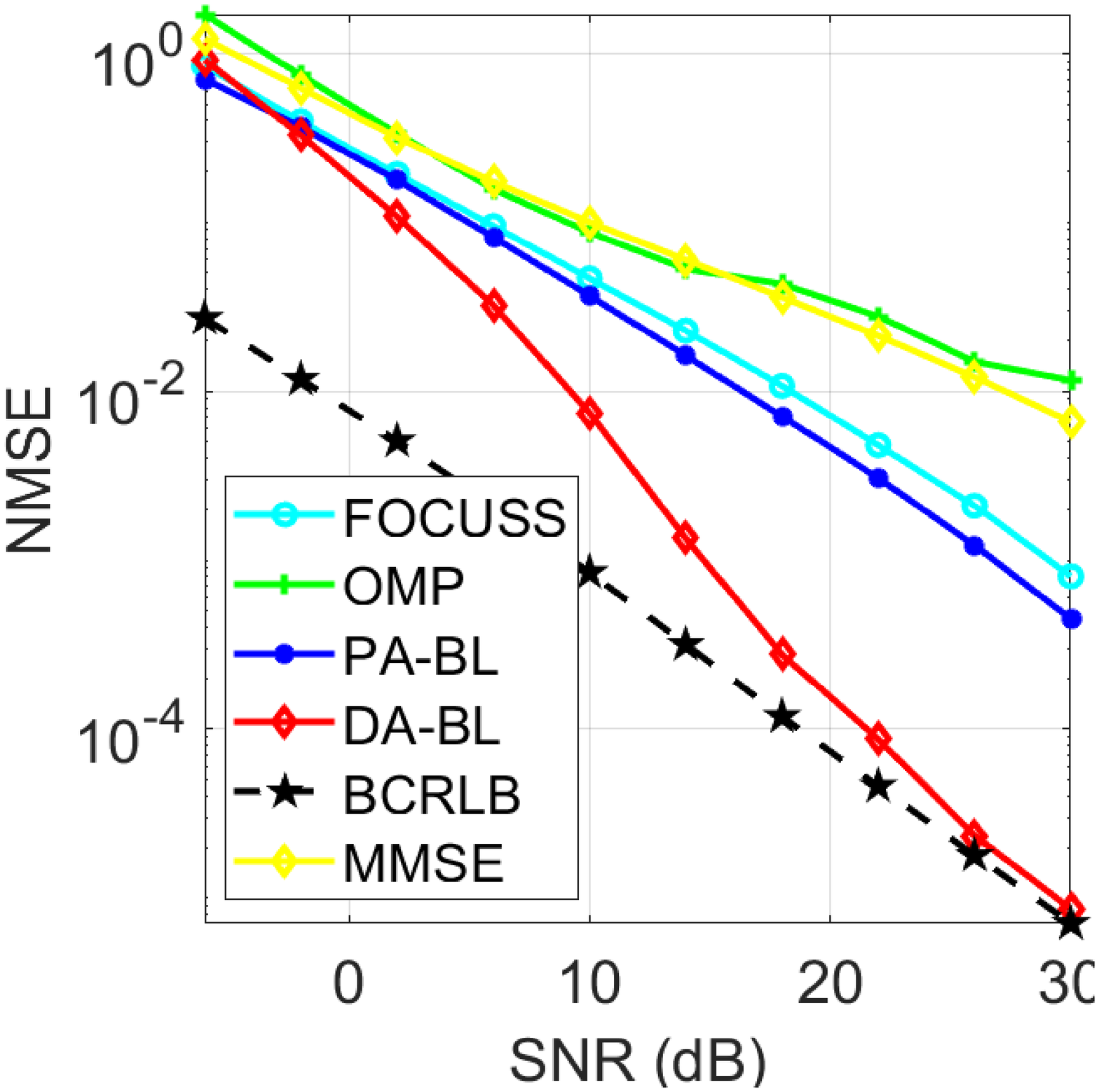}}
\hfil
\subfloat[]{\includegraphics[scale=0.2]{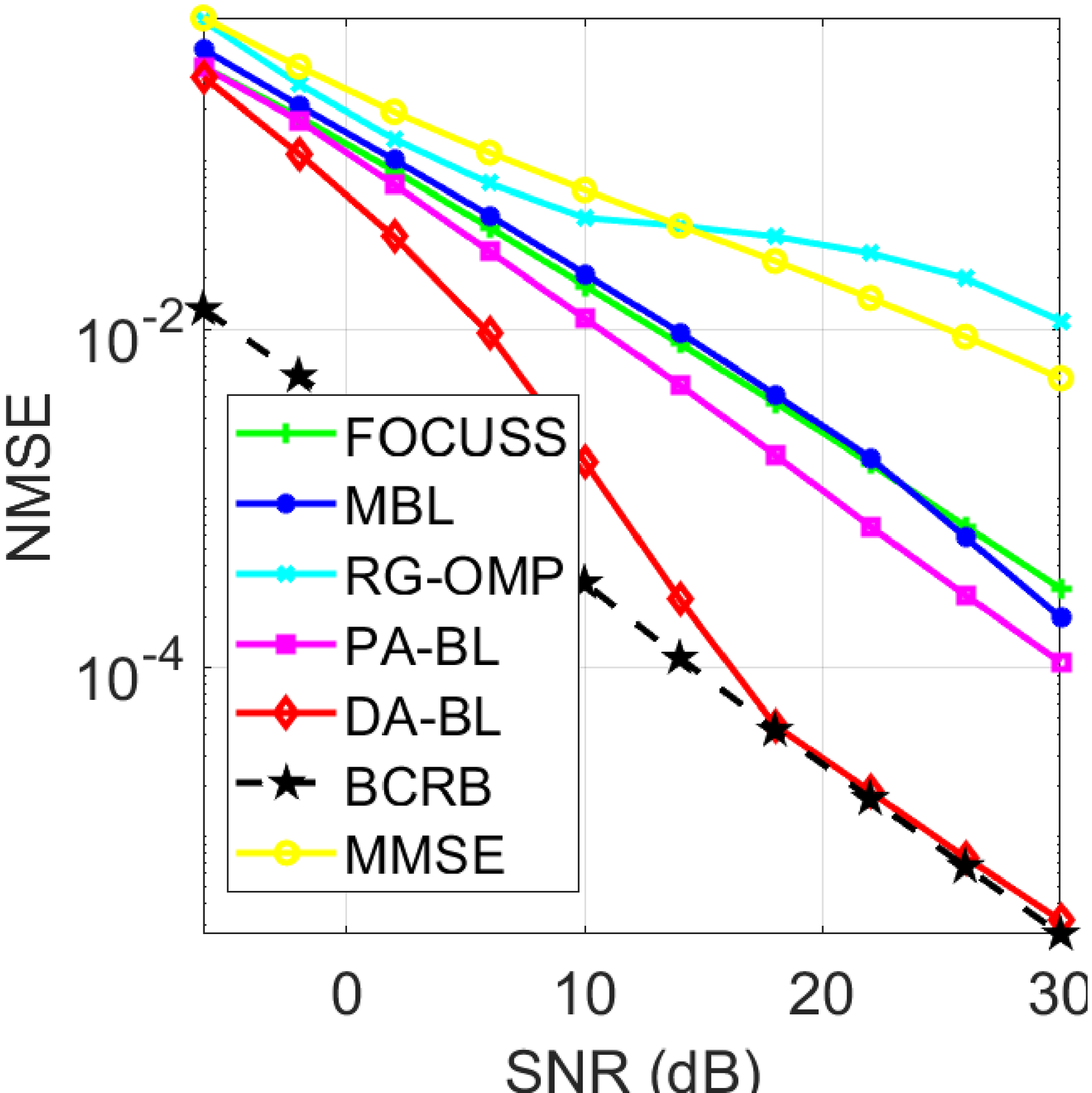}}
\hfil
\subfloat[]
{\includegraphics[scale=0.2]{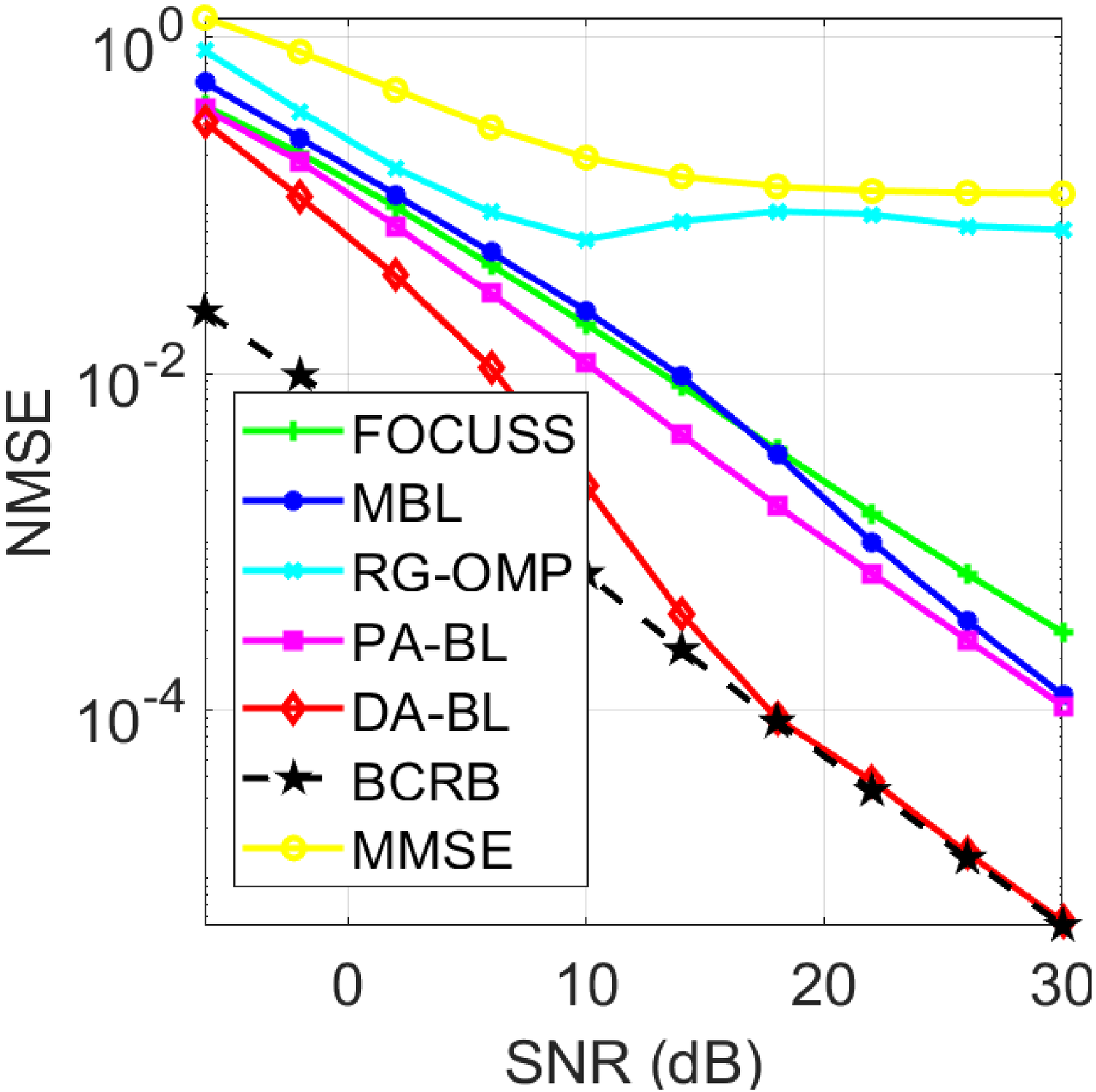}}\\
\subfloat[]
{\includegraphics[scale=0.2]{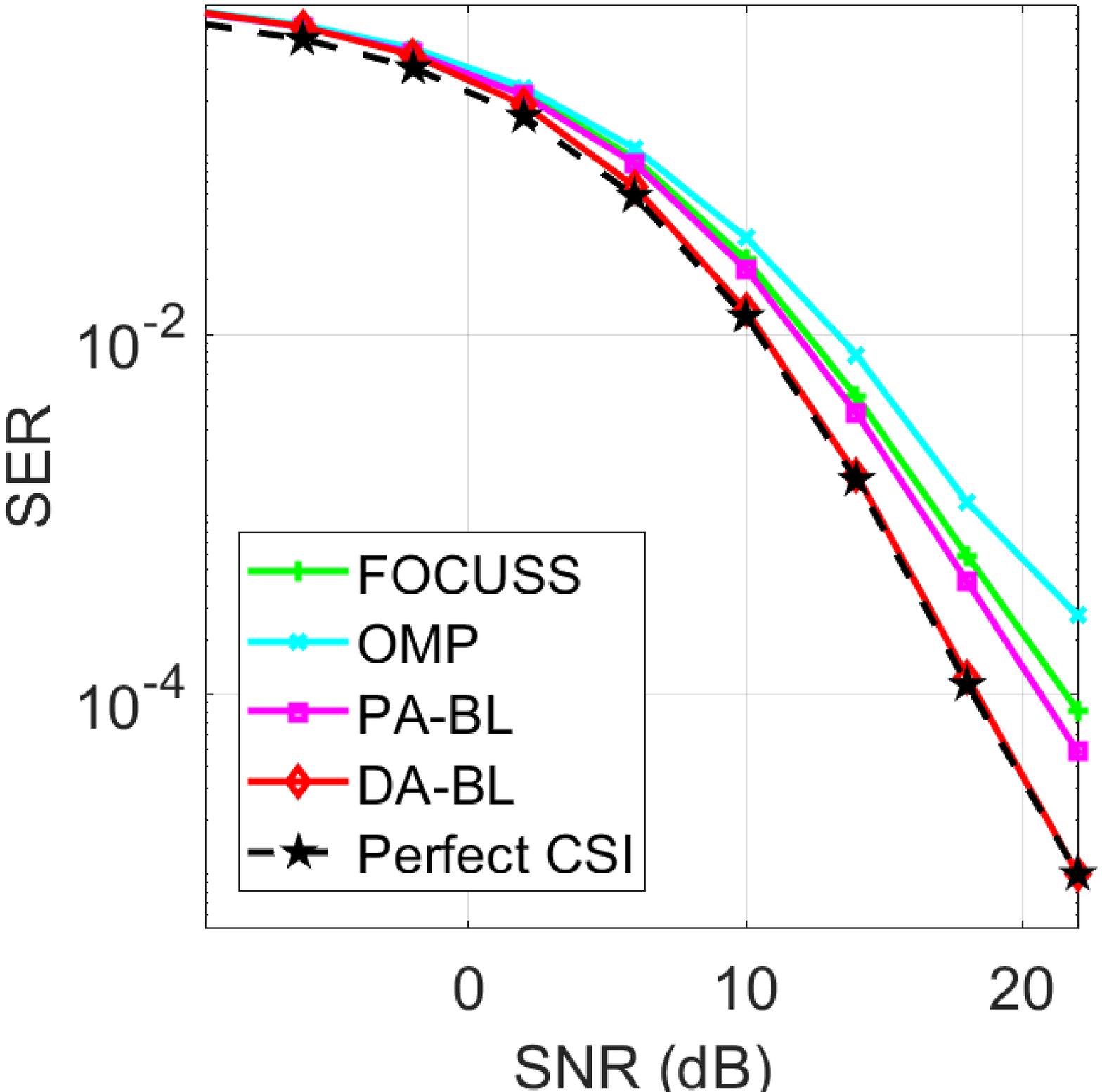}}
\hfil
\subfloat[]
{\includegraphics[scale=0.2]{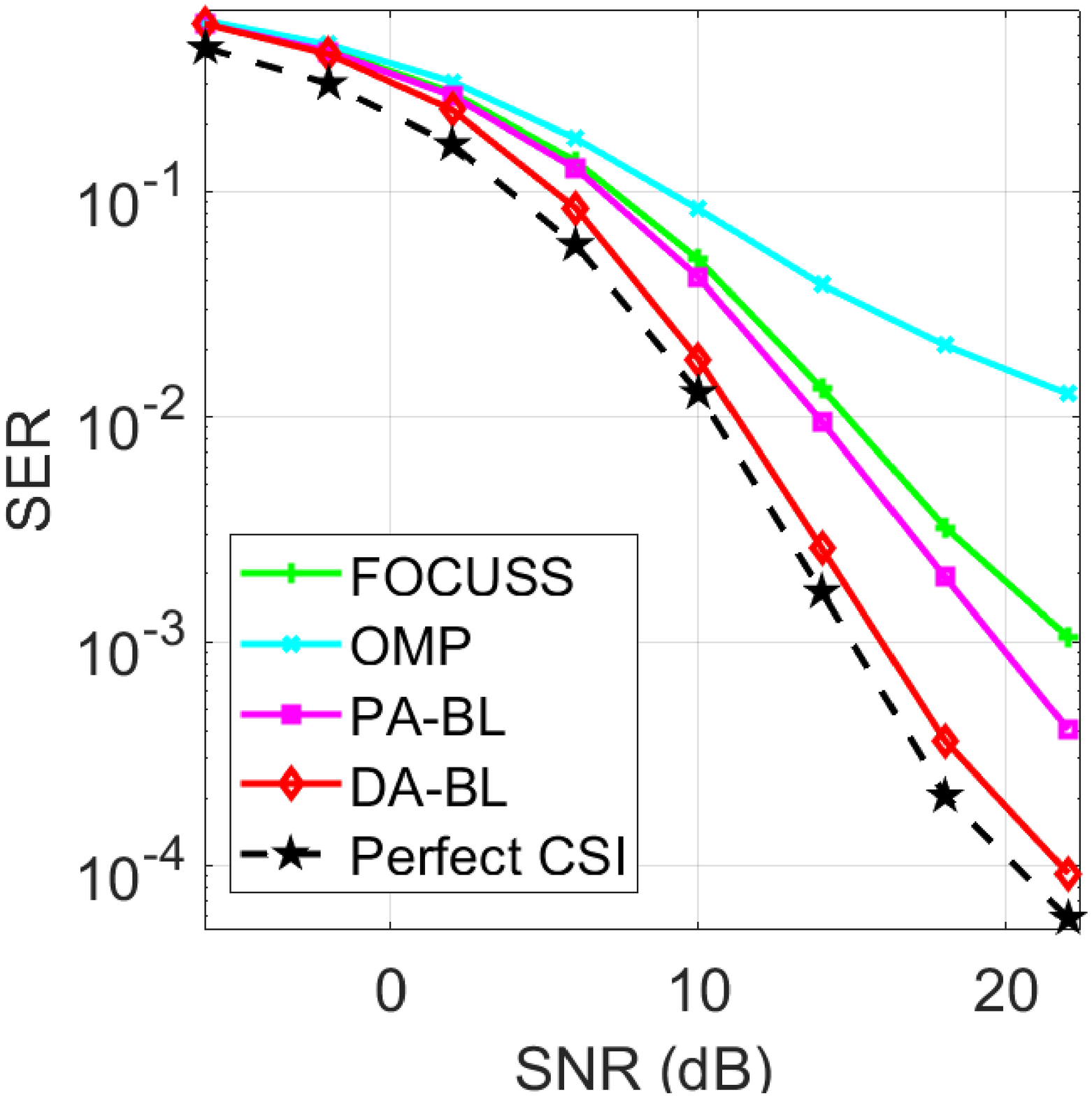}}
\hfil
\subfloat[]
{\includegraphics[scale=0.2]{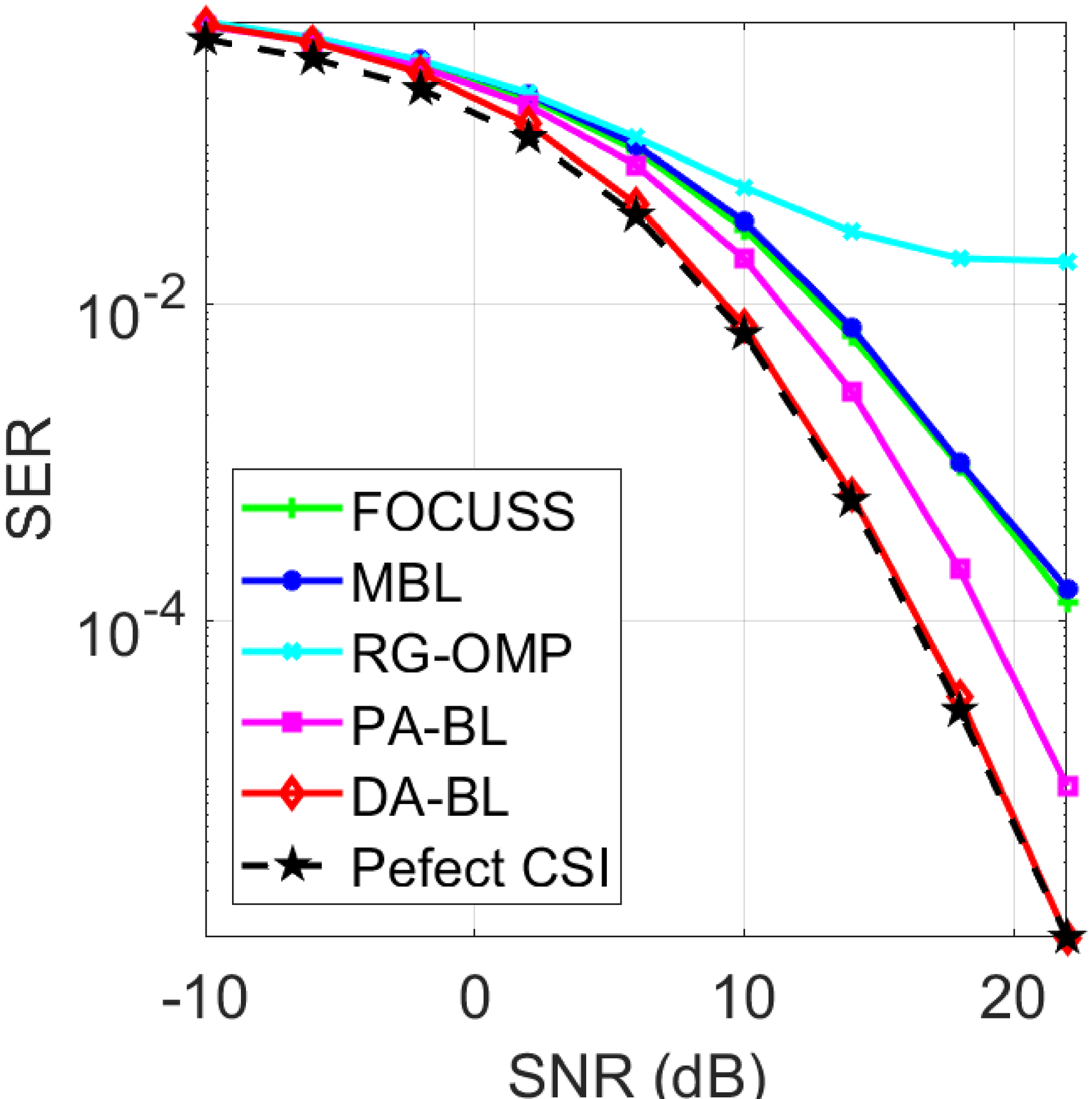}}
\hfil
\subfloat[]
{\includegraphics[scale=0.2]{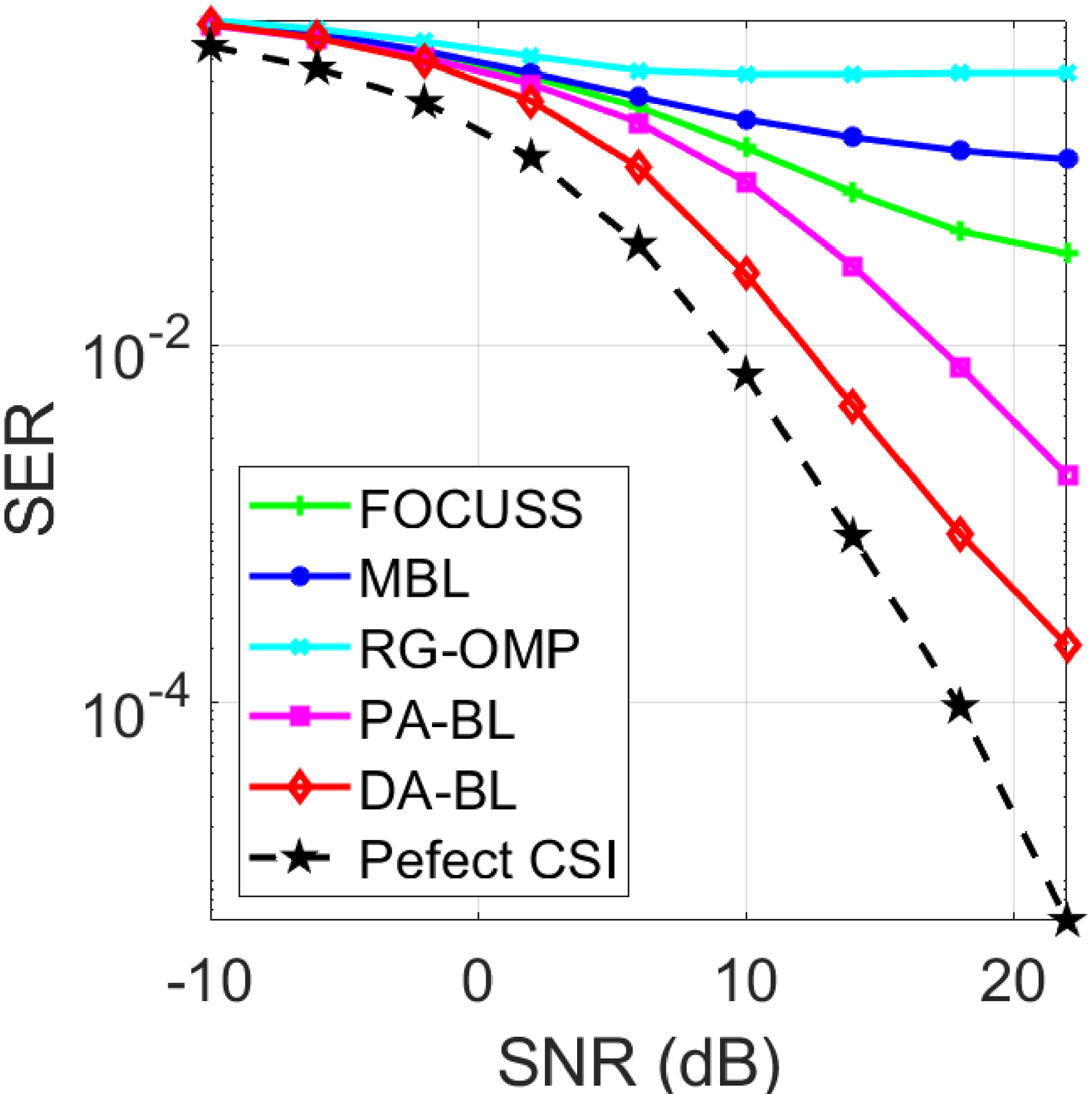}}
\hfil
\caption{NMSE/ SER vs. SNR performance for AP-SIP OTFS modulation (a) NMSE vs. SNR for SISO System 1 (b) NMSE vs. SNR for SISO System 2 (c) NMSE vs. SNR for MIMO System 1 (d) NMSE vs. SNR for MIMO System 2 (e) SER vs. SNR for SISO System 1 (f) SER vs. SNR for SISO System 2 (g) SER vs. SNR for MIMO System 1 (h) SER vs. SNR for MIMO System 2}
\label{fig:plot_system1_system2}
\end{figure*}
Figures \ref{fig:plot_system1_system2}(a), \ref{fig:plot_system1_system2}(b), show the NMSE performance of the proposed DA-BL, PA-BL and other competing schemes, such as OMP \cite{OMP}, FOCUSS \cite{FOCUSS} for AP-SIP-based SISO OTFS system with settings corresponding to System 1 and 2, respectively. {The normalized MSE (NMSE) is defined as
\begin{align}    \mathrm{NMSE}={{||\widehat{\mathbf{h}}_\text{DD}-\mathbf{h}_\text{DD}||}^2}/{{||\mathbf{h}_\text{DD}||}^2},
\label{eq:NMSE}
\end{align} where $\widehat{\mathbf{h}}_\text{DD}=\mathrm{vec}( \widehat{\mathbf{H}}_\text{DD})$. Since $\widehat{\mathbf{H}}_\text{DD}$ depends on the estimated delay and Doppler parameters, as seen from \eqref{eq:h_dd_NMSE}, errors in the estimation of these parameters impact the NMSE defined in \eqref{eq:NMSE}.     }
The figure shows that the BL principle-based PA-BL and DA-BL yield significantly better estimation performance than competing estimation schemes such as OMP, FOCUSS, and MMSE in the SISO OTFS system. The OMP method's subpar performance can be traced back to its dependence on the stopping parameter. On the other hand, convergence issues and sensitivity to the regularisation parameter hinder the performance of FOCUSS\cite{FOCUSS}. Since conventional-MMSE does not leverage DD-domain CSI sparsity, it has the worst NMSE performance. Clearly, the performance of the non-Bayesian sparse estimation schemes OMP, FOCUSS is not as robust as that of the BL approaches, owing to the deficiencies described earlier. The proposed DA-BL is seen to yield the best performance among all the competing sparse estimation schemes. This is due to its ability to leverage the data estimates obtained via the modified LMMSE rule as described in \eqref{eq:MMSE_joint_siso} in addition to the limited pilot overhead. Furthermore, its performance is also close to the BCRB at high SNRs. This is remarkable since the DA-BL achieves this without prior knowledge of the channel's covariance matrix which would be initially important for the conventional LMMSE estimator. Moreover, knowledge of the support of the sparse channel is not required either. This demonstrates that the DA-BL is ideally suited for the practical implementation of OTFS systems, where typically no prior information is available.

As observed from the Fig. \ref{fig:plot_system1_system2}(c), \ref{fig:plot_system1_system2}(d) that the PA-BL is seen to outperform the existing FOCUSS, RG-OMP\cite{suraj2021row_group}, and MBL \cite{suraj2021bayesian}, MMSE approaches. It is also interesting to observe that the PA-BL outperforms the MBL, which can be explained by the fact that the latter only leverages the simultaneous row sparsity present in the multiple measurement vector, while the former exploits the simultaneous row-group sparsity, which leads to its improved performance. The DA-BL algorithm proposed for joint CSI estimation and detection leads to a further improvement as seen for the SISO systems, since it also harnesses the abundant data symbols in addition to the limited number of pilot symbols for improved CSI estimation. Indeed, its performance closely follows the BCRB in the high-SNR regime. 

\begin{figure*}[t]
\centering
\subfloat[]{\includegraphics[scale=0.30]{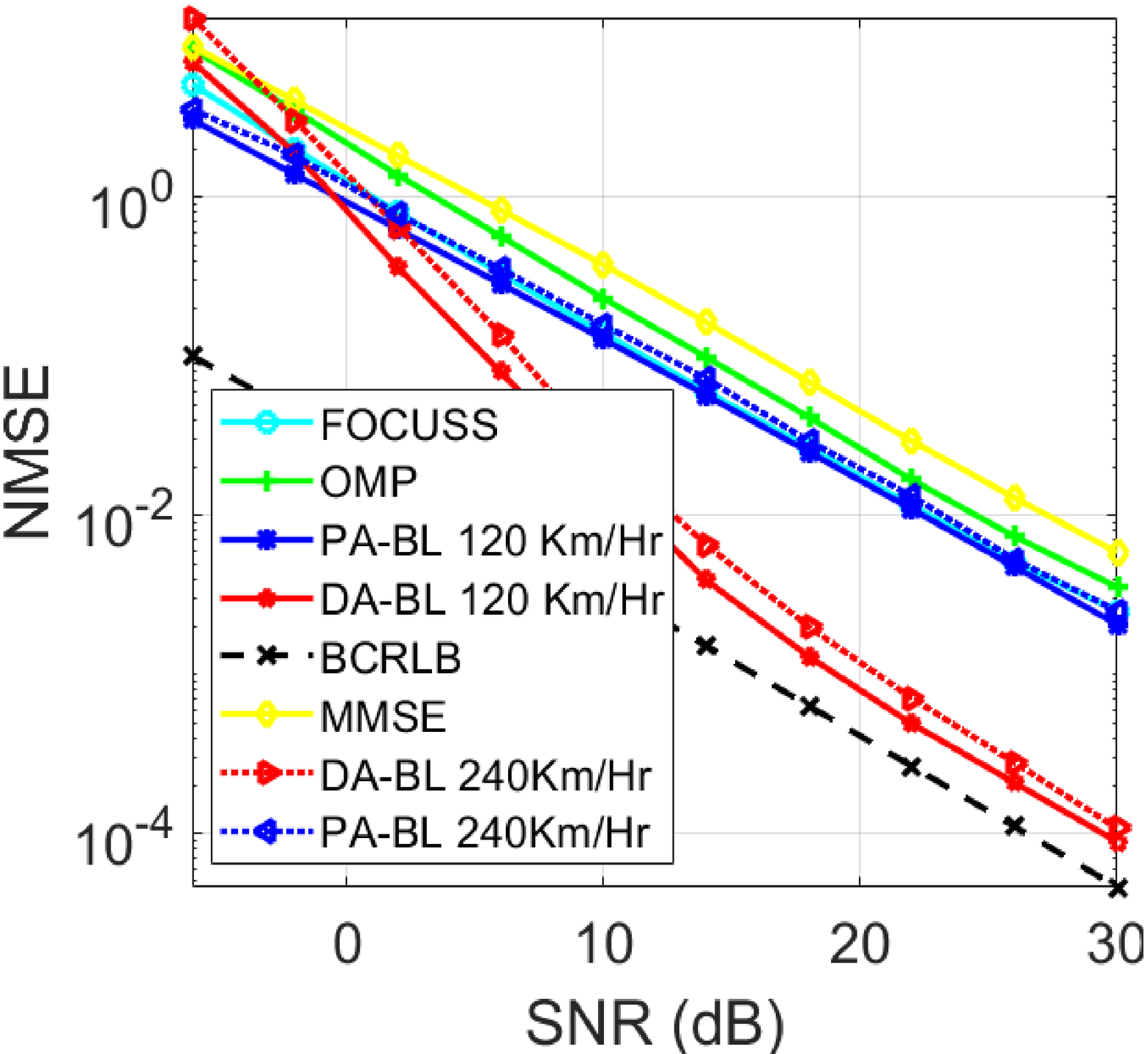}}
\subfloat[]{\includegraphics[scale=0.30]{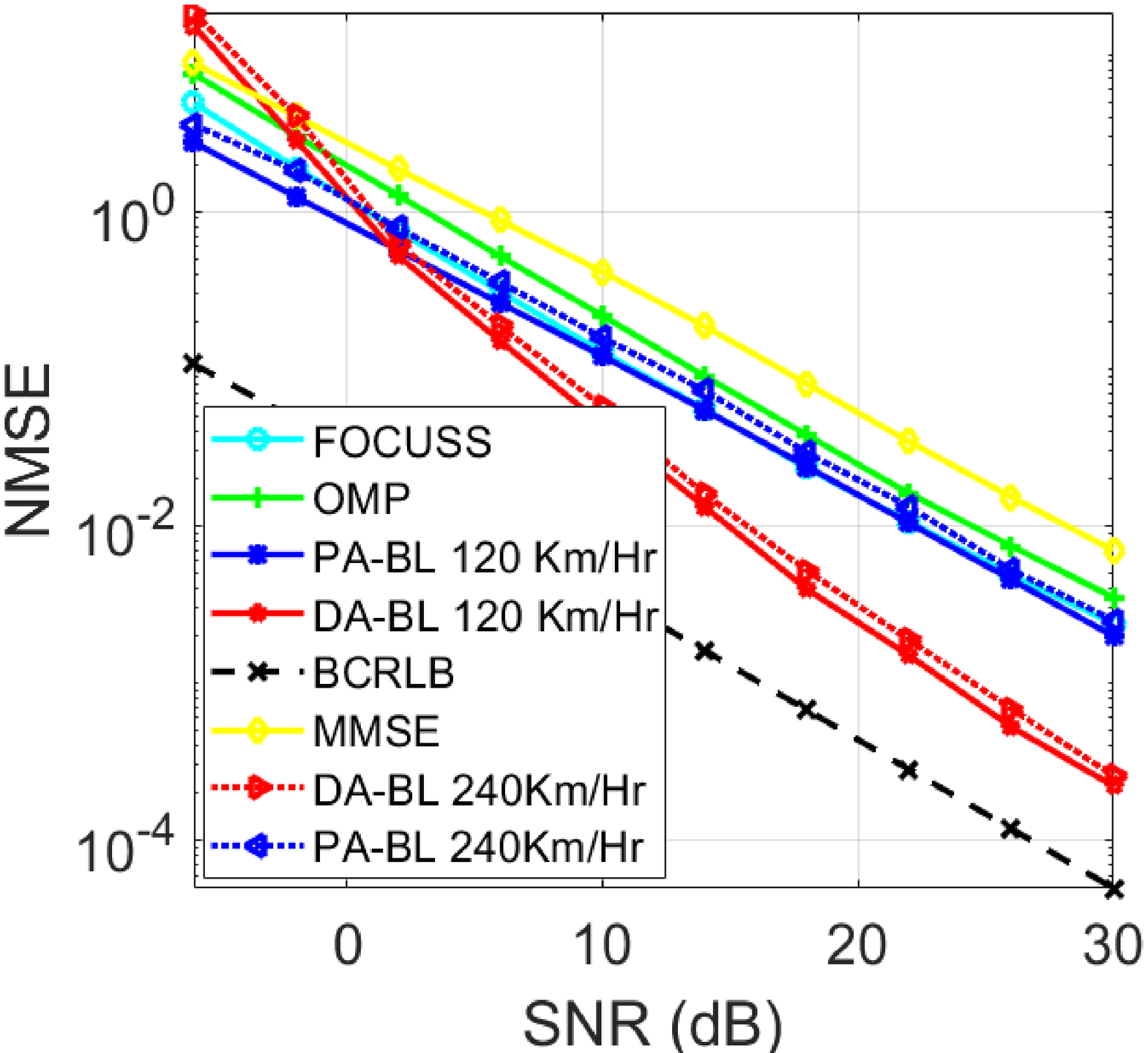}}
\subfloat[]{\includegraphics[scale=0.30]{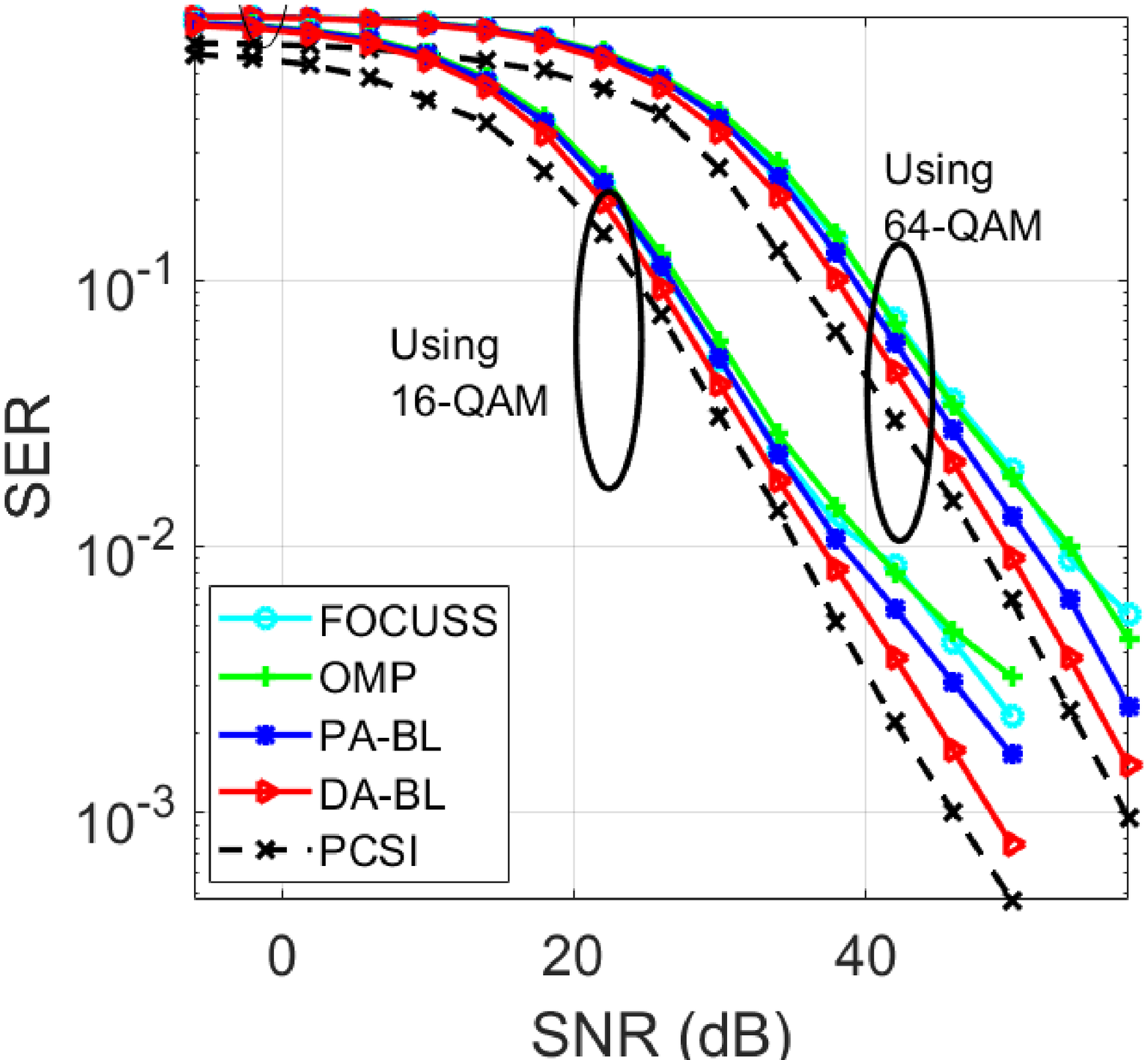}}\\
\subfloat[]{\includegraphics[scale=0.30]{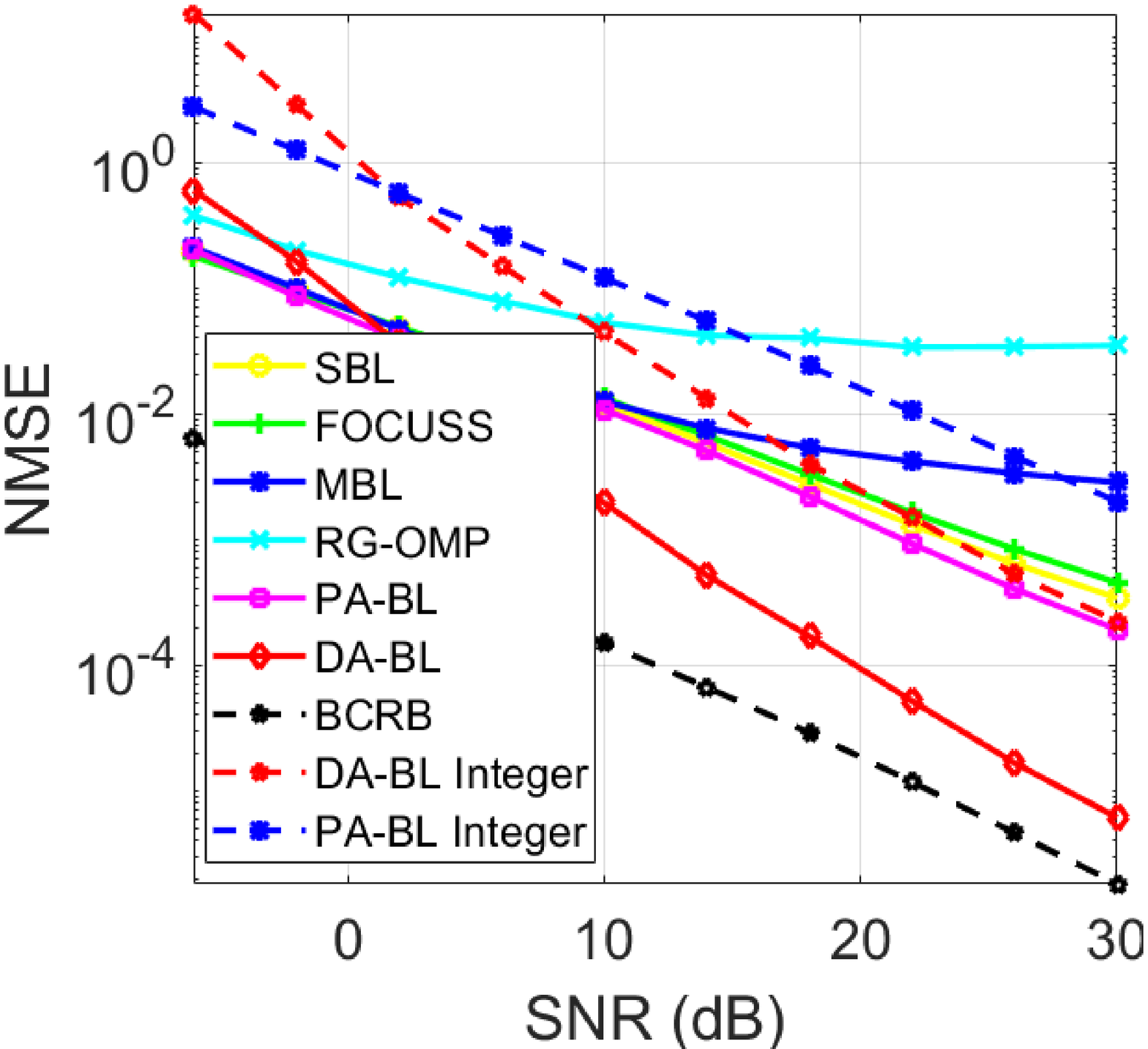}}
\subfloat[]{\includegraphics[scale=0.30]{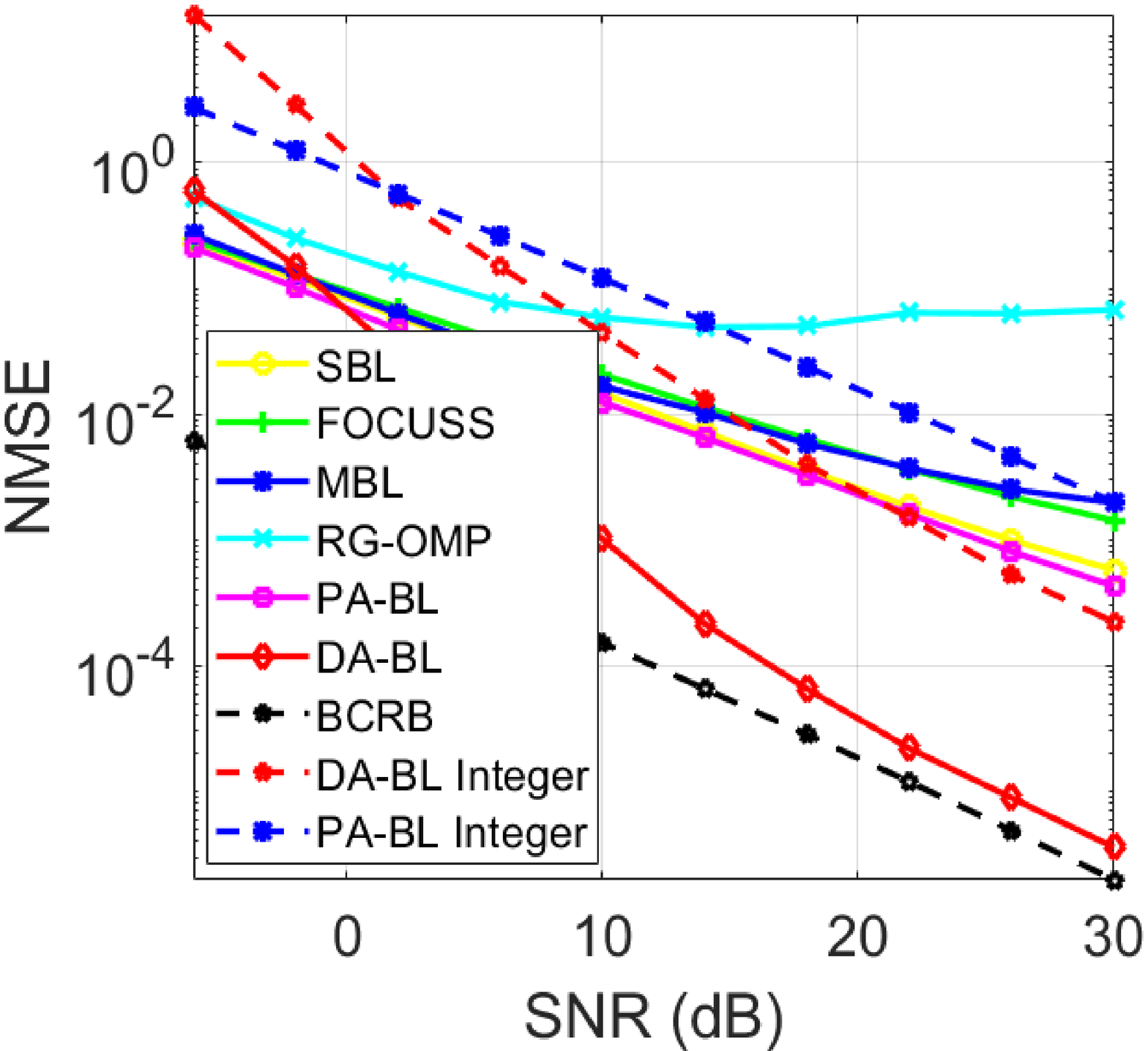}}
\subfloat[]{\includegraphics[scale=0.30]{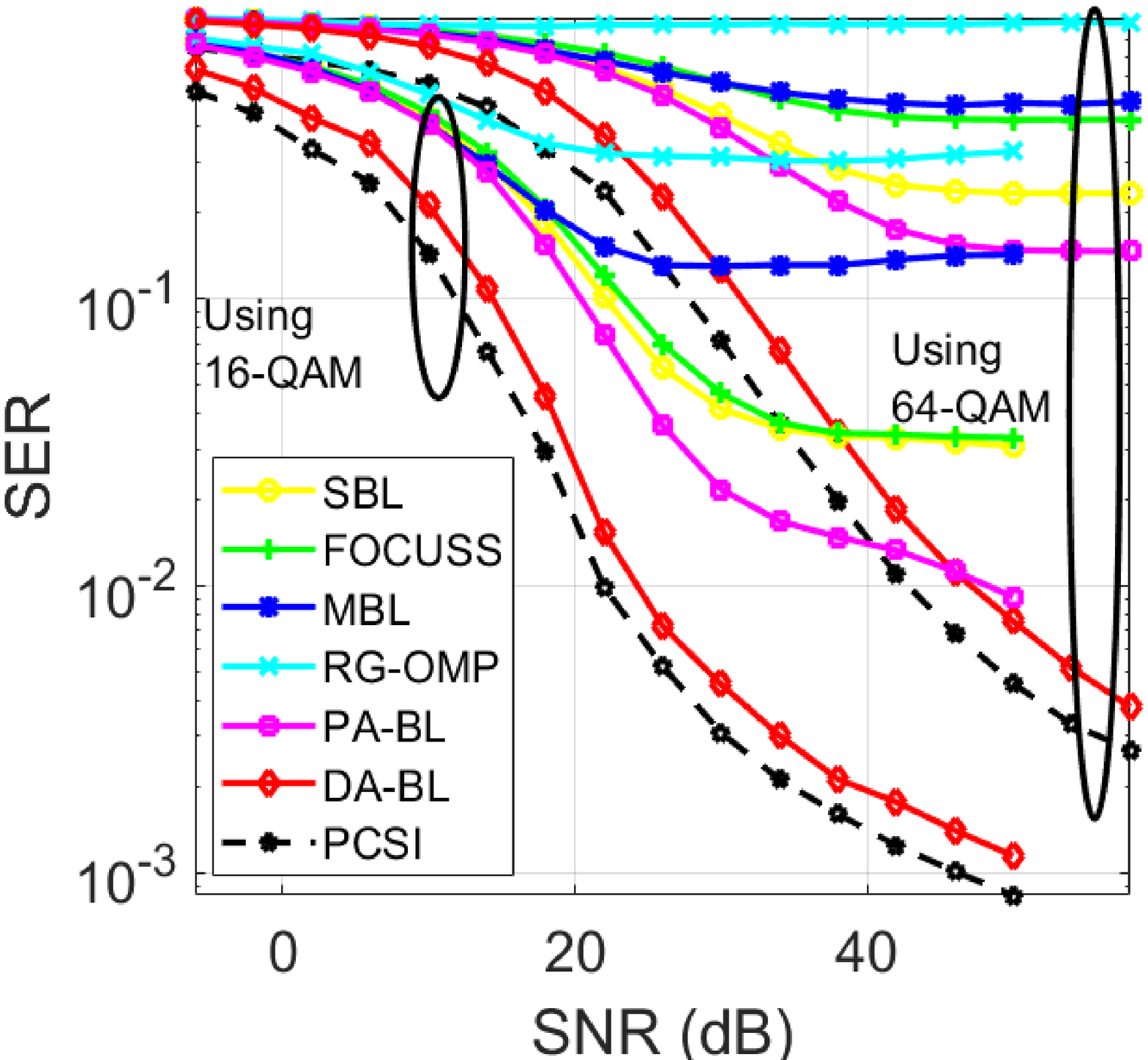}}
\caption{NMSE/ SER vs. SNR perfromance for AP-SIP OTFS modulation, with parameters as per System 3 (a) NMSE vs. SNR for SISO OTFS, 16-QAM, (b) NMSE vs. SNR for SISO OTFS, 64-QAM, (c) SER vs. SNR for SISO OTFS, (d) NMSE vs. SNR for MIMO OTFS, $N_t=N_r=4$, 16-QAM, (e) NMSE vs. SNR for MIMO OTFS, $N_t=N_r=4$, 64-QAM, (f) SER vs. SNR for MIMO OTFS, $N_t=N_r=4$}
\label{fig:plots_system3}
\end{figure*}
\subsection{SER performance analysis}
Figures \ref{fig:plot_system1_system2}(e), \ref{fig:plot_system1_system2}(f) characterize the SER vs. SNR performance of the techniques proposed for the SISO OTFS systems, with setting as per those given for System 1 and 2 respectively, again the corresponding parameters are given in Table \ref{table:2}, \ref{tab:3}. The SER is evaluated for the data symbols that are superimposed onto the pilots by employing the CSI obtained from each of the estimation techniques described previously. The SER performance is also benchmarked against the performance of that of a hypothetical receiver having perfect CSI. Clearly, the SER corresponding to the BL-based schemes such as PA-BL, DA-BL is better than that of the non-BL schemes, i.e., OMP and FOCUSS owing to having better CSI estimates, which is in agreement with the NMSE plots Fig. \ref{fig:plot_system1_system2}(a), \ref{fig:plot_system1_system2}(b). Moreover, the DA-BL technique is once again seen to yield the best performance, with its SER closely following that of the benchmark detector having perfect CSI. This demonstrates the efficacy of the proposed DA-BL procedure conceived for obtaining the CSI estimates that have a high degree of accuracy.

Observe in Fig. \ref{fig:plot_system1_system2}(g), \ref{fig:plot_system1_system2}(h) that the MIMO OTFS System 1 and System 2, respectively, exhibit a trend similar to their SISO OTFS counterparts for $N_r=2, N_t=2$. In line with our previous discussions, it is amply evident that the SER obtained by the BL-based schemes such as MBL, PA-BL, and DA-BL is superior to that of the non-BL schemes such as RG-OMP and FOCUSS. This may be credited to the improved CSI estimates of the former schemes. Furthermore, it is evident that the DA-BL technique provides the best performance, with its SER closely tracking that of the benchmark detector having perfect CSI.
 \begin{figure*}[t]
\centering
\subfloat[]{\includegraphics[scale=0.3]{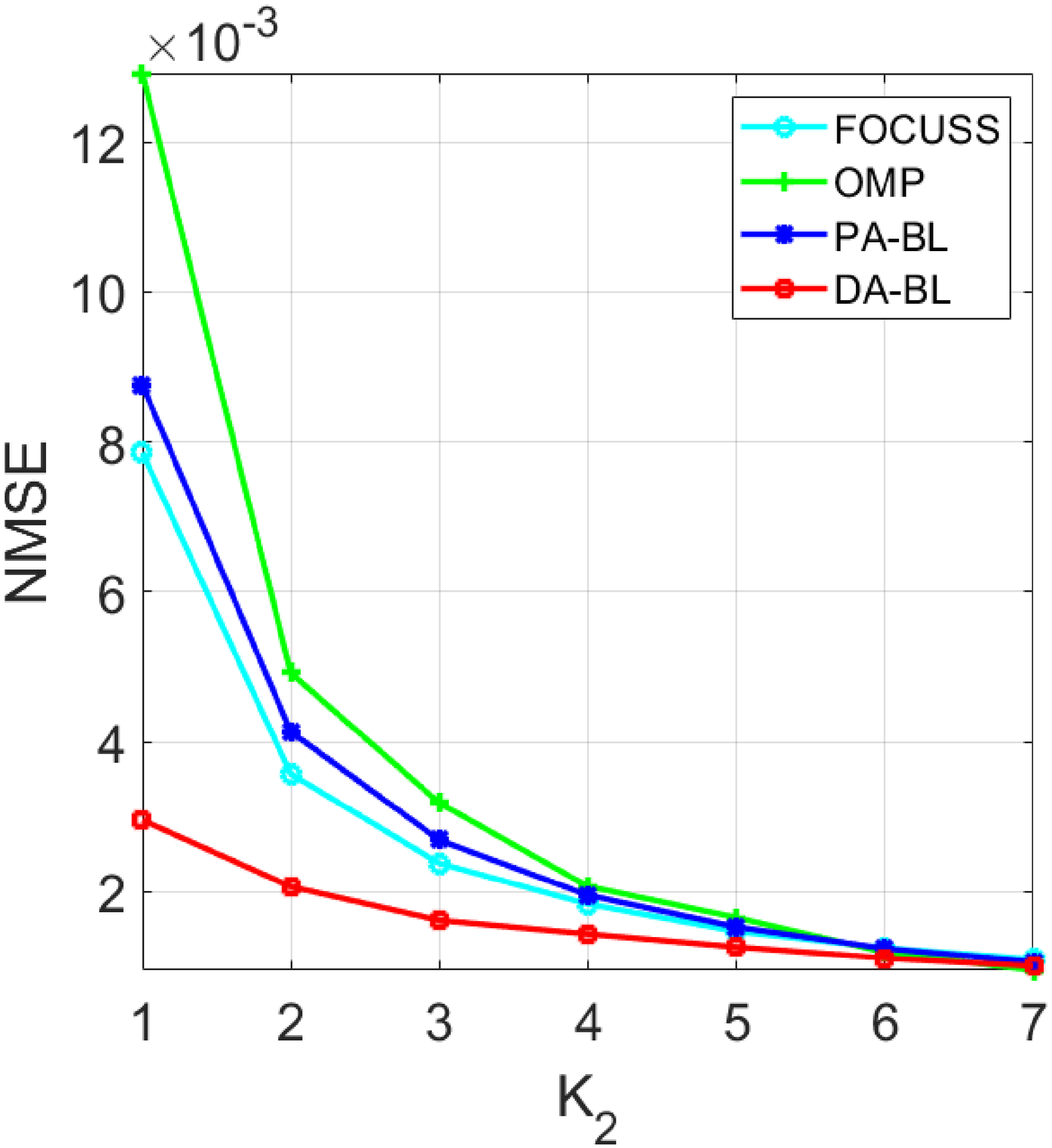}}
\hfil
\subfloat[]{\includegraphics[scale=0.3]{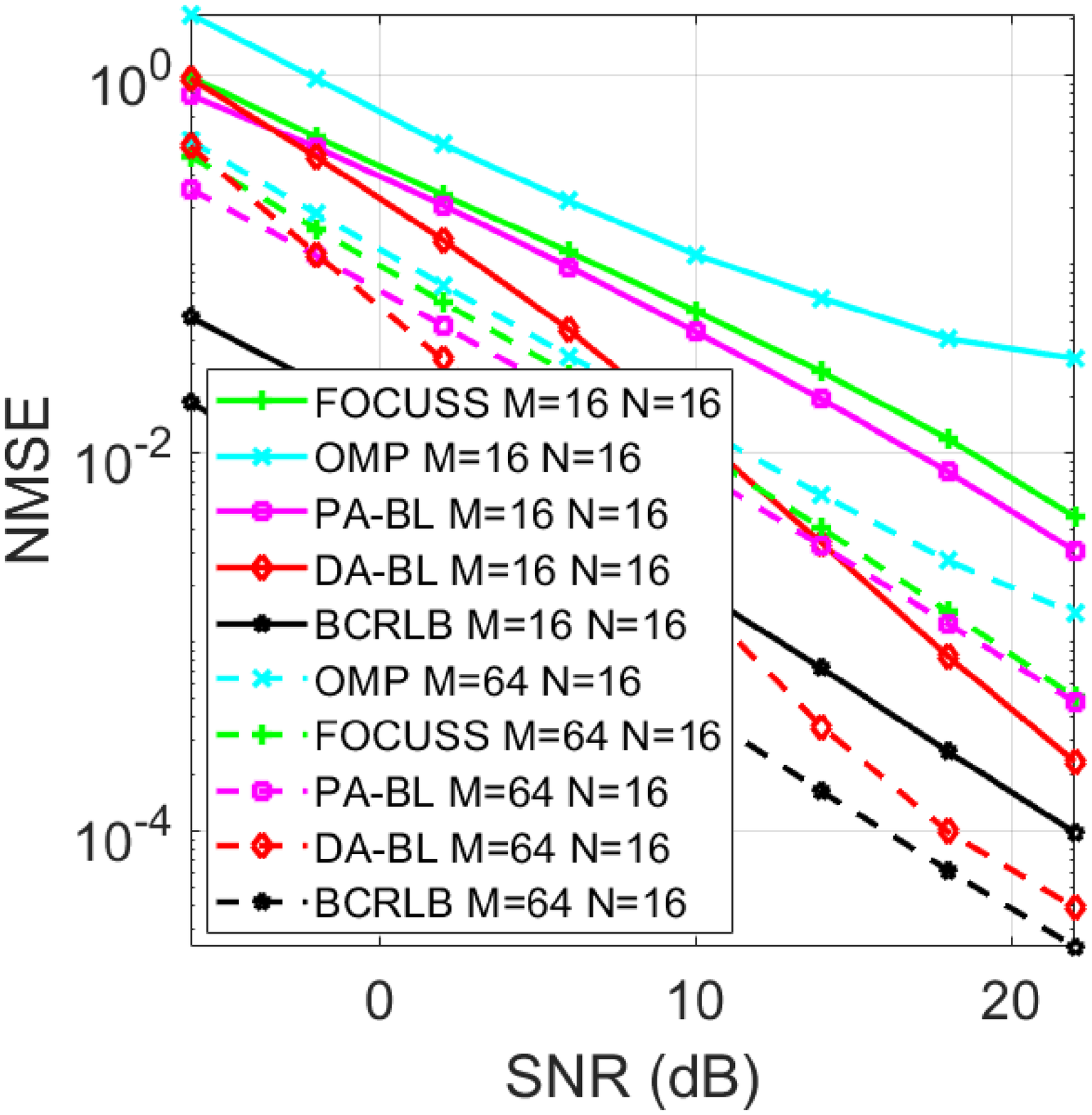}}
\hfil
\subfloat[]{\includegraphics[scale=0.3]{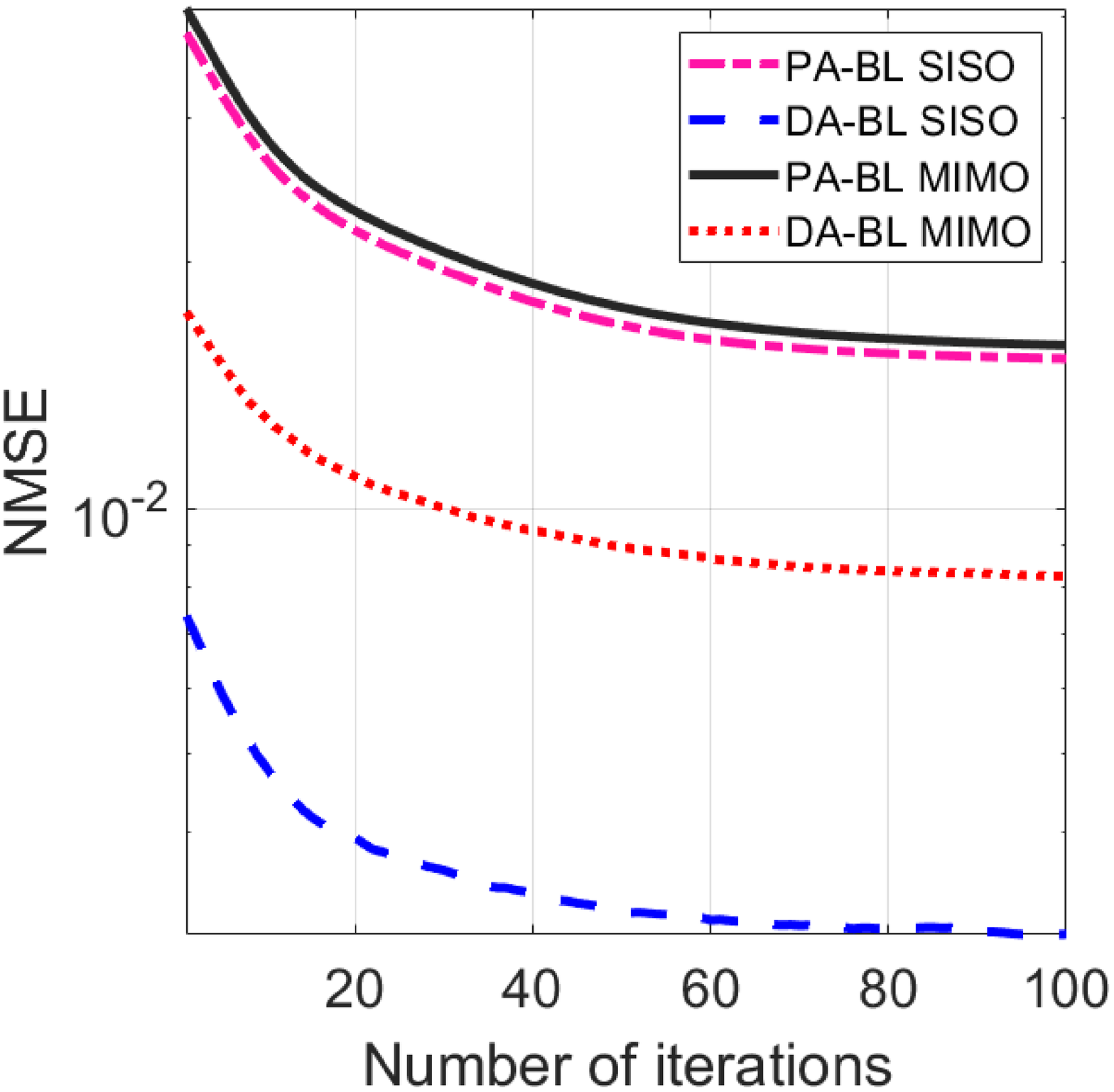}}
\hfil
\caption{(a) NMSE vs. Pilot training duration $(K_2)$ for the AP-SIP SISO OTFS system, $M = 64, N = 8$, (b) NMSE vs. SNR for AP-SIP SISO OTFS system, $M\in \{16,64\}$, $N=16$ (c) NMSE vs. Number of iterations, $SNR=10dB$ (System 1)}
\label{fig:miscellanious_plots}
\end{figure*}   
\subsection{NMSE and SER performance for fractional Doppler}
{
Figures \ref{fig:plots_system3}(a)-\ref{fig:plots_system3}(f) present the NMSE and SER performance of the various algorithms for AP-SIP SISO and MIMO OTFS systems with specifications as per System 3, which incorporates fractional Doppler, where the channel is generated using the EVA model. It is clear that the DA-BL estimator yields an improved performance in comparison to other estimation schemes. The performance can be further enhanced by increasing the number of Doppler bins $\mathbf{G}_\nu$, which improves the grid resolution. Furthermore, it can be observed from Fig. \ref{fig:plots_system3}(a), \ref{fig:plots_system3}(b), that as the speed is increased from 120 Km/Hr to 240 k/Hr, the performance degradation of the proposed algorithms is only minimal. This is owing to the robustness of OTFS modulation to high Doppler shift arising due to mobility.}
\subsection{Other NMSE trends}
Figure \ref{fig:miscellanious_plots}(a) plots the NMSE vs pilot duration $K_2$ for the proposed sparse CSI estimation schemes used by the SISO OTFS system having the parameters of $M=64$, $N=8$, $M_\tau=8$, $N_\nu=8$, $SNR=10dB$. The figure demonstrates that the NMSE decreases as the pilot training duration, i.e., $K_2$ increases, which is along the expected lines.
Once again, the DA-BL scheme outperforms the other schemes, which shows the value of exploiting data symbols. Interestingly, even for a very low value of $K_2$, viz., $K_2=1$ the DA-BL achieves a remarkably accurate CSI estimate, which shows the benefit of data symbols for reducing the pilot overhead.

Figure \ref{fig:miscellanious_plots}(b) compares the performance of the schemes, when the number of subcarriers $M$ is increased from $16$ to $64$, and $N=16$, with the rest of the parameters unchanged. As the number of subcarriers grows, so does the number of pilot symbols, hence enhancing estimation performance. The figure shows the same trend as seen in the previous figures, with the DA-BL having the best performance, followed closely by the PA-BL that only exploits the pilot symbols. 

{Figures \ref{fig:miscellanious_plots}(c), shows the NMSE vs number of iterations for the PA-BL and DA-BL schemes for SISO and MIMO OTFS systems, respectively, with parameters as per System 1. It can be observed that the DA-BL and PA-BL converge at the same rate, due to the fact both are Bayesian learning approaches for CSI estimation. However, the NMSE of the former is considerably lower in comparison to the latter owing to the better CSI estimate that arises from additionally exploiting the data symbols.}
In summary, the capability of jointly exploiting the unknown data symbols and the sparsity of the channel, without prior knowledge of the number of dominant reflectors or channel statistics, renders the proposed DA-BL scheme ideal for implementation in practical SISO- and MIMO-aided OTFS systems.
\section{Summary and conclusions} 
\label{sec:concl_paper}
This paper presented an AP-SIP framework for CSI estimation in CP-aided SISO and MIMO OTFS systems relying on arbitrary Tx-Rx pulse shapes. A key feature of the proposed schemes is that they successfully exploited the DD domain sparsity for improving the accuracy of CSI estimation in comparison to the existing schemes. First, the end-to-end DD-domain model was derived for the decoupled pilot and data symbols, followed by the PA-BL scheme designed for iterative DD-domain sparse CSI estimation. This further evolved to the development of the DA-BL for joint CSI estimation and data detection, which exploits the data symbols for CSI estimation, together with a modified LMMSE rule harnessed for data detection. The AP-SIP framework was subsequently extended to MIMO OTFS systems and a novel technique was developed for affine precoding the DD-domain input matrix that allowed us to decouple the data and pilot responses at the receiver. This was then leveraged to develop the EM algorithm-based PA-BL technique for sparse CSI estimation, which exploits the simultaneous row-group sparsity of the channel using exclusively pilot symbols, and also the DA-BL, which carries out joint MIMO CSI estimation and data detection. Our simulation results demonstrated the improved performance of the proposed CSI estimation strategies in various settings. A performance similar to that of various benchmarks such as the NMSE of the BCRB and the SER performance of ideal OTFS receivers having perfect CSI was attained.
%------------------------------------------------------------
\appendices
%------------------------------------------------------------
\section{MMSE detector}
For data-aided joint AP-SIP SISO OTFS systems, the MMSE detector of $\mathbf{X}_d$ for the $j$th iteration is formulated as
$\mathbf{\widehat{\mathbf{X}}}_d^{(j)}= \mathbf{R}_{XY}\mathbf{R}_{YY}^{-1}\mathbf{Y}_{\text{DD},d}$, where
\begin{align}
\mathbf{R}_{XY}&=\mathbb{E}[\mathbf{X}_{d}\mathbf{Y}_{\text{DD},d}^H]
= \mathbb{E}[\mathbf{X}_{d}\mathbf{X}_{d}^H \widehat{\mathbf{H}}_\text{DD}^H+\mathbf{X}_{d}\mathbf{W}_{\text{DD},d}^H] 
\nonumber\\&= \mathbb{E}[\mathbf{X}_{d}\mathbf{X}_{d}^H] \mathbb{E}[\widehat{\mathbf{H}}_\text{DD}^{(j) H}] = \sigma _d^2(\widehat{\mathbf{H}}_\text{DD}^{(j) H}),\nonumber\\
\mathbf{R}_{YY}&=\mathbb{E}[\mathbf{Y}_{\text{DD},d}\mathbf{Y}_{\text{DD},d}^H]
\nonumber\\&=\mathbb{E}[({\widehat{\mathbf{H}}}_{\text{DD}} \mathbb{E}(\mathbf{X}_d \mathbf{X}_d^H){\widehat{\mathbf{H}}}_{\text{DD}}^H]  +\mathbb{E}[{\mathbf{W}}_{\text{DD},d}{\mathbf{W}}_{\text{DD},d}^H]\nonumber\\
&=\sigma _d^2\mathbb{E}[{\widehat{\mathbf{H}}}_{\text{DD}}{\widehat{\mathbf{H}}}_{\text{DD}}^H] +\sigma^2 ( \mathbf{P}_{\text{rx}} \mathbf{P}_{\text{rx}}^H)
\nonumber\\&=\sigma _d^2({\widehat{\mathbf{H}}}_{\text{DD}}^{(j)}{\widehat{\mathbf{H}}}_{\text{DD}}^{(j) H} + \boldsymbol{\Xi}^{(j)}) +\sigma^2 [( \mathbf{P}_{\text{rx}} \mathbf{P}_{\text{rx}}^H)],
\end{align}
where the data symbols $\mathbf{X}_{d}$, noise $\mathbf{W}_{\text{DD},d}$ and the CSI estimate $\widehat{\mathbf{H}}_\text{DD}$ are independent of each other. Therefore, the MMSE detector for the input $\mathbf{X}_d$ is given by 
\begin{align}
\mathbf{\widehat{\mathbf{X}}}_d^{(j)}=\widehat{\mathbf{H}}_\text{DD}^{(j) H}\left[{\widehat{\mathbf{H}}}_{\text{DD}}^{(j)}{\widehat{\mathbf{H}}}_{\text{DD}}^{(j) H} + \boldsymbol{\Xi}^{(j)} +\frac{\sigma^2}{\sigma _d^2}\left( \mathbf{P}_{\text{rx}} \mathbf{P}_{\text{rx}}^H  \right) \right]^{-1}\mathbf{Y}_{\text{DD},d}.
\end{align}
For a rectangular pulse $\mathbf{P}_{rx}=\mathbf{P}_{tx}=\mathbf{I}_M$. The above expression can be simplified as 
\begin{align*}
\mathbf{\widehat{\mathbf{X}}}_d^{(j)}=\widehat{\mathbf{H}}_\text{DD}^{(j) H}\left[{\widehat{\mathbf{H}}}_{\text{DD}}^{(j)}{\widehat{\mathbf{H}}}_{\text{DD}}^{(j) H} + \boldsymbol{\Xi}^{(j)} +\frac{\sigma^2}{\sigma _d^2}\mathbf{I}_M\right]^{-1}\mathbf{Y}_{\text{DD},d}.
\end{align*}
\section{Calculation of \texorpdfstring{$\boldsymbol{\Xi}$}{} for an AP-SIP OTFS systems}
\label{appendeix_B}
Consider SISO OTFS system, the quantity $\mathbb {E} \lbrace \mathbf {H}_\text{DD}^H \mathbf {H}_\text{DD}\rbrace$ can be expressed as
$
\mathbb {E}\{ \mathbf {H}_\text{DD}^H \mathbf {H}_\text{DD}\} = \widehat{\mathbf {H}}_\text{DD}^H \widehat{\mathbf {H}}_\text{DD}+\boldsymbol{\Xi}
$,
where the matrix $\boldsymbol{\Xi}$  is calculated as follows. Let $\mathbf{h}_\text{DD} \in \mathbb {C}^{M^2 \times 1}$ denote the vectorized equivalent channel defined as $\mathbf {h}_\text{DD}= \mathrm{vec} (\mathbf {H}_\text{DD})$, which can be further simplified as
\begin{align}
    \mathbf{h}_\text{DD}&= \mathrm{vec}\left[\sum_{i=0}^{M_\tau-1}\sum_{J=0}^{G_\nu-1}\mathbf{P}_\text{rx} {h_{i,j}{(\bar{\mathbf{\Pi}}})^i{({\bar{\mathbf{\Delta}}}}_i)^j}\mathbf{P}_\text{tx}\right]
\nonumber\\&= \left[\sum_{i=0}^{M_\tau-1}\sum_{J=0}^{G_\nu-1}{h_{i,j}}\mathbf{\varphi}_i^j\right],
\label{eq:h_dd_NMSE}
\end{align}
where $\mathbf{\varphi}_i^j=\mathrm{vec}\left[\mathbf{P}_\text{rx}({\bar{\mathbf{\Pi}}})^i{({\bar{\mathbf{\Delta}}}_i)^j}\mathbf{P}_\text{tx}\right] \in \mathbb{C}^{M^2 \times 1}$.
The above equation can be recast as $\mathbf{h}_\text{DD}=\boldsymbol{\zeta}\mathbf{h}$, where ${\boldsymbol{\zeta}} \in \mathbb {C}^{M^2\times M_\tau G\nu}$ and is given by
\begin{equation}
 \boldsymbol{\zeta}=\left[\mathbf{\varphi}_{0}^0, \hdots, \mathbf{\varphi}_{0}^{N_\nu-1}, \hdots, \mathbf{\varphi}_{M_\tau-1}^0, \hdots, \mathbf{\varphi}_{M_\tau-1}^{G_\nu-1}\right],
 \label{eq:zeta}
 \end{equation}
 where $\mathbf{h}$ is the sparse channel coefficient vector given by \eqref{eq:sparse_channel_coeeficient_vector}. Let the estimate ${\mathbf{h}}_\text{DD}$ be denoted by $\widehat{\mathbf {h}}_\text{DD}= \mathrm{vec}(\widehat{\mathbf {H}}_\text{DD})$, and the estimate of $\mathbf{h}$ by $\widehat{\mathbf{h}}$. It is readily seen that $\widehat{\mathbf{h}}_\text{DD}=\boldsymbol{\zeta}\widehat{\mathbf{h}}$. The associated error covariance matrix $\boldsymbol{\Sigma }_{h}\in \mathbb {C}^{M^2 \times M^2}$ can be formulated as
$
\boldsymbol{\Sigma }_{h}= {\boldsymbol{\zeta }} \boldsymbol{\Sigma} {\boldsymbol{\zeta}}^H,
$
where $\boldsymbol{\Sigma}\in \mathbb{C}^{M_\tau G_\nu \times M_\tau G_\nu}$ is the error covariance matrix of $\widehat{\mathbf{h}}$. Furthermore, it can be shown that for matrix $\boldsymbol{\Xi}$, element at (p,q) is given by\\
\begin{align*}\boldsymbol{\Xi }(p,q)= \mathrm{Tr}\bigg[ \boldsymbol{\Sigma }_{h} (\tilde{p}-M+1: \tilde{p},\tilde{q}-M+1: \tilde{q}) \bigg].\end{align*}

%\section{Calculation of the Matrix $\boldsymbol{\Xi}$ for data aided AP-SIP MIMO OTFS system}
Consider MIMO OTFS system, the quantity $\mathbb {E} \lbrace \widetilde{\mathbf {H}}^H \widetilde{\mathbf {H}}\rbrace$ can be expressed as
$\mathbb {E} \{ \widetilde{\mathbf {H}}^H \widetilde{\mathbf {H}}\} = \widehat{\mathbf {H}}^H \widehat{\mathbf {H}}+\boldsymbol{\Xi}$,
where the matrix $\boldsymbol{\Xi}$  is calculated as follows. Let $\mathbf{h}_{r,t}^\text{DD} \in \mathbb {C}^{M^2 \times 1}$ denote the vectorized equivalent channel corresponding to the $r$th RA and $t$th TA, which is defined as 
\begin{align}
\mathbf {h}_{r,t}^\text{DD}&= \mathrm{vec} (\mathbf {H}_{r,t}^\text{DD}) =\mathrm{vec}\left[\sum_{i=0}^{M_\tau-1}\sum_{J=0}^{G_\nu-1} {h_{i,j,r,t}\mathbf{P}_\text{rx}({\bar{\mathbf{\Pi}}})^i{({\bar{\mathbf{\Delta}}}}_i)^j}\mathbf{P}_\text{tx}\right] \nonumber\\&= \left[\sum_{i=0}^{M_\tau-1}\sum_{J=0}^{G_\nu-1}{h_{i,j,r,t}}\boldsymbol{\varphi}_i^j\right],
\end{align}
where $\mathbf{\varphi}_i^j=\mathrm{vec}\left[\mathbf{P}_\text{rx}({\bar{\mathbf{\Pi}}})^i{({\bar{\mathbf{\Delta}}}_i)^j}\mathbf{P}_\text{tx}\right] \in \mathbb{C}^{M^2 \times 1}$. The above equation can be expressed as $\mathbf{h}_{r,t}^\text{DD}=\boldsymbol{\zeta}\mathbf{h}_{r,t}$, where ${\boldsymbol{\zeta}} \in \mathbb {C}^{M^2\times M_\tau G\nu}$ is given by \eqref{eq:zeta}
and $\mathbf{h}_{r,t}$ is the sparse channel coefficient vector given by \eqref{eq:sparse_channel_coefficient_vector_MIMO_r_t}.
Vectorizing $\widetilde{\mathbf{H}}_\text{DD}$, one obtains $\widetilde{\mathbf{h}}_\text{DD} = \mathrm{vec}(\widetilde{\mathbf{H}}_\text{DD})=[(\mathrm{vec} (\mathbf{H}_{1,1}^\text{DD}))^T\hdots (\mathrm{vec}(\mathbf{H}_{N_r,N_t}^\text{DD}))^T]^T=\big[(\boldsymbol{\zeta}\mathbf{h}_{1,1})^T, (\boldsymbol{\zeta}\mathbf{h}_{1,2})^T\hdots(\boldsymbol{\zeta}\mathbf{h}_{N_r,N_t})^T\big]^T$, which can be further simplified as
\begin{align} 
   \widetilde{\mathbf{h}}_\text{DD} &= ({\mathbf{I}_{N_r N_t}}
     \otimes {\boldsymbol{\zeta}})\left[(\mathbf{h}_{1,1})^T, (\mathbf{h}_{1,2})^T\hdots(\mathbf{h}_{N_r,N_t})^T\right]^T \nonumber\\&=({\mathbf{I}_{N_r N_t}}
     \otimes {\boldsymbol{\zeta}})
    \widetilde{\mathbf{h}}
    \label{eq:MIMO_hdd_h}
\end{align}
The estimate of $\widetilde{\mathbf{h}}_\text{DD}$, denoted by $\widehat{\mathbf{h}}_\text{DD}$, is given as $\widehat{\mathbf {h}}_\text{DD}= \mathrm{vec}(\widehat{\mathbf {H}}_\text{DD})$. Let the estimate of $\widetilde{\mathbf{h}}$ be $\widehat{\mathbf {h}}$. The relationship between $\widehat{\mathbf{h}}_\text{DD}$ and $\widehat{\mathbf{h}}$ is given by $\widehat{\mathbf{h}}_\text{DD}=({\mathbf{I}_{N_r N_t}}
     \otimes {\boldsymbol{\zeta}})\widehat{\mathbf{h}}$ with the associated error covariance matrix $\boldsymbol{\Sigma }_{h}\in \mathbb {C}^{M^2 N_r N_t \times M^2 N_r N_t}$ determined as
$
\boldsymbol{\Sigma }_{h}= ({\mathbf{I}_{N_r N_t}}
     \otimes {\boldsymbol{\zeta}}) ({\mathbf{I}_{N_r}}
     \otimes\boldsymbol{\Sigma}) ({\mathbf{I}_{N_r N_t}}
     \otimes {\boldsymbol{\zeta}})^H,
$
where $\boldsymbol{\Sigma}\in \mathbb{C}^{M_\tau G_\nu N_t \times M_\tau G_\nu N_t}$ is the error covariance matrix obtained from the E-step of Algorithm-4. It can be shown that for matrix $\boldsymbol{\Xi}$, element at (p,q) is given as
$\boldsymbol{\Xi }(p,q)= \mathrm{Tr}\left[ \boldsymbol{\Sigma }_{h} (\tilde{p}-MN_r+1: \tilde{p},\tilde{q}-MN_t+1: \tilde{q}) \right],$
where $\tilde{p}=pMN_r$, and $\tilde{q}=qMN_t$.
\bibliographystyle{IEEEtran} % Choose the bibliography style you prefer
\bibliography{IEEEabrv} % Replace "your-bib-file" with the actual name of your .bib file (without the extension)

\begin{thebibliography}{10}
\providecommand{\url}[1]{#1}
\csname url@samestyle\endcsname
\providecommand{\newblock}{\relax}
\providecommand{\bibinfo}[2]{#2}
\providecommand{\BIBentrySTDinterwordspacing}{\spaceskip=0pt\relax}
\providecommand{\BIBentryALTinterwordstretchfactor}{4}
\providecommand{\BIBentryALTinterwordspacing}{\spaceskip=\fontdimen2\font plus
\BIBentryALTinterwordstretchfactor\fontdimen3\font minus
  \fontdimen4\font\relax}
\providecommand{\BIBforeignlanguage}[2]{{%
\expandafter\ifx\csname l@#1\endcsname\relax
\typeout{** WARNING: IEEEtran.bst: No hyphenation pattern has been}%
\typeout{** loaded for the language `#1'. Using the pattern for}%
\typeout{** the default language instead.}%
\else
\language=\csname l@#1\endcsname
\fi
#2}}
\providecommand{\BIBdecl}{\relax}
\BIBdecl

\bibitem{high_speed1}
F.~Hasegawa, A.~Taira, G.~Noh, B.~Hui, H.~Nishimoto, A.~Okazaki, A.~Okamura,
  J.~Lee, and I.~Kim, ``High-speed train communications standardization in
  {3GPP 5G NR},'' \emph{IEEE Communications Standards Magazine}, vol.~2, no.~1,
  pp. 44--52, 2018.

\bibitem{high_speed2}
Y.~Liu, C.-X. Wang, and J.~Huang, ``Recent developments and future challenges
  in channel measurements and models for {5G} and beyond high-speed train
  communication systems,'' \emph{IEEE Communications Magazine}, vol.~57, no.~9,
  pp. 50--56, 2019.

\bibitem{high_speed3}
J.~Zhang, T.~Chen, S.~Zhong, J.~Wang, W.~Zhang, X.~Zuo, R.~G. Maunder, and
  L.~Hanzo, ``Aeronautical $ad~hoc$ networking for the
  {Internet-Above-the-Clouds},'' \emph{Proceedings of the IEEE}, vol. 107,
  no.~5, pp. 868--911, 2019.

\bibitem{hadani2017orthogonal}
R.~Hadani, S.~Rakib, M.~Tsatsanis, A.~Monk, A.~J. Goldsmith, A.~F. Molisch, and
  R.~Calderbank, ``Orthogonal time frequency space modulation,'' in \emph{2017
  IEEE Wireless Communications and Networking Conference (WCNC)}.\hskip 1em
  plus 0.5em minus 0.4em\relax IEEE, 2017, pp. 1--6.

\bibitem{application2019otfs_radar}
P.~Raviteja, K.~T. Phan, Y.~Hong, and E.~Viterbo, ``Orthogonal time frequency
  space ({OTFS}) modulation based radar system,'' in \emph{2019 IEEE Radar
  Conference (RadarConf)}, 2019, pp. 1--6.

\bibitem{application2021otfs_iot}
Y.~Ma, G.~Ma, N.~Wang, Z.~Zhong, and B.~Ai, ``{OTFS-TSMA for Massive Internet
  of Things in High-Speed Railway},'' \emph{IEEE Transactions on Wireless
  Communications}, pp. 1--1, 2021.

\bibitem{applicationotfs_v2x}
J.~Cheng, C.~Jia, H.~Gao, W.~Xu, and Z.~Bie, ``{OTFS} based receiver scheme
  with multi-antennas in high-mobility {V2X} systems,'' in \emph{2020 IEEE
  International Conference on Communications Workshops (ICC Workshops)}, 2020,
  pp. 1--6.

\bibitem{monk2016otfs}
A.~Monk, R.~Hadani, M.~Tsatsanis, and S.~Rakib, ``{OTFS} - orthogonal time
  frequency space,'' 2016.

\bibitem{mehrotra_anand}
A.~Mehrotra, R.~K. Singh, S.~Srivastava, and A.~K. Jagannatham, ``{Channel
  Estimation Techniques for CP-Aided OTFS Systems Relying on Practical Pulse
  Shapes},'' in \emph{2022 IEEE International Conference on Signal Processing
  and Communications (SPCOM)}, 2022, pp. 1--5.

\bibitem{Ramachandran2020OTFSAN}
M.~K. Ramachandran, G.~D. Surabhi, and A.~Chockalingam, ``{OTFS}: {A} new
  modulation scheme for high-mobility use cases,'' \emph{Journal of the Indian
  Institute of Science}, vol. 100, pp. 315 -- 336, 2020.

\bibitem{MIMO_OTFS_CHOCKALINGAM}
M.~Kollengode~Ramachandran and A.~Chockalingam, ``{MIMO-OTFS in High-Doppler
  Fading Channels: Signal Detection and Channel Estimation},'' in \emph{2018
  IEEE Global Communications Conference (GLOBECOM)}, 2018, pp. 206--212.

\bibitem{raviteja2018embedded}
P.~Raviteja, K.~T. Phan, Y.~Hong, and E.~Viterbo, ``Embedded delay-{Doppler}
  channel estimation for orthogonal time frequency space modulation,'' in
  \emph{2018 IEEE 88th Vehicular Technology Conference (VTC-Fall)}.\hskip 1em
  plus 0.5em minus 0.4em\relax IEEE, 2018, pp. 1--5.

\bibitem{shan_otfs_antenna_array}
Y.~Shan and F.~Wang, ``Low-complexity and low-overhead receiver for {OTFS} via
  large-scale antenna array,'' \emph{IEEE Transactions on Vehicular
  Technology}, vol.~70, no.~6, pp. 5703--5718, 2021.

\bibitem{prem2021superimposed}
H.~B. Mishra, P.~Singh, A.~K. Prasad, and R.~Budhiraja, ``Iterative channel
  estimation and data detection in {OTFS} using superimposed pilots,'' in
  \emph{2021 IEEE International Conference on Communications Workshops (ICC
  Workshops)}, 2021.

\bibitem{yuan2021superimposed}
W.~Yuan, S.~Li, Z.~Wei, J.~Yuan, and D.~W.~K. Ng, ``Data-aided channel
  estimation for {OTFS} systems with a superimposed pilot and data transmission
  scheme,'' \emph{IEEE Wireless Communications Letters}, vol.~10, no.~9, pp.
  1954--1958, 2021.

\bibitem{data_aided_OFDM}
I.~Khan, A.~Zaib, S.~Khattak, and S.~Azmat, ``Data aided channel estimation for
  ofdm wireless systems using reliable carriers,'' in \emph{2019 International
  Conference on Electrical, Communication, and Computer Engineering (ICECCE)},
  2019.

\bibitem{suraj2021bayesian}
S.~Srivastava, R.~K. Singh, A.~K. Jagannatham, and L.~Hanzo, ``Bayesian
  learning aided sparse channel estimation for orthogonal time frequency space
  modulated systems,'' \emph{IEEE Transactions on Vehicular Technology},
  vol.~70, no.~8, pp. 8343--8348, 2021.

\bibitem{suraj2021row_group}
------, ``Bayesian learning aided simultaneous row and group sparse channel
  estimation in orthogonal time frequency space modulated {MIMO} systems,''
  \emph{IEEE Transactions on Communications}, pp. 1--1, 2021.

\bibitem{Data_aided_MIMO_OFDM}
S.~Srivastava, J.~Nath, and A.~K. Jagannatham, ``Data aided quasistatic and
  doubly-selective csi estimation using affine-precoded superimposed pilots in
  millimeter wave mimo-ofdm systems,'' \emph{IEEE Transactions on Vehicular
  Technology}, vol.~70, no.~7, pp. 6983--6998, 2021.

\bibitem{affine_precoding_MIMO}
N.~N. Tran, D.~H. Pham, H.~D. Tuan, and H.~H. Nguyen, ``Orthogonal affine
  precoding and decoding for channel estimation and source detection in {MIMO}
  frequency-selective fading channels,'' \emph{IEEE Transactions on Signal
  Processing}, vol.~57, no.~3, pp. 1151--1162, 2009.

\bibitem{superimposed_affine_precoding}
N.~N. Tran, H.~D. Tuan, and H.~H. Nguyen, ``Superimposed training designs for
  spatially correlated {MIMO-OFDM} systems,'' in \emph{IEEE GLOBECOM 2008 -
  2008 IEEE Global Telecommunications Conference}, 2008, pp. 1--6.

\bibitem{ofdmbasedotfs2018}
A.~Farhang, A.~RezazadehReyhani, L.~E. Doyle, and B.~Farhang-Boroujeny, ``Low
  complexity modem structure for {OFDM}-based orthogonal time frequency space
  modulation,'' \emph{IEEE Wireless Communications Letters}, vol.~7, no.~3, pp.
  344--347, 2018.

\bibitem{Ahmad2018MIMO_OFDMbasedOTFS}
A.~RezazadehReyhani, A.~Farhang, M.~Ji, R.~R. Chen, and B.~Farhang-Boroujeny,
  ``Analysis of discrete-time {MIMO OFDM}-based orthogonal time frequency space
  modulation,'' in \emph{2018 IEEE International Conference on Communications
  (ICC)}, 2018, pp. 1--6.

\bibitem{ai_chokalingam}
S.~R. Mattu and A.~Chockalingam, ``{Learning based Delay-Doppler Channel
  Estimation with Interleaved Pilots in OTFS},'' in \emph{2022 IEEE 96th
  Vehicular Technology Conference (VTC2022-Fall)}, 2022, pp. 1--6.

\bibitem{ai2}
X.~Zhang, W.~Yuan, C.~Liu, F.~Liu, and M.~Wen, ``{Deep Learning with a
  Self-Adaptive Threshold for OTFS Channel Estimation},'' in \emph{2022
  International Symposium on Wireless Communication Systems (ISWCS)}, 2022, pp.
  1--5.

\bibitem{ai3}
------, ``{Deep Learning with a Self-Adaptive Threshold for OTFS Channel
  Estimation},'' in \emph{2022 International Symposium on Wireless
  Communication Systems (ISWCS)}, 2022, pp. 1--5.

\bibitem{raviteja2018practical}
P.~Raviteja, Y.~Hong, E.~Viterbo, and E.~Biglieri, ``Practical pulse-shaping
  waveforms for reduced-cyclic-prefix {OTFS},'' \emph{IEEE Transactions on
  Vehicular Technology}, vol.~68, no.~1, pp. 957--961, 2018.

\bibitem{raviteja2018interference}
P.~Raviteja, K.~T. Phan, Y.~Hong, and E.~Viterbo, ``Interference cancellation
  and iterative detection for orthogonal time frequency space modulation,''
  \emph{IEEE Transactions on Wireless Communications}, vol.~17, no.~10, pp.
  6501--6515, 2018.

\bibitem{ramachandran2020otfs}
M.~Ramachandran, G.~Surabhi, and A.~Chockalingam, ``{OTFS}: A {N}ew
  {M}odulation scheme for {H}igh-{M}obility {U}se {C}ases,'' \emph{Journal of
  the Indian Institute of Science}, pp. 1--22, 2020.

\bibitem{Hanzo_book}
D.~Greenwood and L.~Hanzo, ``Characterisation of mobile radio channels: 2nd,''
  in \emph{Modern Mobile Radio Communications: Second and Third Generation
  Cellular and WATM Systems}, R.~Steele and L.~Hanzo, Eds.\hskip 1em plus 0.5em
  minus 0.4em\relax John Wiley \& Sons, 1999, pp. 91--185, chapter: 2 Address:
  Chichester, UK.

\bibitem{tse_viswanath}
D.~Tse and P.~Viswanath, \emph{Fundamentals of Wireless Communication}.\hskip
  1em plus 0.5em minus 0.4em\relax Cambridge University Press, 2005.

\bibitem{raviteja_embedded}
P.~Raviteja, K.~T. Phan, and Y.~Hong, ``Embedded pilot-aided channel estimation
  for otfs in delay–doppler channels,'' \emph{IEEE Transactions on Vehicular
  Technology}, vol.~68, no.~5, pp. 4906--4917, 2019.

\bibitem{Proakis2007}
Proakis, \emph{Digital Communications 5th Edition}.\hskip 1em plus 0.5em minus
  0.4em\relax McGraw Hill, 2007.

\bibitem{cramer_bound_book}
H.~L. V. T. K.~L. Bell, \emph{''Bayesian Bounds for Parameter Estimation and
  Nonlinear Filtering/Tracking''}.\hskip 1em plus 0.5em minus 0.4em\relax Wiley
  IEEE press, 2007.

\bibitem{EVA}
\BIBentryALTinterwordspacing
{3rd Generation Partnership Project; Technical Specification Group Radio Access
  Network}, ``{{Evolved Universal Terrestrial Radio Access (E-UTRA); User
  Equipment (UE) Radio Transmission and Reception}},'' {3GPP}, {Technical
  Specification} {36.101}, 2012. [Online]. Available:
  \url{{https://www.3gpp.org}}
\BIBentrySTDinterwordspacing

\bibitem{OMP}
T.~T. Cai and L.~Wang, ``Orthogonal matching pursuit for sparse signal recovery
  with noise,'' \emph{IEEE Transactions on Information Theory}, vol.~57, no.~7,
  pp. 4680--4688, 2011.

\bibitem{FOCUSS}
I.~Gorodnitsky and B.~Rao, ``Sparse signal reconstruction from limited data
  using {FOCUSS}: a re-weighted minimum norm algorithm,'' \emph{IEEE
  Transactions on Signal Processing}, vol.~45, no.~3, pp. 600--616, 1997.

\end{thebibliography}
%\bibliographystyle{IEEEtran}
%\bibliography{IEEEabrv.bib}
\end{document}